\documentclass[english]{iopart}
\usepackage[T1]{fontenc}
\usepackage[latin9]{inputenc}
\usepackage{geometry}
\usepackage{babel}
\usepackage{graphicx}
\usepackage{esint}
\usepackage[unicode=true,pdfusetitle,
 bookmarks=true,bookmarksnumbered=false,bookmarksopen=false,
 breaklinks=false,pdfborder={0 0 1},backref=false,colorlinks=false]
 {hyperref}

\makeatletter

\providecommand{\tabularnewline}{\\}

\usepackage{iopams}
\usepackage{setstack}

\usepackage{graphicx,color,curves,epic}

\makeatother

\begin{document}
\global\long\def\i{\imath}

\global\long\def\plll{\mathcal{P}_{\mathrm{LLL}}}

\global\long\def\TT{\mathrm{TT}}

\global\long\def\T{\mathcal{T}}

\global\long\def\S{\mathcal{S}}

\global\long\def\Td{\mathcal{T}\mbox{-}}

\global\long\def\Sd{\mathcal{S}\mbox{-}}

\global\long\def\elliptic#1#2#3{\vartheta_{#1}\left(\left.#2\vphantom{#3}\right|#3\right)}

\global\long\def\ellipticweight#1#2{\kappa\left[\left.#1\vphantom{#2}\right|#2\right]}

\global\long\def\shortelliptic#1#2{\vartheta_{#1}\left(#2\right)}

\global\long\def\ellipticgeneralized#1#2#3#4{\vartheta\left[\begin{array}{c}
 #1\\
 #2 
\end{array}\right]\left(\left.#3\vphantom{#4}\right|#4\right)}

\global\long\def\ket#1{\left|#1\right\rangle }

\global\long\def\bra#1{\left\langle #1\right|}

\global\long\def\braket#1#2{\left\langle #1\left|\vphantom{#1}#2\right.\right\rangle }

\global\long\def\ketbra#1#2{\left|#1\vphantom{#2}\right\rangle \left\langle \vphantom{#1}#2\right|}

\global\long\def\braOket#1#2#3{\left\langle #1\left|\vphantom{#1#3}#2\right|#3\right\rangle }

\global\long\def\mbf#1{\mathbf{#1}}

\newcommand{\ie}[0]{\emph{i.e.}\@ }

\newcommand{\eg}[0]{\emph{e.g.}\@ }

\title{Success and failure of the plasma analogy\\for Laughlin states on a torus} 

\author{Mikael Fremling}

\address{Department of Physics, Stockholm University, SE-106 91 Stockholm, Sweden} 

\address{Department of Mathematical Physics, National University of Ireland, Maynooth, Ireland} 

\ead{mikael.fremling@nuim.ie} 
\begin{abstract}
We investigate the nature of the plasma analogy for the Laughlin wave
function on a torus describing the quantum Hall plateau at $\nu=\frac{1}{q}$.
We first establish, as expected, that the plasma is screening if there
are no short nontrivial paths around the torus. We also find that
when one of the handles has a short circumference -- \ie the thin-torus
limit -- the plasma no longer screens. To quantify this we compute
the normalization of the Laughlin state, both numerically and analytically.
In the thin torus limit, the analytical form of the normalization
simplify and we can reconstruct the normalization and analytically
extend it back into the 2D regime. We find that there are geometry
dependent corrections to the normalization, and this in turn implies
that the plasma in the plasma analogy is not screening when in the
thin torus limit. Despite the breaking of the plasma analogy in this
limit, the analytical approximation is still a good description of
the normalization for all tori, and also allows us to compute hall
viscosity at intermediate thickness.
\end{abstract}
\maketitle

\section{Introduction}

The Laughlin wave function is the drosophila of representative trial
wave functions for the fractional quantum Hall effect (FQHE). It was
introduced in a planar geometry by Laughlin\cite{Laughlin_1983},
generalized to the sphere by Haldane\cite{Haldane_1983}, to the torus
by Haldane and Rezayi\cite{Haldane_1985a}, and has has been extensively
studied on these geometries ever since. Most remarkably, its elementary
excitations -- its quasi-particles -- have fractional charges\cite{Laughlin_1983},
and are believed to obey anyonic statistics\cite{Arovas_1984}. The
latter, although not proven experimentally, is strongly suggested
by extensive numerical studies\cite{Kjonsberg_1998,Kjonsberg_1999}.
Recently, matrix product states have been successfully used to probe
the anyonic statistics\cite{Zaletel_2013,Cincio_2013}.

The theoretical argument for fractional statistics is based on the
so called plasma analogy, first introduced by Laughlin\cite{Laughlin_1983}.
The main observation is that, introducing a particular factor dependent
on the quasi-particle positions into the normalization of the wave
functions, the corresponding full normalization constant can be expressed
as the partition function for a classical screening plasma. In this
analogy the quasi-particles, at positions $\eta_{i}$, appear as test
charges with a strength one $q^{\mathrm{th}}$ of the electron charge.
As the plasma is screening, the partition function does not depend
on the positions of the quasi-particles, provided they are separated
much further than the screening length. From this, one concludes that
the full $\eta_{i}$-dependence of the normalization is accounted
for, and from this it is straightforward to calculate the fractional
statistics parameter.

Another way of saying this, is that the \textquotedbl{}holonomy\textquotedbl{}
\ie the phase related to adiabatically dragging one quasi-particle
at $\eta_{1}$ around another at $\eta_{2}$ will equal the \textquotedbl{}monodromy\textquotedbl{},
\ie the explicit phase obtained from the normalization factor. For
this to occur, no additional contribution from any Berry phase related
to the adiabatic change must be accumulated.

Since screening is essential for the argument of vanishing extra Berry
phases, we may ask the contrary question: When does the plasma not
screen? Can the screening properties of the plasma change if the quasi-particles
move too close to the edge of the sample or if there are non-trivial
loops on the spatial surface with short length scales\cite{Kjonsberg_1998,Kjonsberg_1999}?
A related question that is also important for the Laughlin states
is the following: How does the free energy of the plasma depend on
the topology, geometry and size of the surface on which the QH liquid
resides? This is particularly interesting in the case of a \textit{torus}
where the normalization can depend on the aspect ratio of the two
principal axes, $\tau$. $\tau$ is a complex valued parameter that
can be changed while still keeping the density and number of electrons
fixed. In the context of the plasma analogy, this corresponds to asking
if the plasma is screening for all $\tau$.

Recently the $\tau\mbox{-}$dependence of the normalized Laughlin
wave function, $\psi_{L}$, has been determined using techniques from
conformal field theory\cite{Read_2009}. Later numerical work\cite{Read_2011}
confirmed that there where no extra $\tau\mbox{-}$dependent contributions
to this normalization. The numerical analysis was however only performed
for special values of $\tau$, so the question remains if there is
need for corrections when considering general $\tau$.

It is fruitful to think of $\tau$ as a parameter that can be adiabatically
changed, in the same way as the quasi-particle positions can be adiabatically
dragged around each other. Under such an adiabatic change analogous
questions appear as for the quasi-particles, \ie is the Berry phase
the same as the monodromy of changing $\tau$. In fact, for the Laughlin
state this computation can be considered even more interesting as
changing $\tau$ can transform the $q$ degenerate Laughlin states
into each other, giving rise to Berry matrices. These matrices, which
are well defined even without the presence of quasi-particles, would
form a non-abelian representation of the modular group, the mapping
class group of the torus, while exchanges of quasi-holes only give
rise to phase factors.

In this paper we study the normalization properties of $\psi_{L}$,
with no quasi-particles present. For general $\tau$ we obtain the
normalization as a finite -- but intractably large -- sum in powers
of $e^{\i\pi\tau}$. We expand $\psi_{L}$ in $\tau$ by rewriting
it in a basis of single particle orbitals \ie a Fock basis. Progress
has been made on the plane and sphere by identifying the Laughlin
states with Jack polynomials\cite{Bernevig_2008}. The Jack polynomials
can be recursively calculated\cite{Lapointe_2000} by starting with
the root partition in the orbital occupation basis and applying squeezing
rules. The root partition is the configuration of electrons where
the distance between all occupied orbitals is maximized. For the Laughlin
state at $\nu=\frac{1}{3}$, the root partition is the orbital occupation
pattern $\ldots1001001001\ldots$.

On the torus, no known squeezing rules exist and the Laughlin state
can not be written in terms of Jack polynomials. On the other hand,
there is a well defined limit, the Tau-Thouless (TT) limit\cite{Bergholtz_2008},
in which the Laughlin state is precisely a single slater determinant
corresponding to the root partition. From the TT-limit it is possible
to generate the Laughlin state at a general aspect ratio $\tau$ using
a differential equation in this parameter\cite{Zhou_2013}. The differential
operator acts on the Fock expansion of the Laughlin state. In the
TT-limit the analytical expression for the normalization also simplifies
greatly can be expanded perturbatively in powers of $e^{\i\pi\tau}$.

To find the Fock expansion of the wave function is useful since it
would allow for a direct computation of the normalization of the Laughlin
state as well as many other interesting quantities! It is conjectured
that for a torus large enough, the $\tau\mbox{-}$dependence of the
Laughlin state is fully captured by the normalization factor proposed
by Read\cite{Read_2009}. In that case, there would be no extra dependence
on the precise geometry of the torus, just as the normalization is
insensitive to the positions of well separated quasi-holes of the
Laughlin state\cite{Laughlin_1983}, if the factors mentioned before
are taken into account. 

For small system sizes the normalization of the Laughlin state can
be computed exactly for any value of $\tau$. The rough picture is
that when the torus is large enough in both directions then there
is no $\tau\mbox{-}$dependence in the normalization. However, if
one of the torus handles becomes too thin there is $\tau\mbox{-}$dependence
in the normalization. See \eg figure (\ref{fig:N_0_in_tau_plane})
in Section \ref{sec:LLL-on-Torus}. This is consistent with the intuitive
picture that at some length scale the screening properties of the
Laughlin plasma analogy break down.

In this paper we first analytically extract the Fock coefficients
for the Laughlin state in a generic torus geometry $\tau$. We obtain
a sum that is finite, but untractably large for practical purposes.
From the analytical expressions we approximate the leading behavior
of the Fock coefficients in the TT-limit, and using this leading order
expansion we comment on the validity of the plasma analogy in the
TT-limit and on the universality of the Hall viscosity. We find that
the plasma analogy is no longer screening when the circumference of
the torus becomes sufficiently small, and we compute the leading corrections
to the normalization in this limit. We conclude that although the
plasma is screening in the thermodynamic limit, it will always cease
to to screen if there is a loop around the torus shorter than some
characteristic length scale. Intuitively, this distance, which we
find to be roughly $6$ magnetic lengths, should be on par with the
screening length of the plasma.

This paper is organized as follows. Section \ref{sec:LLL-on-Torus}
contains a brief summary of relevant notation. In Section \ref{sec:Laughlin-Expansion}
we introduce the analytical Fock expansion for the Laughlin state
on the torus. In Section \ref{sec:TT-limit-analysis} we use the Fock
expansion to study the Laughlin state in the TT-limit. We address
both the proper norm in the TT-limit as well as extrapolate back to
the thick torus. We also show that the quantum Hall viscosity of the
Laughlin state is different in the TT-limit than in the thermodynamic
limit. Finally, in Section \ref{sec:Hierachy-expand} we sketch the
route to obtain a Fock expansion for the chiral Haldane-Halperin hierarchy
states\cite{Haldane_1983,Halperin_1983} constructed using CFT techniques\cite{Fremling_2014}.

\section{Lowest Landau Level Wave Functions\label{sec:LLL-on-Torus}}

For self consistency, but also to define our notation, we give the
basic formula for the lowest Landau level (LLL) wave functions and
translation operators on the torus.

\begin{figure}
\begin{centering}
\setlength{\unitlength}{1.5cm}
\begin{picture}(4,3)(0,-.5)
\put(1.2,1.2){\huge{$\tau=\frac{L_\Delta+\i L_y}{L_x}$}}
%%Axes
\put(1.0,-0.5){\huge{$L_x=L$}}
\put(-0.6,1){\huge{$L_y$}}
\put(.2,2.3){\huge{$L_\Delta$}}
\put(0,0){\line(1,0){3}}
\put(0,0){\line(1,2){1}}
\put(3,0){\line(1,2){1}}
\put(1,2){\line(1,0){3}}
\dashline{0.2}(0,0)(0,2)
\dashline{0.2}(0,2)(1,2)
%%%Unit vetors
\put(3.6,.2){\vector(1,0){.5}}
\put(3.6,.2){\vector(0,1){.5}}
\put(4.2,.1){$\tilde x$}
\put(3.6,.8){$\tilde y$}
\put(0.2,0.1){\vector(1,0){.5}}
\put(0.2,0.1){\vector(1,2){.333}}
\put(0.8,0.05){$x$}
\put(0.6,0.7){$y$}
\end{picture}
\par\end{centering}

\caption{The relationship between the Cartesian coordinates $\left(\tilde{x},\tilde{y}\right)$
and the dimensionless coordinates $\left(x,y\right)$. In the figure
one can also see that $\tau_{1}=\frac{L_{\Delta}}{L_{x}}$ is interpreted
as the skewness, and $\tau_{2}=\frac{L_{y}}{L_{x}}$ as the aspect
ratio, of the torus. The area of the torus is fixed to be $L_{x}L_{y}=L^{2}\tau_{2}=2\pi N_{\phi}\ell_{B}$.\label{fig:Geometry}}
\end{figure}
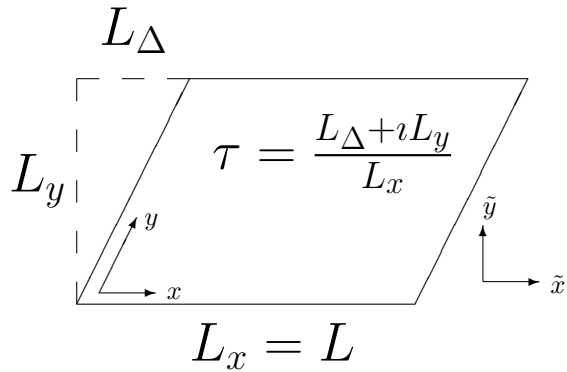

In this paper we will exclusively work in the dimensionless coordinates
where $0\leq x,y<1$, defined on a unit square. They are related to
the dimensionful coordinate $z$ as $z=L\left(x+\tau y\right)$. See
Figure \ref{fig:Geometry}. The complex parameter $\tau=\tau_{1}+\i\tau_{2}$
parametrizes the torus geometry and is defined in the complex upper
half plane. 

The single particle Hamiltonian is the Landau Hamiltonian 
\begin{eqnarray*}
H & = & \frac{1}{2m}\frac{1}{L^{2}}\left(p_{x}-eA_{x}\right)^{2}+\frac{1}{2m\tau_{2}^{2}L^{2}}\left(p_{y}-\tau_{1}p_{x}+\tau_{1}A_{x}\right)^{2}.
\end{eqnarray*}
In these units the vector potential is $\left(A_{x},A_{y}\right)=\left(2\pi N_{\phi}\ell_{B}^{2}By,0\right)=\left(\tau_{2}L^{2}By,0\right)$.
We have here chosen to use, what we call the $\tau$-gauge, as in
the physical coordinates $z=\tilde{x}+\i\tilde{y}$ the vector potential
is perpendicular to $\vec{\tau}=\left(\tau_{1},\tau_{2}\right)$.
The physical area of the torus is $A=\tau_{2}L^{2}=2\pi N_{\phi}\ell_{B}^{2}$
and is penetrated by $N_{\phi}$ magnetic fluxes; we also define $\epsilon=\frac{1}{N_{\phi}}$.
We will throughout this paper let the magnetic length be $\ell_{B}=1$.
The Landau Hamiltonian can be brought to the form of a harmonic oscillator
$H=\hbar\omega\left(a^{\dagger}a+\frac{1}{2}\right)$ by introducing
ladder operators 

\begin{eqnarray*}
a & = & \sqrt{2}\left(\partial_{\bar{z}}+\frac{\tau}{2}Ly\right)\\
a^{\dagger} & = & -\sqrt{2}\left(\partial_{z}-\frac{\bar{\tau}}{2}Ly\right).
\end{eqnarray*}
The energy levels are called Landau levels (LL) and are all $N_{\phi}\mbox{-}$fold
degenerate. Acting with $a$ on a state in the lowest Landau level
(LLL) gives the equation $\left(\partial_{\bar{z}}+\frac{\tau}{2}Ly\right)\psi_{\mathrm{LLL}}=0$.
The solution to this equation shows that all LLL wave functions have
the form 
\[
\psi_{\mathrm{LLL}}=e^{\i\pi\tau N_{\phi}y^{2}}f\left(z\right),
\]
where $f\left(z\right)$ is a holomorphic function. 

The translation operators that commute with the Hamiltonian are finite
magnetic translation operators. There are two minimal translation
operators $t_{x}$ and $t_{y}$, that move a state a finite distance
along the principal axes of the torus, \ie\emph{ }in the directions
$x$ and $y$. The distance translated by $t_{x}$ is $\epsilon$
and so $t_{x}^{N_{\phi}}$ makes a full revolution around the torus.
The periodic boundary conditions are then defined through the equation
\[
t_{a}^{N_{\phi}}\ket{\psi}=e^{\i\phi_{a}}\ket{\psi},
\]
 where the $\phi_{a}$ can be thought of as fluxes threading the two
handles of the torus. The translation operators expressed in the $\tau$-gauge
are $t_{x}=e^{\epsilon\partial_{x}}$ and $t_{y}=e^{\epsilon\partial_{y}+\i2\pi x}$.
From these two minimal translations we construct the full set of translation
operators as 
\[
t_{m,n}=e^{\epsilon m\partial_{x}+\epsilon n\partial_{y}+\i2\pi nx}.
\]

The minimal translations are identified as $t_{x}=t_{1,0}$, $t_{y}=t_{0,1}$.
The operator $t_{m,n}$ moves a wave function a distance $z\to z+\epsilon L\left(m+\tau n\right)$
and has commutation relation $t_{m,n}t_{m^{\prime},n^{\prime}}=e^{\i2\pi\epsilon\left(mn^{\prime}-m^{\prime}n\right)}t_{m^{\prime},n^{\prime}}t_{m,n}$.

We choose to diagonalize the single particle wave functions in the
LLL with respect to $t_{x}$. For periodic boundary conditions ($\phi_{x}=\phi_{y}=0$)
the single particle basis states are 
\begin{eqnarray}
\eta_{k}\left(z\right) & = & \frac{1}{\sqrt{L\sqrt{\pi}}}e^{\i\pi\tau N_{\phi}y^{2}}\ellipticgeneralized{\epsilon k}0{N_{\phi}\frac{z}{L}}{N_{\phi}\tau}\label{eq:Basis_wave_function}\\
 & = & \sqrt{\frac{1}{\pi}\sqrt{\frac{\tau_{2}}{2N_{\phi}}}}\sum_{t\in\mathbb{Z}}e^{\i2\pi\left(N_{\phi}t+k\right)x}e^{\i\pi\tau N_{\phi}\left(y+t+\epsilon k\right)^{2}},\nonumber 
\end{eqnarray}
with properties $t_{x}\eta_{k}=e^{\i2\pi\epsilon k}$ and $t_{y}\eta_{k}=\eta_{k+1}$.
The eigenstates for generic boundary conditions are trivially obtained
by acting with $t_{-\frac{\phi_{y}}{2\pi},\frac{\phi_{x}}{2\pi}}$
on $\eta_{k}$. There are $N_{\phi}$ single particle orbitals, one
for each flux quantum. Since the LLL has a non-commutative geometry
there is a connection between the momentum $k$ and the expectation
value of the position in $y\mbox{-}$direction. This relation is simply
$\left\langle y\right\rangle =-\epsilon k$. As such, the momentum
label can alternatively be thought of as a physical displacement label
of a one-dimensional system. This view is especially fruitful in the
TT-limit where the width of the orbitals is much smaller than the
inter-orbital distance, $\sigma_{y}\ll\epsilon$. Here $\sigma_{y}=\sqrt{\left\langle y^{2}\right\rangle -\left\langle y\right\rangle ^{2}}$
is the standard deviation measure of uncertainty in the $y\mbox{-}$direction.

The $\vartheta\mbox{-}$function appearing in the above formula is
defined as 
\begin{equation}
\ellipticgeneralized abz{\tau}=\sum_{t=-\infty}^{\infty}e^{\i\pi\tau\left(t+a\right)^{2}}e^{\i2\pi\left(t+a\right)\left(z+b\right)},\label{eq:def_theta_function}
\end{equation}
 and is frequently encountered when considering wave functions on
the torus.

Note that, in (\ref{eq:Basis_wave_function}) due to the holomorphic
structure of $\psi_{\mathrm{LLL}}$, the exponentials linear in $x$
and $y$ are locked to form the object $e^{\i2\pi kz}$. We will utilize
this fact when extracting Fock coefficients for the Laughlin state
in the next section. For future convenience we also define the short
hand notation 
\begin{eqnarray}
\zeta_{k}\left(z\right) & = & e^{\i2\pi kx}e^{\i\pi\tau N_{\phi}\left(y+\frac{k}{N_{\phi}}\right)^{2}}\label{eq:Basis_fourier_terms}\\
 & = & e^{\i2\pi k\frac{z}{L}}e^{\i\pi\tau N_{\phi}y^{2}}e^{\i\pi\tau\frac{k^{2}}{N_{\phi}}},\nonumber 
\end{eqnarray}
 for the terms entering (\ref{eq:Basis_wave_function}). This allows
us to write the basis wave functions (\ref{eq:Basis_wave_function})
as 
\begin{equation}
\eta_{k}\left(z\right)=\frac{1}{\sqrt{L\sqrt{\pi}}}\sum_{t\in\mathbb{Z}}\zeta_{k+N_{\phi}t}\left(z\right).\label{eq:Basis_fourier_sum}
\end{equation}

In this manner we can isolate the Fourier factors $\zeta_{k}\left(z\right)$
that make up the single particle orbitals $\eta_{k}$. Note that if
we set $\tau=\i\frac{L_{y}}{L}$ and $p_{k}=\frac{2\pi}{L}k$, then
$\zeta_{k}\left(z\right)$ can be identified with the single particle
eigenstates on a cylinder as $\zeta_{k}\left(z\right)=e^{\i p_{k}\tilde{x}}e^{-\frac{1}{2}\left(\tilde{y}+p_{k}\right)^{2}}$
in the ordinary Landau gauge and physical coordinates.

\subsection{The Fock basis}

In this section we summarize the essential features of the occupation
basis, or Fock basis, for many-body quantum Hall states. The single
particle orbitals defined in equation (\ref{eq:Basis_wave_function})
are labeled by the momentum label $k$. A Fock state is defined as
a state with a definite set \textbf{$\mathbf{k}$ }of single particle
orbitals $\eta_{k}$ occupied. We write this state as 
\begin{equation}
\mathfrak{F}_{\mathbf{k}}\left(z\right)=\mathcal{A}\left[\prod_{i=1}^{N_{e}}\eta_{k_{i}}\left(z_{i}\right)\right],\label{eq:Fock-Basis}
\end{equation}
 where $\mathcal{A}$ is an anti-symmetrization operator over the
different coordinates. The anti-symmetrization operator is efficiently
implemented by constructing the slater determinant of $\eta_{k_{i}}\left(z_{j}\right)$.
A generic many-body state can thus be written as 
\begin{equation}
\psi_{\mathrm{MB}}\left(z\right)=\sum_{\mathbf{k}}a_{\mathbf{k}}\mathfrak{F}_{\mathbf{k}}\left(z\right),\label{eq:Fock-Expansion-Generic}
\end{equation}
 where the sum is constrained to $1\leq k_{1}<k_{2}<\ldots<k_{N_{e}}\leq N_{\phi}$.

For bosons, the anti-symmetrization operator is replaced by a symmetrization
operator $\S$ and the determinant becomes a permanent, which is harder
to evaluate. Also, the set of momentum labels need not all be distinct,
so the $<$ is replaced by $\leq$ in the sum over \textbf{$\mathbf{k}$}.

\subsection{The Laughlin wave function\label{sub:The-Laughlin-wave}}

The original construction of the torus Laughlin state goes back to
Haldane and Rezayi\cite{Haldane_1985a} by generalizing the short
distance behavior on the plane to also be true on the torus. The complication
on the torus was to find a proper center of mass (CoM) piece $\mathcal{F}_{s}\left(Z,\tau\right)$
that together with the Jastrow factor and Gaussian piece would give
the desired single particle periodic boundary conditions. The resulting
wave function for the Laughlin state at $\nu=\frac{1}{q}$ is

\begin{equation}
\psi_{s}\left(z\right)=\mathcal{N}\left(\tau\right)e^{\i\pi\tau N_{\phi}\sum_{i}y_{i}^{2}}\prod_{i<j}\elliptic 1{\frac{z_{ij}}{L}}{\tau}^{q}\mathcal{F}_{s}\left(\frac{Z}{L},\tau\right),\label{eq:wfn_Laughlin}
\end{equation}
 where $Z=\sum_{i}z_{i}$ and $z_{ij}=z_{i}-z_{j}$. The function
$\vartheta_{1}$ is related to the generalized $\vartheta\mbox{-}$function
in (\ref{eq:def_theta_function}) as $\elliptic 1z{\tau}=\ellipticgeneralized{\frac{1}{2}}{\frac{1}{2}}z{\tau}$,
and has zeros at $z=n+m\tau$.

The label $s$ enumerates the $q$ different degenerate ground states
that exists on the torus\cite{Haldane_1985b}. These states only differ
in the CoM function $\mathcal{F}_{s}$, given as 
\begin{equation}
\mathcal{F}_{s}\left(Z,\tau\right)=\ellipticgeneralized{\frac{s}{q}}{\frac{N_{e}-1}{2}}{qZ}{q\tau}.\label{eq:CoM_Laughlin}
\end{equation}

The $q$ states are all related by rigid magnetic translations of
all the particles. As such, the different states are transformed into
each other as $\prod_{j=1}^{N_{e}}t_{y}^{\left(j\right)}\mathcal{\psi}_{s}=\mathcal{\psi}_{s+1}$.
Note that $\mathcal{\psi}_{s+q}=\mathcal{\psi}_{s}$ since under a
rigid translation of $q$ steps $Z\to Z+\tau q\frac{N_{e}}{N_{\phi}}=Z+\tau$,
and $\vartheta$ is quasi-periodic under this shift. The CoM label
$s$ is related to the total momentum of the state as $K_{\mathrm{total}}=N_{e}s\,\mathrm{mod}\,N_{\phi}$,
which also shows that there are precisely $q$ independent values
of $s$. To specify periodic boundary conditions, $s$ is an (half
) integer when $N_{e}$ is (even) odd.

The normalization factor $\mathcal{N\left(\tau\right)}$ comes from
constructing the Laughlin state using conformal field theory and is
\begin{equation}
\mathcal{N}\left(\tau\right)=\frac{\left[\sqrt{\tau_{2}}\eta\left(\tau\right)^{2}\right]^{\frac{qN_{e}}{2}}}{\eta\left(\tau\right)^{\frac{qN_{e}\left(N_{e}-1\right)}{2}+1}},\label{eq:Laughlin_Norm_CFT}
\end{equation}
 where $\eta\left(\tau\right)$ is Dedekind's $\eta\mbox{-}$function.
This normalization was first introduced by Read\cite{Read_2009} and
later generalized to the full chiral abelian hierarchy in \cite{Fremling_2014}.
The normalization coefficient $\mathcal{N}\left(\tau\right)$ has
a particular $\tau$-dependence that ensures that $\psi_{s}$ transforms
as 
\begin{eqnarray}
\psi_{s} & \overset{\S}{\to} & \sum_{s^{\prime}}S_{s,s^{\prime}}\psi_{s^{\prime}}\nonumber \\
\psi_{s} & \overset{\T}{\to} & \sum_{s^{\prime}}T_{s,s^{\prime}}\psi_{s^{\prime}}\label{eq:ST-transforms}
\end{eqnarray}
 under the modular $\S$ transformation $\tau\to-\frac{1}{\tau}$
and $\T$ transformation $\tau\to\tau+1$. The matrices $S_{s,s^{\prime}}=\frac{1}{\sqrt{q}}e^{-\i2\pi\frac{ss^{\prime}}{q}}$
and $T_{s,s^{\prime}}=\delta_{s,s^{\prime}}e^{\i2\pi\left(\frac{s^{2}}{2q}-\frac{1}{24}\right)}$
are the modular $S$ and $T$ matrices of the CFT for the Laughlin
state.

The modular covariance of (\ref{eq:ST-transforms}) is related of
the redundancy in the parametrization of the torus using $\tau$.
All $\tau$ related by $\S$ and $\T$ transformations are identical
and this has to be reflected in the transformation properties of the
wave functions under such transformations. In particular, the space
spanned by the $q\mbox{-}$fold degenerate $\psi_{s}$ should be invariant.

\begin{figure}
\begin{centering}
\begin{tabular}{c}
\includegraphics[width=0.72\columnwidth]{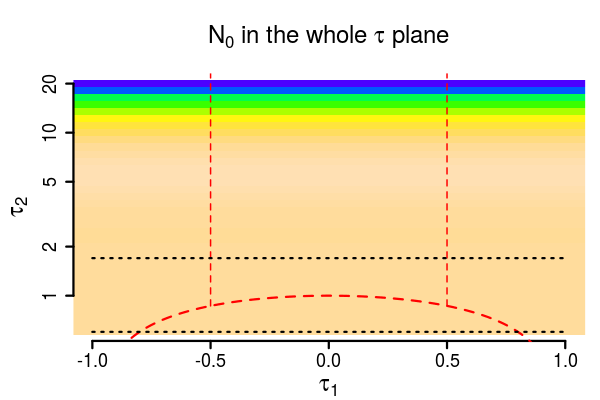}\includegraphics[width=0.12\columnwidth]{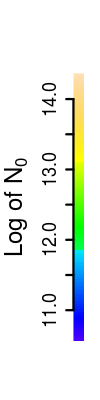}\tabularnewline
$a)$\tabularnewline
\includegraphics[width=0.72\columnwidth]{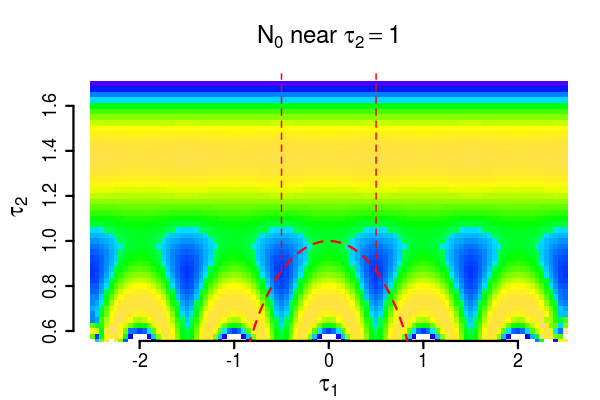}\includegraphics[width=0.12\columnwidth]{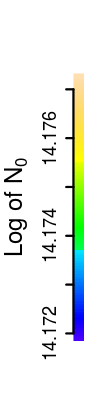}\tabularnewline
$b)$\tabularnewline
\tabularnewline
\end{tabular}
\par\end{centering}

\caption{Scan of the true normalization $\mathcal{N}_{0}$ of the Laughlin
wave function (\ref{eq:wfn_Laughlin}) in the whole $\tau\mbox{-}$plane.
The red dashed lines mark the fundamental domain and the black dotted
lines mark the region zoomed in on in $b)$. Both panels are for $N_{e}=6$
electrons.\protect \\
$a)$ The normalization $\mathcal{N}_{0}$ is independent of $\tau_{1}$
and depends only weakly on $\tau_{2}$ in a large region of the $\tau\mbox{-}$plane.\protect \\
$b)$ Focusing on a small region near $0.6<\tau_{2}<1.6$, small $\tau\mbox{-}$dependent
finite size modulations of $\mathcal{N}_{0}$ can be seen, but these
appear only in the 4:th decimal of $\mathcal{N}_{0}$. These decline
at larger system sizes that are now shown here. Note how the the pattern
of modulations conform to the invariance under modular transformations
$\tau\to\tau+1$ and $\tau\to-\frac{1}{\tau}$ in (\ref{eq:Normalization_Modular_invariance}).\label{fig:N_0_in_tau_plane}}
\end{figure}

The real space wave function $\psi_{s}$ is however not properly normalized.
There exists an extra -- unknown -- normalization factor $\mathcal{N}_{0}$
that can \emph{not} be obtained by CFT arguments. The normalized Laughlin
wave function $\psi_{L}$ is related to $\psi_{s}$ as

\begin{equation}
\psi_{L}=\mathcal{N}_{0}^{-1}\psi_{s},\label{eq:Normalized_Laughlin}
\end{equation}
and the general consensus is that in the thermodynamic limit (at fixed
$\tau$) $\mathcal{N}_{0}$ is constant -- independent of $\tau$
-- and only depends on the particle number $N_{e}$. To say that $\mathcal{N}_{0}$
is constant is analogous to claiming that the Laughlin plasma\cite{Laughlin_1983}
is in a screening phase. Several works have added to the consensus
of screening\cite{Read_2011,Bernevig_2012} but they have not considered
the TT-limit. 

From the analysis of the modular properties of $\psi_{s}$ we can
reassuringly see that the proposed normalization (\ref{eq:Laughlin_Norm_CFT})
ensures that the density of $\psi_{s}$ is modularly covariant. This
means that 
\[
\left|\psi_{s}\left(z,\tau\right)\right|=\left|\psi_{s}\left(z,\tau+1\right)\right|=\left|\psi_{s}\left(\frac{\tau}{\left|\tau\right|}z,-\frac{1}{\tau}\right)\right|,
\]
 which is a also a property that we request from $\psi_{L}$. As a
byproduct we note that the unknown normalization factor $\mathcal{N}_{0}$
must satisfy the same modular covariance

\begin{equation}
\left|\mathcal{N}_{0}\left(\tau\right)\right|=\left|\mathcal{N}_{0}\left(\tau+1\right)\right|=\left|\mathcal{N}_{0}\left(-\frac{1}{\tau}\right)\right|.\label{eq:Normalization_Modular_invariance}
\end{equation}

Although unknown, $\mathcal{N}_{0}$ can be computed numerically for
small systems by using that $\psi_{L}$ is the normalized ground state
of the Haldane pseudo-potential\cite{Haldane_1983}. By comparing
the analytical $\psi_{s}$ in (\ref{eq:wfn_Laughlin}) to the numerically
obtained $\psi_{L}$, $\mathcal{N}_{0}$ can be extracted. In Figure
\ref{fig:N_0_in_tau_plane}, $\mathcal{N}_{0}$ is shown for $N_{e}=6$
particles in the entire $\tau\mbox{-}$plane. As expected $\mathcal{N}_{0}$
is constant and independent of $\tau$ in a large section of the $\tau\mbox{-}$plane.
However, when $\tau_{2}\gtrsim10$ then $\mathcal{N}_{0}$ develops
clear $\tau\mbox{-}$dependence, see Figure \ref{fig:N_0_in_tau_plane}a.
This change in $\mathcal{N}_{0}$ signals that we are entering the
region where the thinness of the torus becomes noticeable. Physically
this means that one of the torus handles is so thin that the plasma
stops screening. We will return to this in Section \ref{sec:TT-limit-analysis}
where we will use the analytic Fock expansion developed in Section
\ref{sec:Laughlin-Expansion} to analyze the behavior of $\mathcal{N}_{0}$
in the TT-limit.

\section{Normalization and Berry Phases}

We can indirectly gauge the $\tau\mbox{-}$dependence of $\mathcal{N}_{0}$
by computing the Berry curvature of (\ref{eq:wfn_Laughlin}) as a
function of the geometry parameter $\tau$. Computing the Berry curvature
is analogous to computing the quantum hall viscosity\cite{Avron_1995,Read_2009}.
Just as in Refs. \cite{Read_2009} and \cite{Fremling_2014} we make
the ansatz that the normalized Laughlin wave function can be written
on the form 
\begin{equation}
\psi_{L}=\mathcal{N}_{0}^{-1}\tau_{2}^{P}\hat{\psi}\left(\left\{ z\right\} ;\tau\right),\label{eq:power-anzats}
\end{equation}
 where $\hat{\psi}$ is holomorphic in $z$ and $\tau$, and $\mathcal{N}_{0}$
is constant. The power of $\tau_{2}$ is then $P=\frac{qN_{e}}{4}$.
Note that (\ref{eq:wfn_Laughlin}) has precisely this form. After
a short calculation, one finds that the Berry potential $A_{\tau}=\braket{\psi_{L}}{\partial_{\tau}\psi_{L}}$
is $A_{\tau}=A_{\bar{\tau}}=-\frac{P}{2\tau_{2}}=-\frac{qN_{e}}{8\tau_{2}}$.
In terms of $A_{\tau_{1}}$and $A_{\tau_{2}}$ this means that 
\begin{eqnarray}
A_{\tau_{1}} & = & A_{\tau}+A_{\bar{\tau}}=-\frac{P}{\tau_{2}}=-\frac{qN_{e}}{4\tau_{2}}\nonumber \\
A_{\tau_{2}} & = & \i A_{\tau}-\i A_{\bar{\tau}}=0\label{eq:Berry_Connection}
\end{eqnarray}

We note that $A_{s}$ is a gauge dependent quantity and will be sensitive
to the $\tau\mbox{-}$dependence of the phase of $\psi_{s}$. Nevertheless
it is a good first test to detect when $\mathcal{N}_{0}$ might not
be constant.

We use importance sampled Monte Carlo integration to evaluate the
Berry connection $A_{\tau_{1}}$ and $A_{\tau_{2}}$ numerically as

\begin{eqnarray*}
\tilde{A}_{\tau_{1}} & = & \lim_{\epsilon\to0}\frac{1}{\epsilon}\Im\left(\braket{\psi_{s}\left(\tau\right)}{\psi_{s}\left(\tau+\epsilon\right)}\right)\\
\tilde{A}_{\tau_{2}} & = & \lim_{\epsilon\to0}\frac{1}{\epsilon}\Im\left(\braket{\psi_{s}\left(\tau\right)}{\psi_{s}\left(\tau+\i\epsilon\right)}\right).
\end{eqnarray*}

\begin{figure}
\begin{centering}
\begin{tabular}{cc}
\includegraphics[width=0.45\columnwidth]{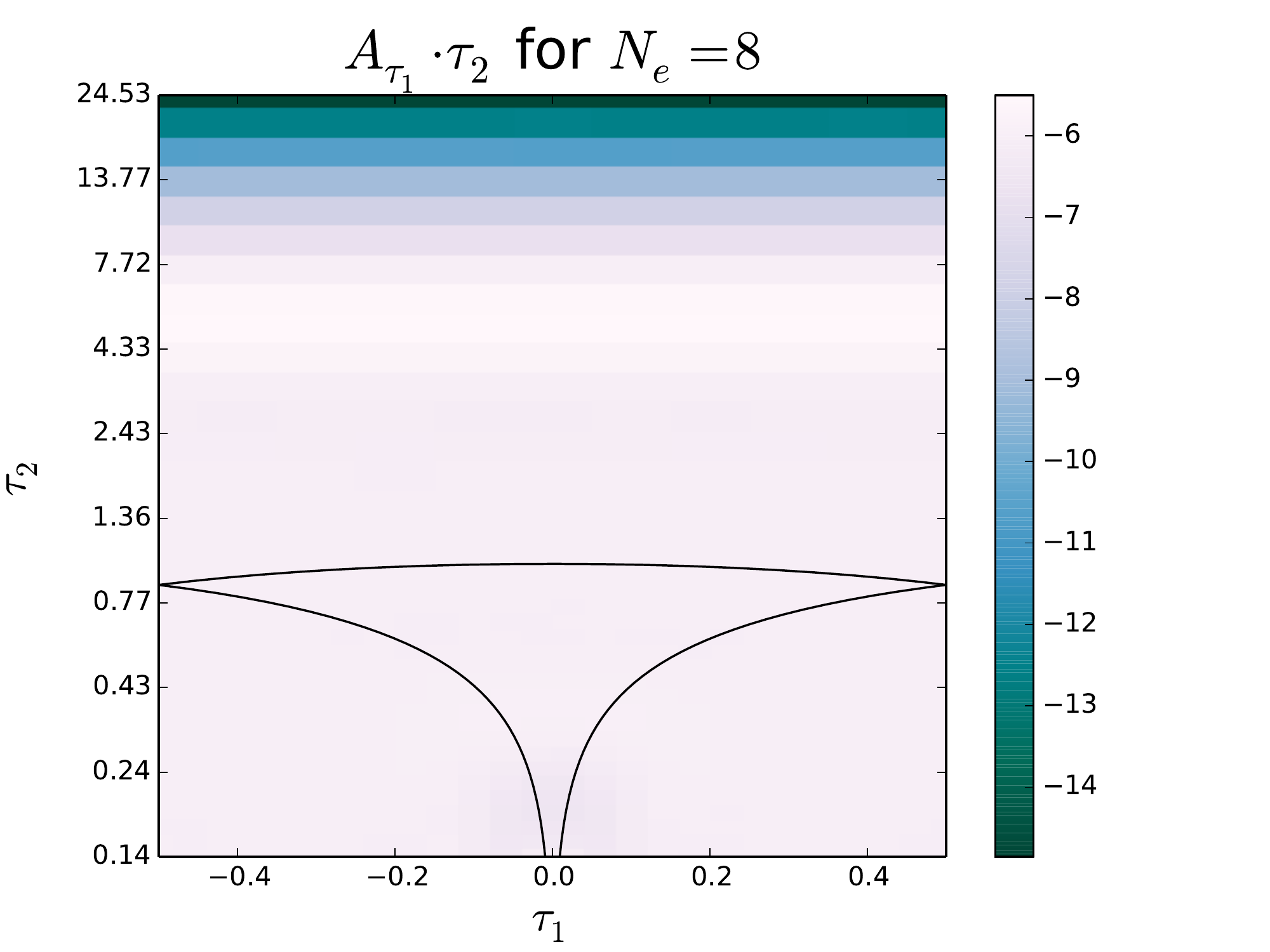} & \includegraphics[width=0.45\columnwidth]{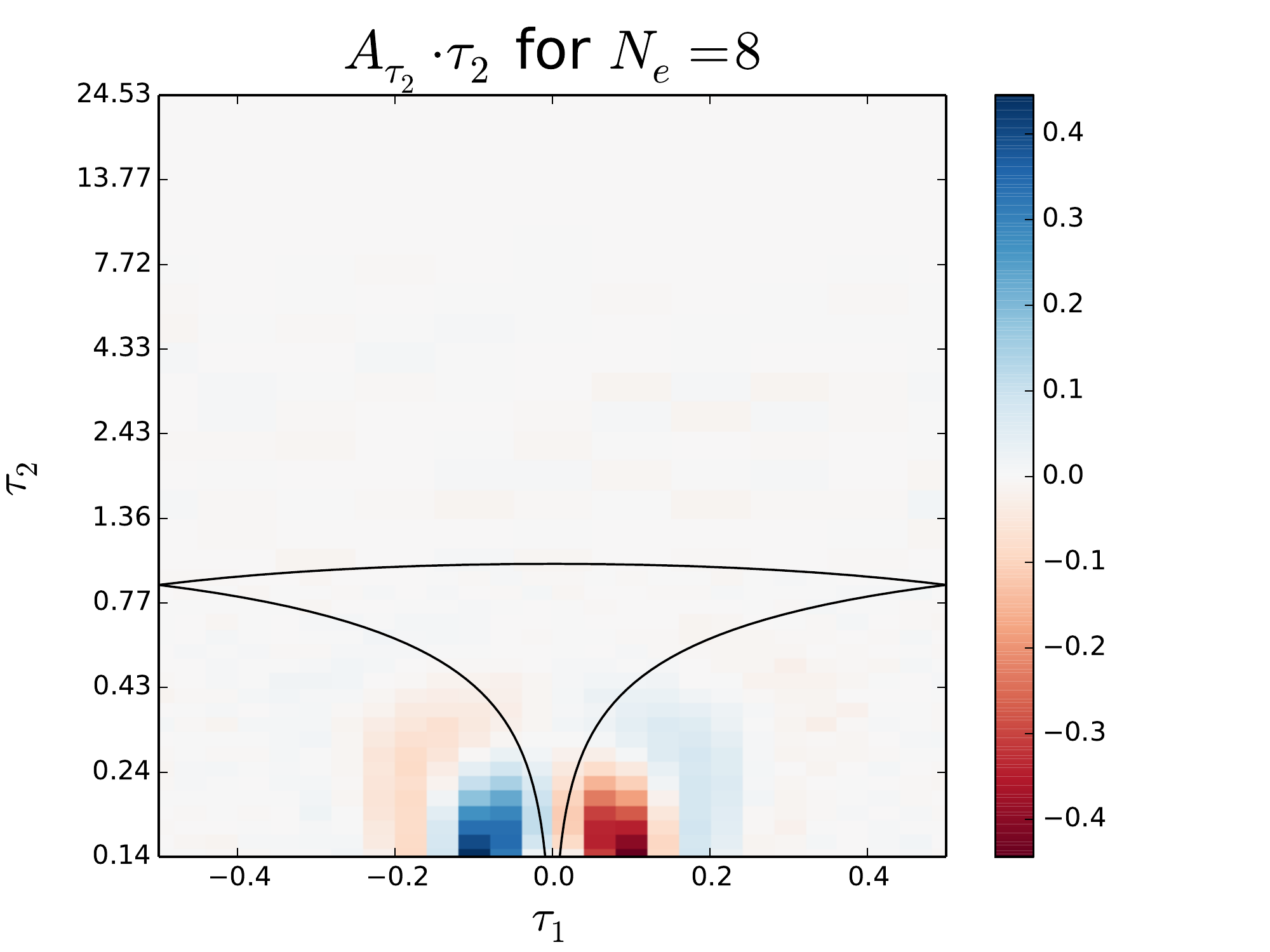}\tabularnewline
$a)$ & $b)$\tabularnewline
\end{tabular}
\par\end{centering}

\caption{Berry connection in a large part of the $\tau$ plane, for $N_{e}=8$
particles, for illustrative purposes. The black lines show the boundaries
of the fundamental domain and a $S\mbox{-}$transformed ($\tau\mbox{\ensuremath{\to}}-\frac{1}{\tau}$)
version thereof. Note that the $\tau_{2}$ axis is logarithmic. \protect \\
$a)$ Plot of $\tau_{2}\tilde{A}_{\tau_{1}}$. When $\mathcal{N}_{0}$
is constant one expects a constant $\tau_{2}\tilde{A}_{\tau_{1}}=-\frac{qN_{e}}{4}=-6$
which can be observed in a region around $\tau\sim\protect\i$. When
$\tau_{2}\to\infty$ then $\tilde{A}_{\tau_{1}}\tau_{2}$ diverges.
\protect \\
$b)$ Plot of $\tau_{2}\tilde{A}_{\tau_{2}}$. Here it is clear that
when $\tau_{2}>0$ then $\tilde{A}_{\tau_{2}}\approx0$ for most tori.
We note however that for small values of $\tau_{2}$ then $\tilde{A}_{\tau_{2}}\propto\tau_{1}$
(for small $\tau_{1}$). This region can be understood by noting that
it is the $\protect\S\mbox{-}$transformed image of the region $\tau_{2}\gtrsim5$
for $\tilde{A}_{\tau_{1}}$.\protect \\
Taking both $\tilde{A}_{\tau_{1}}$and $\tilde{A}_{\tau_{2}}$ into
consideration we conclude that the normalization $\mathcal{N}_{0}$
is constant for a region around $\tau\approx\protect\i$ but deviates
when $\tau_{2}\to\infty$ or when $\left|\tau\right|$ is small. \label{fig:Berry_Connection}}
\end{figure}

In Figure \ref{fig:Berry_Connection} we scan $\tilde{A}_{s}$ over
a large part of the $\tau\mbox{-}$plane. For illustrative purposes
we show a relatively small system of $N_{e}=8$ electrons in the figure.
The features in this picture are also present in larger systems but
will occur at different values of $\tau$. In the figure we can see
that $\tilde{A}_{\tau_{1}}\approx-\frac{qN_{e}}{4\tau_{2}}$ in a
large region around $\tau=\i$ but that it deviates significantly
from the expected value when $\tau_{2}\to\infty$. For $N_{e}=8$
particles this deviation begins at around $\tau_{2}\gtrsim5$ and
is almost independent of $\tau_{1}$.

We also see -- Figure \ref{fig:Berry_Connection}b) -- that $\tilde{A}_{\tau_{2}}\approx0$
everywhere except in an approximate circle of radius $r\approx0.1$
centered at $\tau=\i r$. This non-zero value can be understood by
appealing to modular covariance and noting that the this region is
the modular image of the region $\tau_{2}>\frac{1}{2r}\approx5$ for
$\tilde{A}_{\tau_{1}}$ in Figure \ref{fig:Berry_Connection}a). 

Since the Berry connection is not a gauge invariant quantity, it can
be difficult to see what part of the deviation in $\tilde{A}_{s}$
with respect to $A_{s}$ is due to an actual $\tau\mbox{-}$dependence
of the normalization, and what is due to a simple $\tau\mbox{-}$dependence
of the phase (at fixed $z$). For that purpose, we also compute the
Berry curvature which is a gauge invariant quantity. Under the assumption
of a constant $\mathcal{N}_{0}$ the Berry curvature is

\[
\mathcal{F}_{\tau_{1}\tau_{2}}=-\i2\mathcal{F}_{\tau\bar{\tau}}=\partial_{\tau_{1}}A_{\tau_{2}}-\partial_{\tau_{2}}A_{\tau_{1}}=-\frac{qN_{e}}{4\tau_{2}^{2}}.
\]

The curvature is closely related to the Hall viscosity \cite{Avron_1995,Read_2009}
which can be used as a probe to distinguish different topological
phases at the same filling fraction. Hall viscosity is conjectured
to be a probe containing the same information as the shift $\S$\cite{Read_2009}.
On a spherical geometry the Laughlin state is characterized by the
shift $\S_{\mathrm{sphere}}=q$ in the relation 
\[
N_{\phi}=\nu^{-1}N_{e}-\S_{\mathrm{sphere}},
\]

between the number of flux quantum and the number of electrons. The
shift appears as a consequence of the curvature of the sphere and
the non-zero orbital spin $\bar{s}$ of the electrons. When the electrons
move over the surface of the sphere a Berry phase is accumulated giving
rise to a extra effective magnetic flux, $\S_{\mathrm{sphere}}=2\bar{s}$\cite{Read_2008,Read_2011}.

On the torus there is no curvature, so here $\S_{\mathrm{torus}}=0$
and $N_{\phi}=\nu^{-1}N_{e}$ precisely. However, as $\S_{\mathrm{sphere}}$
is a topological characteristic of the Laughlin state, the same information
should exist also on the torus, the question is how it manifests itself.
Read\cite{Read_2009} showed that the topological information could
be extracted by studying the Hall viscosity\cite{Avron_1995} $\eta^{H}$
of the quantum fluid. Read conjectured that the viscosity in the thermodynamic
limit (on any geometry, but especially on a torus) should be 
\begin{equation}
\eta^{H}=\frac{1}{4}\hbar\bar{n}\S_{\mathrm{sphere}}=\frac{1}{2}\hbar\bar{n}\bar{s},\label{eq:Read_conj}
\end{equation}
 where here $\bar{n}$ is the electron number density. In a later
paper Read \& Rezayi\cite{Read_2011} numerically showed this result
to hold for the Laughlin state (amongst other things), when $\tau$
was close to $\tau=\i$.

To analytically compute viscosity for a many body state is usually
difficult, but it is simplified dramatically if the real space wave
function can written as 
\begin{equation}
\psi_{\mathrm{MB}}\left(z\right)=\tau_{2}^{P}f\left(z;\tau\right),\label{eq:non-holomophic_psi}
\end{equation}
 where $f$ is holomorphic in $\tau$. Under this assumption the viscosity
can be computed to be $\eta^{H}=\frac{eB\nu}{2\pi}\frac{P}{N_{e}}$\cite{Read_2009}.
This can be reformulated as the intensive quantity of effective average
spin $\bar{s}$ as 
\begin{equation}
\bar{s}=\frac{2P}{N_{e}}.\label{eq:Effective_sbar}
\end{equation}

The Laughlin wave function $\psi_{s}$ in (\ref{eq:wfn_Laughlin})
is precisely of the form (\ref{eq:non-holomophic_psi}) where $\mathcal{N}\left(\tau\right)$
contains the factor $\tau_{2}^{\frac{qN_{e}}{4}}$, so $\bar{s}=\frac{q}{2}$
for $\psi_{s}$. In a similar manner the viscosity for the rest of
the chiral Haldane-Halperin\cite{Haldane_1983,Halperin_1983} hierarchy
on the torus\cite{Fremling_2014} can also be computed. In doing so
a viscosity is obtained that also agrees with Read's conjecture (\ref{eq:Read_conj}).

The numerical work by Read \& Rezayi\cite{Read_2011} showed that
$\bar{s}=\frac{q}{2}$ and supports that Read's conjecture holds in
the thermodynamic limit for the nearly square tori. This lends indirect
evidence to the screening properties of the plasma analogy. However,
in other numerical studies of the viscosity \cite{Zhou_2013,Fremling_2014,Tuegel_2015}
it is clear that $\bar{s}$ has significant $\tau\mbox{-}$dependence,
if the torus is asymmetric enough.

\begin{figure}
\begin{centering}
\begin{tabular}{cc}
\includegraphics[width=0.45\columnwidth]{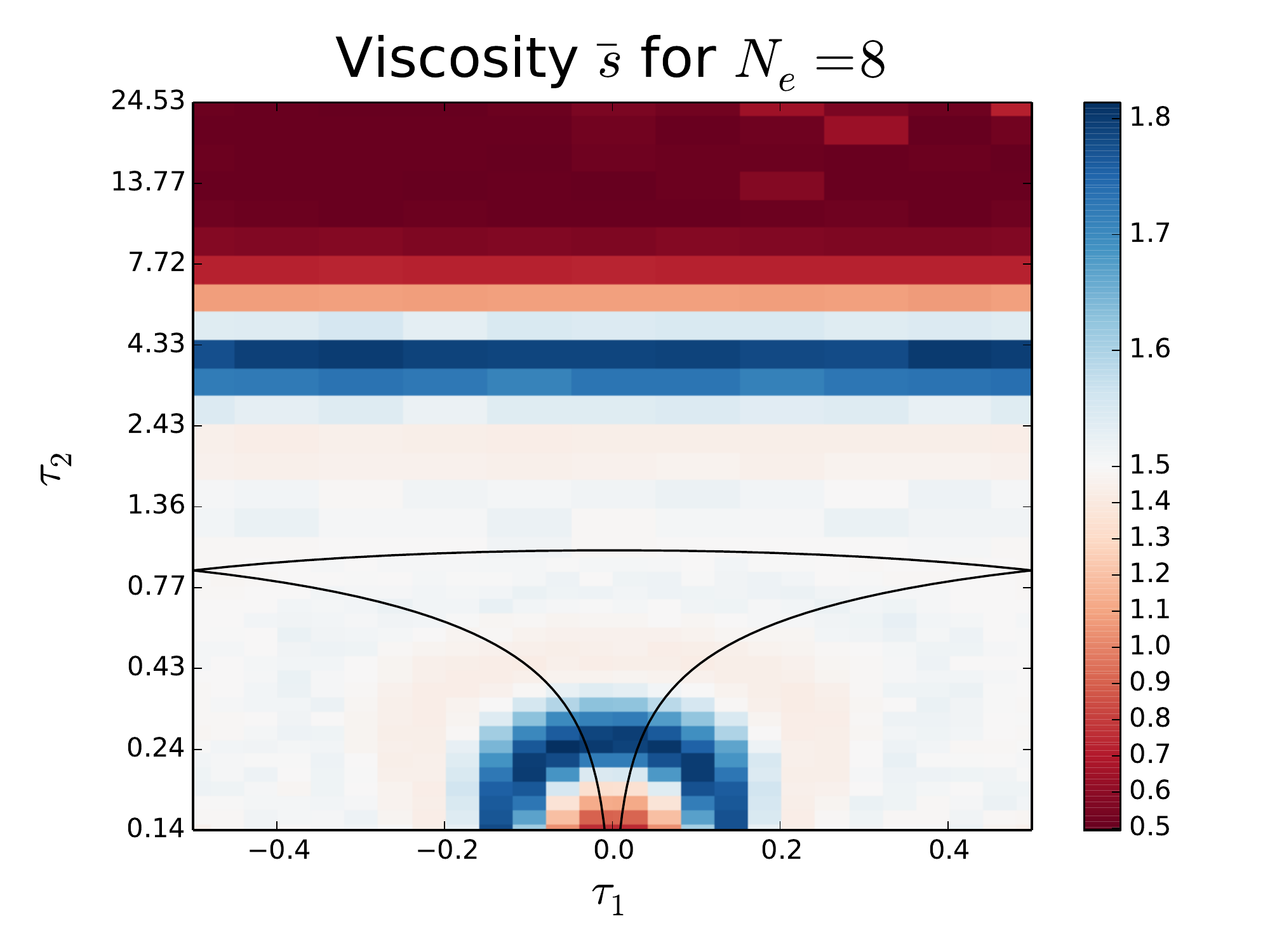} & \includegraphics[width=0.45\columnwidth]{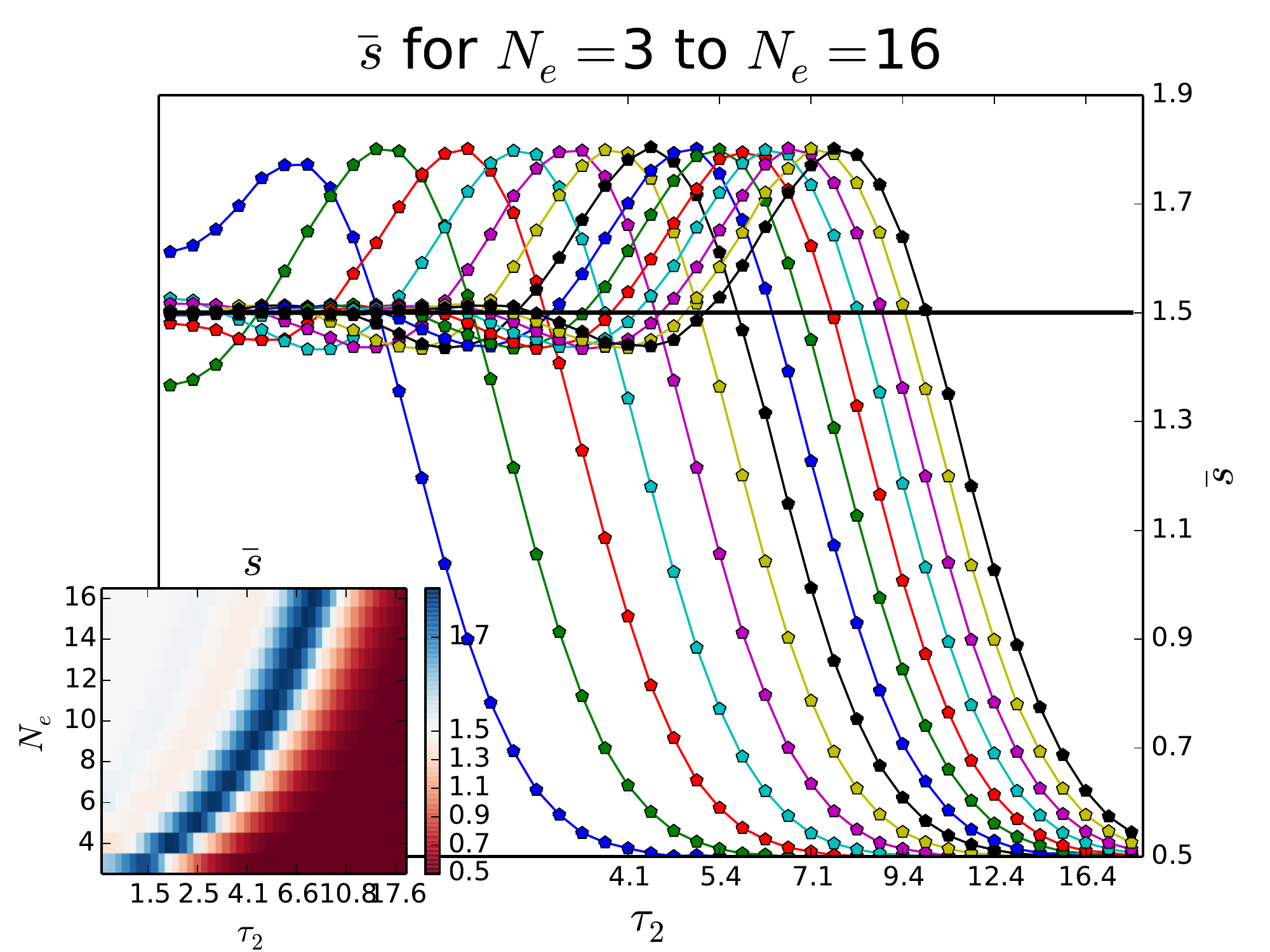}\tabularnewline
$a)$ & $b)$\tabularnewline
\end{tabular}
\par\end{centering}

\caption{$a)$ The viscosity units of the average orbital spin $\bar{s}$ plotted
in the entire $\tau\mbox{-}$plane for $N_{e}=8$. Just as for the
Berry connection $A_{\tau_{1}}$and $A_{\tau_{2}}$ we see the expected
curvature $\mathcal{F}_{\tau_{1}\tau_{2}}\approx-\frac{qN_{e}}{4\tau_{2}^{2}}$
for a large region of the $\tau\mbox{-}$plane. Note that as $\mathcal{F}_{\tau_{1}\tau_{2}}$
is a gauge invariant quantity the deviation from $-\frac{qN_{e}}{4\tau_{2}^{2}}$
when $\tau_{2}>r=3$ is nicely reproduced in the circle $\left|\tau-\frac{\protect\i}{2r}\right|<\frac{1}{2r}$.\protect \\
$b)$ The average orbital spin $\bar{s}$ plotted for $\tau=\protect\i\tau_{2}$
and several different $N_{e}$.\textbf{ }The Inset shows same data
as a color plot. In the figure it is clearly visible how the plateau
at $\bar{s}=\frac{3}{2}$ widens monotonically with system size. We
also see that at large $\tau_{2}$ all system sizes transition to
$\bar{s}=\frac{1}{2}$ which is expected for Fock state.\label{fig:Berry_Curvature}}
\end{figure}

In Figure \ref{fig:Berry_Curvature}$a)$ we plot the effective shift
$\bar{s}=\frac{2\tau_{2}\mathcal{F}_{\tau_{1}\tau_{2}}}{N_{e}}=\frac{q}{2}$
in the $\tau\mbox{-}$plane for $N_{e}=8$ particles. Again the the
same features as for $A_{\tau_{1}}$ and $A_{\tau_{2}}$ are present.
However, one important difference is that in the TT-limit $\bar{s}$
stabilizes at $\frac{1}{2}$ instead of the expected $\bar{s}=\frac{3}{2}$.
Note that the features at large $\tau_{2}$ are modularly mapped to
the ringlike structure at small $\left|\tau\right|$.

In Figure \ref{fig:Berry_Curvature}$b)$ we get further indication
that the transition from $\bar{s}=\frac{3}{2}$ to $\bar{s}=\frac{1}{2}$
is a generic feature independent of system size. In the plot we see
that the transition to $\bar{s}=\frac{1}{2}$ happens for all systems
sizes examined, but the point of transition scales as $\tau_{2}\propto\sqrt{N_{e}}$
(see inset).

This TT-limit value of $\bar{s}=\frac{1}{2}$ can be understood if
the state at $\tau_{2}\to\infty$ is described not by a strongly correlated
fluid, but rather by a single slater determinant, \emph{\ie}\textbf{\emph{
}}a Fock state\emph{ }$\psi_{s}=\mathfrak{F}_{\mathbf{k}}\left(z\right)$
as in (\ref{eq:Fock-Basis}). We will expand on this observation further
in Section \ref{sec:TT-limit-analysis}. For now we simply conclude
that in a region around $\tau\approx\i$ the Berry curvature has the
desired properties. The number of MC data points is here $10^{6}$
for all system sizes. 

Returning to the Berry connection, we can now also compute the Berry
phases associated with the modular $\Td$transform $\tau\to\tau+1$
and $\Sd$transform $\tau\to-\frac{1}{\tau}$. Starting with the $\Td$transform,
we find that the accumulated Berry phase for a straight path from
$\tau$ to $\tau+1$ is 
\begin{equation}
\phi_{\T}=\int_{0}^{1}d\tau_{1}\,\braket{\psi_{L}}{\partial_{\tau_{1}}\psi_{L}}=A_{\tau_{1}}=-\frac{qN_{e}}{4\tau_{2}}.\label{eq:Berry_T_CFT}
\end{equation}
where we assume that $\mathcal{N}_{0}$ is constant. 

The accumulated Berry phase $\phi_{\T}$ can also be approximated
numerically by discretizing the path from $\tau$ to $\tau+1$ into
$n$ steps and computing the cumulative overlap along the path as
\[
\tilde{\phi}=\int ds\,\braket{\psi\left(\tau\right)}{\partial_{s}\psi\left(\tau\right)}\approx\Im\left\{ \prod_{j=1}^{n}\braket{\psi_{L}\left(\tau_{j}\right)}{\psi_{L}\left(\tau_{j+1}\right)}\right\} .
\]

\begin{figure}
\begin{centering}
\includegraphics[width=0.7\columnwidth]{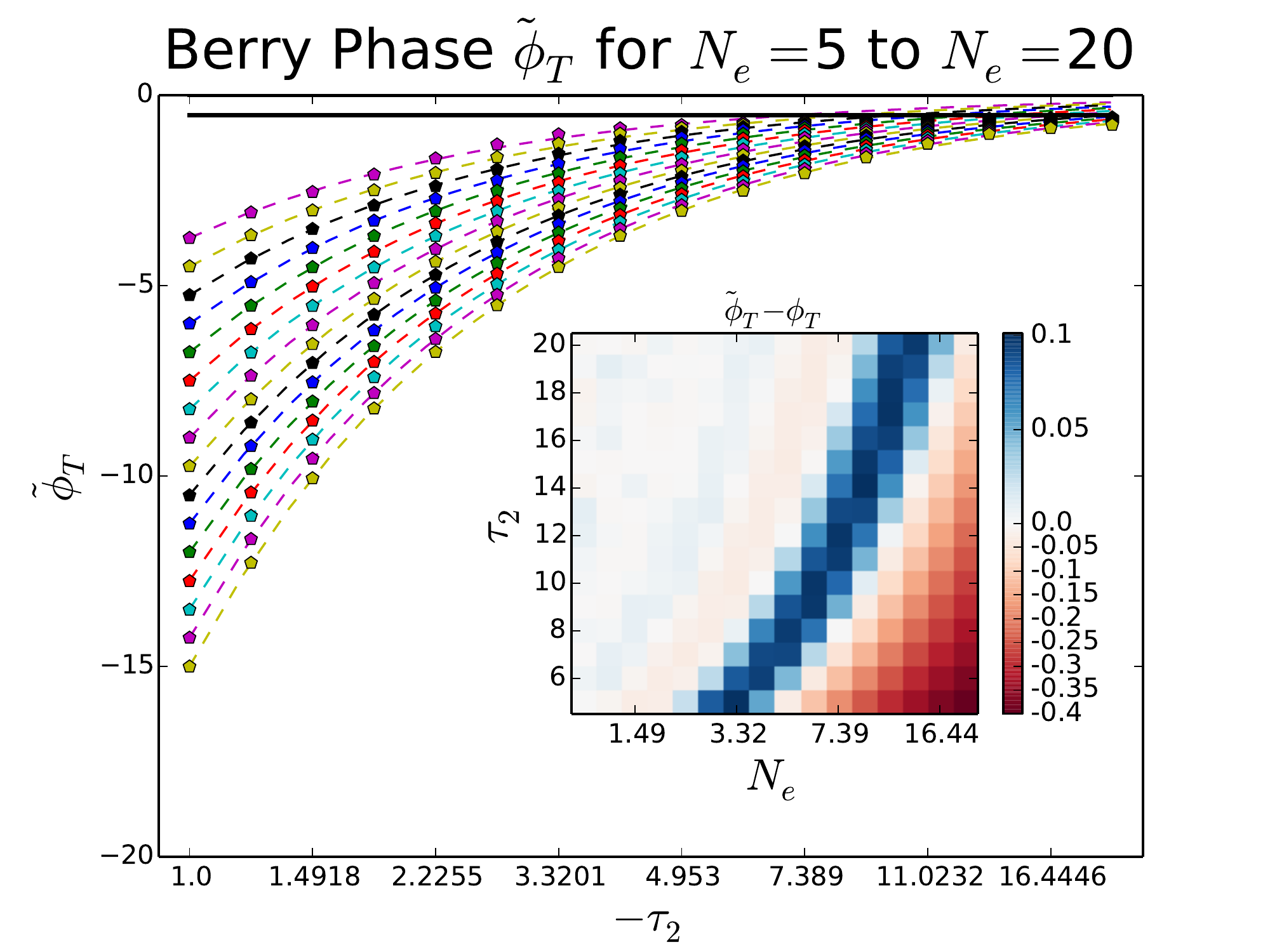}
\par\end{centering}

\caption{The Berry phase $\tilde{\phi}_{\protect\T}$ accumulated by $\psi_{L}$
under a $\protect\T$ transform from $\tau$ to $\tau+1$, as a function
of $\tau_{2}$ for $\tau_{1}=0$ and several $N_{e}$. Dashed lines
show the expected values $\phi_{\protect\T}$ given by (\ref{eq:Berry_T_CFT}).
Inset shows the same data as the difference $\tilde{\phi}_{\protect\T}-\phi_{\protect\T}$.
The path $\tau\left(s\right)$ is discretized in $100$ steps with
$7\times10^{4}$ MC points at each step.\protect \\
It is clear that $\tilde{\phi}_{\protect\T}$ agrees well with (\ref{eq:Berry_T_CFT})
when $\tau_{2}$ is not to large. However for large $\tau$ there
is a clear deviation between $\tilde{\phi}_{\protect\T}$ and $\phi_{\protect\T}$
which is \emph{not} an effect of the discretization of the path. Note
that as $\tau_{2}\to\infty$ then $\tilde{\phi}_{\protect\T}\to\frac{\pi}{6}$
instead of $\phi_{\protect\T}\to0$.\label{fig:Modular_T_transforms}}
\end{figure}

A comparison of the numerical value $\tilde{\phi}_{\T}$ and theoretical
value $\phi_{\T}$ can be seen in Figure \ref{fig:Modular_T_transforms}
for $5\leq N_{e}\leq20$ and $1\leq\tau_{2}\leq20$. The numeric data
is represented by points and the theoretic expectation is shown by
by dashed lines of the same colors as the points. One can clearly
see that for small $\tau_{2}$, the numeric phase $\tilde{\phi}_{\T}$
agrees very well with (\ref{eq:Berry_T_CFT}). It is also possible
to see that at larger $\tau_{2}$ the the accumulated phase $\tilde{\phi}_{\T}$
saturates at $\tilde{\phi}_{\T}\approx0.5$ instead of $\phi_{\T}\to0$
as expected. In Section \ref{sub:Modular-Transformation-Phases}\textbf{
}we will show that this is actually constant is actually $\tilde{\phi}_{\T}=\frac{\pi}{6}$
in the limit $\tau_{2}\to\infty$. In the inset in the same figure,
one sees that the $\tau_{2}$ at which $\tilde{\phi}_{\T}$ deviates
from (\ref{eq:Berry_T_CFT}) grows monotonically with $N_{e}$. This
hints that in the thermodynamic limit the transition will be at $\tau_{2}\to\infty$.

We may strengthen the picture that the normalization is constant for
$\tau$ near $\tau\approx\i$ by also considering the Berry phase
accumulated under an $\Sd$transform. For the $\Sd$transform one
has to be a bit careful, as there is no canonical path between $\tau$
and $-\frac{1}{\tau}$. Depending on the value of $\tau$ some paths
would be numerically more stable that others. We choose the a path
that is self-dual, \emph{i.e. }where $\tau\left(s\right)=\frac{-1}{\tau\left(1-s\right)}$.
One such path which is particularly nice is 
\begin{equation}
\tau\left(s\right)=\tau\left(\frac{-1}{\tau^{2}}\right)^{s}=r^{1-2s}e^{\i\theta\left(1-2s\right)+\i s\pi},\label{eq:self-dual-path}
\end{equation}
 where $\tau\left(0\right)=re^{\i\theta}$. Using that $A_{\tau_{1}}=-\frac{qN_{e}}{4\tau_{2}}$,
$A_{\tau_{2}}=0$ and $\frac{\partial\tau_{1}}{\partial s}=-2\ln r\cdot\tau_{1}\left(s\right)+\left(2\theta-\pi\right)\tau_{2}\left(s\right)$,
the Berry phase becomes simply

\begin{eqnarray}
\phi_{\S} & = & \int_{0}^{1}ds\left(-2\ln r\cdot\tau_{1}\left(s\right)+\left(2\theta-\pi\right)\tau_{2}\left(s\right)\right)A_{\tau_{1}}\nonumber \\
 & = & -\frac{qN_{e}}{4}\int_{0}^{1}ds\left(-2\ln r\cdot\frac{\tau_{1}\left(s\right)}{\tau_{2}\left(s\right)}+\left(2\theta-\pi\right)\right)\nonumber \\
 & = & \frac{qN_{e}}{4}\left(\pi-2\theta\right)\label{eq:Berry_S_CFT}
\end{eqnarray}
 where we in the last step use that $\int_{0}^{1}ds\frac{\tau_{1}\left(s\right)}{\tau_{2}\left(s\right)}=0$.
Again we assume that $\mathcal{N}_{0}$ is constant. 

\begin{figure}
\begin{centering}
\includegraphics[width=0.7\columnwidth]{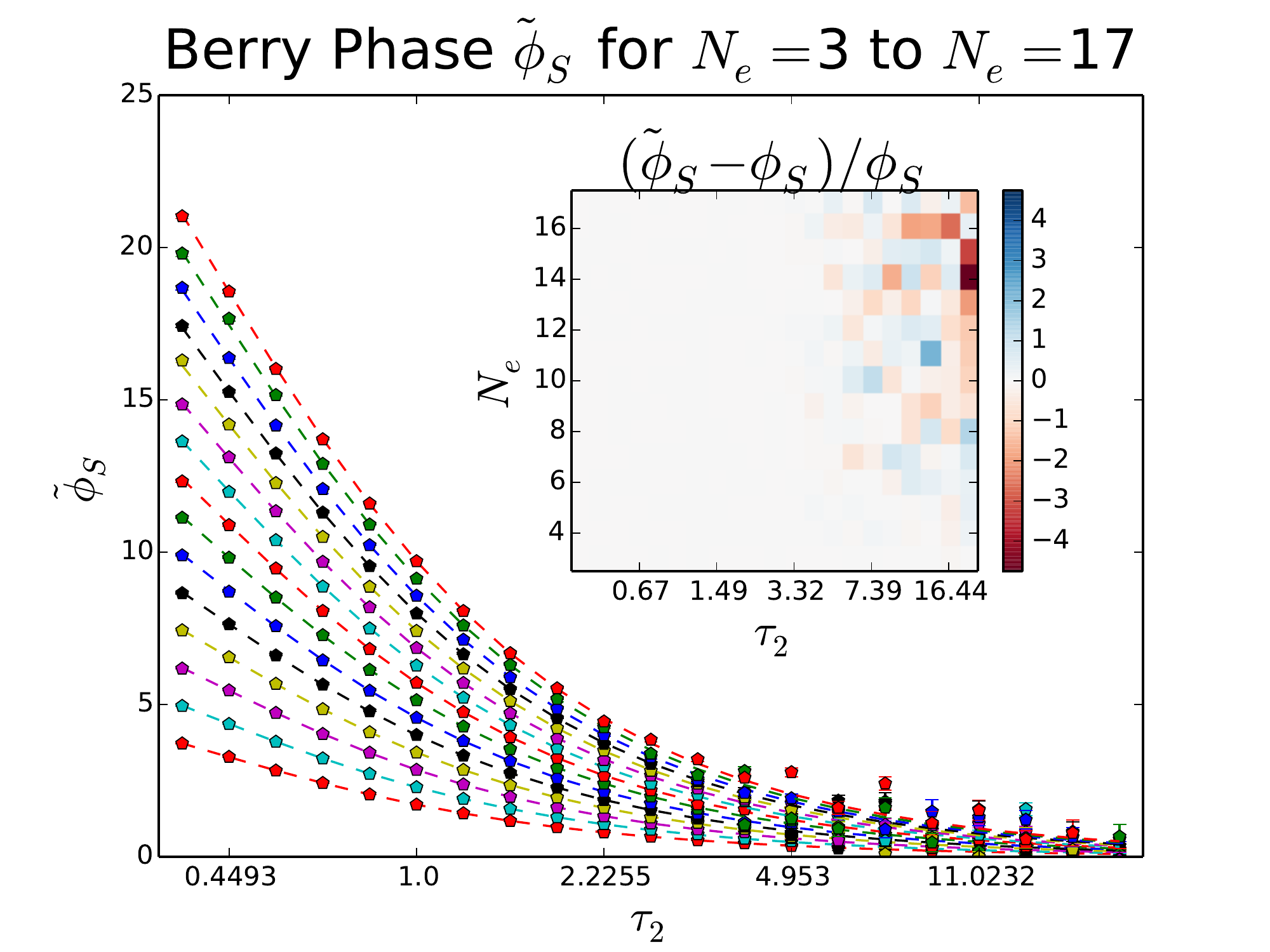}
\par\end{centering}

\caption{The Berry phase $\tilde{\phi}_{\protect\S}$ accumulated under an
$\protect\S$ transform from $\tau$ to $\frac{-1}{\tau}$ as a function
of $\tau_{2}$ for $\tau_{1}=0.4$ and several $N_{e}$. The dashed
lines show the expected values $\phi_{\protect\S}$ given by (\ref{eq:Berry_S_CFT}).
The inset shows the same data as relative deviation $\left(\tilde{\phi}_{\protect\S}-\phi_{\protect\S}\right)/\phi_{\protect\S}$.
The path $\tau\left(s\right)$ is discretized in $500$ steps with
$10^{5}$ MC-samples at each step.\protect \\
In the main plot the numeric $\tilde{\phi}_{\protect\S}$ follows
the theoretically expected $\phi_{\protect\S}$ for all $\tau_{2}$
and system sizes considered. The inset plot shows no structure in
the relative difference between $\tilde{\phi}_{\protect\S}$ and $\phi_{\protect\S}$,
and hints that this is dominated by numerical noise. The main reason
for this noise is that the length of the path increases with $\tau_{2}$
such that the discretization error grows. We note that $\tilde{\phi}_{\protect\S}$
agrees well with (\ref{eq:Berry_S_CFT}) as long as $\tau_{2}$ is
in a (large) region around $\tau_{2}=1$. \label{fig:Modular_S_transforms}}
\end{figure}

In Figure \ref{fig:Modular_S_transforms} we plot the phase $\tilde{\phi}_{\S}$
accumulated by performing a $\Sd$transform along the self dual path
(\ref{eq:self-dual-path}) starting at $\tau=0.4+\i\tau_{2}$. We
see that this particular path gives a $\tilde{\phi}_{\S}$ that agrees
well with the expected $\phi_{\S}$ as long as $\tau_{2}\lesssim6$
for all system sizes considered. Beyond this aspect ration it is possible
to see clear deviations from (\ref{eq:Berry_S_CFT}). However, the
deviation can be mostly attributed no Monte Carlo noise (upper left
in inset). That the noise levels are bigger at larger $\tau_{2}$
is only natural as the length path through $\tau\mbox{-}$space will
increase as $\left|\tau-\i\right|$ grows.

Taking all the data into account (especially from $A_{s}$ and $\mathcal{F}_{\tau_{1}\tau_{2}}$)
we conclude that the normalization $\mathcal{N}_{0}$ is constant
in a large region around $\tau\approx\i$. We also conclude that the
region of constant $\mathcal{N}_{0}$ increases with $N_{e}$. However
we also see that if $\tau_{2}$ is sufficiently large, then $\mathcal{N}_{0}$
develops a clear $\tau\mbox{-}$dependence and the ansatz (\ref{eq:power-anzats})
with $P=\frac{qN_{e}}{4}$ is changed to $P=\frac{N_{e}}{4}$. This
new power of $\tau_{2}$ is consistent with $\psi_{L}$ being described
by only one Fock state.

To gain more understanding of the nature of the transition to this
Fock state, we will in the next sections expand $\psi_{s}$ in a Fock
basis and analytically send $\tau_{2}\to\infty$.

\section{Expanding Laughlin in a Fock Basis\label{sec:Laughlin-Expansion}}

In this section we will rewrite the real space Laughlin wave function
(\ref{eq:wfn_Laughlin}) in the single particle orbital basis of (\ref{eq:Fock-Basis}),
\ie the Fock basis. We do this to gain analytical control over the
coefficients $a_{\mathbf{k}}$ of the Laughlin state. Later, in Section
\ref{sec:TT-limit-analysis} we will use this analytic knowledge to
make statements regarding the plasma analogy in the limit of a thin
torus. In Section \ref{sec:Hierachy-expand} we will also outline
how to perform the Fock expansion for the full chiral Haldane-Halperin
hierarchy\cite{Haldane_1983,Halperin_1983,Fremling_2013} on the torus.

We begin the next part \ref{sub:Main-Results} by displaying and commenting
on the main results. Readers who are interested in the derivation
of these results may consult the supplementary material. In the following
part \ref{sub:Numerically-evaluating}, we comment on the numerical
efficiency of the Fock expansion and compare it to other methods of
obtaining the same information.

\subsection{Main results and formulas\label{sub:Main-Results}}

In this part we present the main results and formulas. In what follows
we will consider only the fermionic Laughlin wave function unless
otherwise specified. The fermionic and bosonic Laughlin wave functions
are so closely related that all results here carry over to the bosonic
case with only minor modifications. Where there are differences these
will be pointed out.

The Fock expansion of any state will in its most general form involve
a sum over all partitions of the momentum numbers $k_{1},\ldots,k_{N_{e}}$
with the total momentum $K_{\mathrm{total}}=\sum_{i=1}^{N_{e}}k_{i}\,\mathrm{mod}\,N_{\phi}$.
For periodic boundary conditions $k_{i}\in\mathbb{Z}$ are integers
and $K_{\mathrm{total}}=\frac{1}{2}qN_{e}\left(N_{e}-1\right)+mN_{e}\,\mathrm{mod}\,qN_{e}$
for $m\in\mathbb{Z}$. Since a Fock basis expansion could potentially
contain all of these partitions it could become a very large sum.
A convenient way to organize this sum is by changing to a variable
offset from $k_{i}$ which we call\textbf{ }$\mathbb{T}_{i}$. The
relation between the two is

\begin{equation}
k_{i}=\mathbb{T}_{i}+s\,\mathrm{mod}\,N_{\phi},\label{eq:relation_k_T}
\end{equation}
 where the $\mathbb{T}_{i}$ are chosen such that $\sum_{i=1}^{N_{e}}\mathbb{T}_{i}=0$.
The variable $s$ is precisely the center of mass momentum label in
(\ref{eq:wfn_Laughlin}). Because $k_{i}$ is an integer, the index
$\mathbb{T}_{i}$ will be an (half) integer when $N_{e}$ is (even)
odd, to match $s$.

Using the auxiliary variable $\mathbb{T}_{i}$ the Laughlin wave function
can be rewritten as

\begin{eqnarray}
\psi_{s} & = & \mathcal{N}\left(\tau\right)\sum_{\left\{ \mathbb{T}_{i}\in\mathbb{Z}_{N_{\phi}}+q\frac{N_{e}-1}{2}\right\} }\mathcal{Z}\left(\mathbb{T}\right)\prod_{i=1}^{N_{e}}\eta_{\mathbb{T}_{i}+s}\left(z_{i}\right),\label{eq:MAIN_psi_laughlin_Z}
\end{eqnarray}
 which makes explicit the single particle orbital basis. Here, $\eta_{k_{i}}\left(z_{i}\right)$
are the single particle orbitals from (\ref{eq:Basis_wave_function}).
Note that no explicit anti-symmetrization is needed as the coefficients
$\mathcal{Z}\left(\mathbb{T}\right)$ are all fully antisymmetric
by construction.

As the basis states $\eta_{k}\left(z\right)$ are labeled modulo $N_{\phi}$
the sum over $\mathbb{T}$ is finite. The half-integer offset to $\mathbb{T}$
is a remnant from the expansion of the Jastrow factor. Note that the
CoM index $s$ only enters the relation between $k_{i}$ and $\mathbb{T}_{i}$
as a constant shift. This shift serves to relabel the basis states
of the expansion. The same relabeling can be achieved by acting with
the $t_{y}$ translation operator as $t_{y}^{s}\eta_{\mathbb{T}_{i}}=\eta_{\mathbb{T}_{i}+s}$,
which shows again that all the $q$ different Laughlin states are
related by rigid magnetic translations. 

The factor $\mathcal{Z}\left(\mathbb{T}\right)$ is anti-symmetric
in the arguments $\mathbb{T}$ and is related to the Fock coefficient
$a_{\mathbf{k}}$ in (\ref{eq:Fock-Expansion-Generic}). As it stands
(\ref{eq:MAIN_psi_laughlin_Z}) is not of the form of (\ref{eq:Fock-Expansion-Generic}).
However as $\mathcal{Z}\left(\mathbb{T}\right)$ is anti-symmetric,
(\ref{eq:MAIN_psi_laughlin_Z}) can be recast on the form (\ref{eq:Fock-Expansion-Generic})
with the identification\textbf{ }$a_{\mathbf{k}}=\sqrt{N_{e}!}\mathcal{N}\left(\tau\right)\mathcal{Z}\left(\mathbb{T}\right)$.

The Fock expansion coefficient $\mathcal{Z}\left(\mathbb{T}\right)$
is given by

\begin{eqnarray}
\mathcal{Z}\left(\mathbb{T}\right) & = & \sum_{\left\{ \tilde{T}_{ij}\in\mathbb{Z}+\frac{q}{2}\right\} }\exp\left\{ \i\pi\tau\frac{1}{qN_{e}}\sum_{i<j<k}\left(\tilde{T}_{ij}+\tilde{T}_{jk}+\tilde{T}_{ki}\right)^{2}\right\} \prod_{i<j}\left(\tilde{Z}_{\tilde{T}_{ij}}^{\left(q\right)}e^{\i\pi\tilde{T}_{ij}}\right),\label{eq:MAIN_Z_laughlin}
\end{eqnarray}
 where the $\tilde{T}_{ij}$ are for technical reasons (see the supplemental
material) anti-symmetric half-integer numbers $\tilde{T}_{ij}=-\tilde{T}_{ij}$.
The sum over all the $\tilde{T}_{ij}$ is constrained to fulfill 
\begin{equation}
\mathbb{T}_{i}=\sum_{j=1}^{N_{e}}\tilde{T}_{ij},\label{eq:T_from_T_ij}
\end{equation}
 such that each particle has the desired momentum $\mathbb{T}_{i}$.

To obtain the expression for $\mathcal{Z}\left(\mathbb{T}\right)$
one has to expand the $\vartheta_{1}\mbox{-}$functions in the Jastrow
factors in terms of Fourier -- or momentum -- components $e^{\i2\pi\tilde{T}_{ij}z_{ij}}$.
This makes it possible to extract the total momentum of a specific
particle $z_{i}$ as $e^{\i2\pi z_{i}\sum_{j=1}^{N_{e}}\tilde{T}_{ij}}$
by multiplying together all the contributions from all the pairs of
particles. Thus $\tilde{T}_{ij}$ appear naturally in $\mathcal{Z}\left(\mathbb{T}\right)$
as they enumerate all the ways of forming the total momentum $\mathbb{T}_{i}$
from all the pairwise contributions $\tilde{T}_{ij}$.

In the above formula \ref{eq:MAIN_Z_laughlin} the $\tilde{T}_{ij}$
are offset from integer by $\frac{q}{2}$ and subject to (\ref{eq:T_from_T_ij}).
This is a direct consequence of the fact that each $\vartheta_{1}\mbox{-}$function
in the Jastrow factor is taken to the power of $q$. The higher power
in the Jastrow factors is also noticeable in the factors $\tilde{Z}_{\tilde{T}_{ij}}^{\left(q\right)}$.
The $\tilde{Z}_{\tilde{T}_{ij}}^{\left(q\right)}$ can be thought
of as structure factors for the $q^{\mathrm{th}}$ power of a $\vartheta\mbox{-}$function.
It enters the Fourier expansion of the $\vartheta\mbox{-}$function
as 

\begin{eqnarray}
\left(\ellipticgeneralized abz{\tau}\right)^{q} & = & \sum_{\tilde{T}\in\mathbb{Z}+aq}e^{\i\pi\tau\frac{1}{q}\tilde{T}^{2}}e^{\i2\pi\tilde{T}\left(z+b\right)}\tilde{Z}_{\tilde{T}}^{\left(q\right)}\label{eq:Theta_Z_expansion}
\end{eqnarray}
 and was discussed briefly in Ref. \cite{Fremling_2013}. The sum
over $\tilde{T}$ is infinite and offset from the integers by $aq$.
Details of $\tilde{Z}_{\tilde{T}_{ij}}^{\left(q\right)}$ as well
as explicit expressions for it are given in \ref{App:Z_n_recursion}.

The factor $e^{\i\pi\tilde{T}_{ij}}$ in \ref{eq:MAIN_Z_laughlin}
also comes from the Jastrow factor. It is this piece that ensures
that $\mathcal{Z}\left(\mathbb{T}\right)$ is anti-symmetric with
respect to $\mathbb{T}$. The anti-symmetry comes about since an interchange
of $\mathbb{T}_{i}$ and $\mathbb{T}_{j}$ will only affect the factor
$e^{\i\pi\tilde{T}_{ij}}$. It does so by sending $e^{\i\pi\tilde{T}_{ij}}\to e^{-\i\pi\tilde{T}_{ij}}$.
Since $\tilde{T}_{ij}$ is a half-integer then $\tilde{T}_{ij}-\tilde{T}_{ji}\in2\mathbb{Z}+q$
is an odd (even) integer and $e^{-\i\pi\tilde{T}_{ij}}=\left(-1\right)^{q}e^{\i\pi\tilde{T}_{ji}}$.

See Figure \ref{fig:Triangles}b for and interpretation of the sum
$\sum_{i<j<k}\left(\tilde{T}_{ij}+\tilde{T}_{jk}+\tilde{T}_{ki}\right)^{2}$
as sum over triangles with corners in $\left(i,j\right)$, $\left(j,k\right)$
and $\left(i,k\right)$.

\begin{figure}
\begin{centering}
\begin{tabular}{cc}
\begin{tabular}{cccccc|c}
 &  &  &  &  & \multicolumn{1}{c}{} & $\sum_{j=1}^{N}\tilde{T}_{ij}$\tabularnewline
\cline{7-7} 
 & {\large{}0} & {\large{}$\tilde{T}_{12}$} & {\large{}$\tilde{T}_{13}$} & {\large{}$\cdots$} & {\large{}$\tilde{T}_{1N}$} & {\large{}$\mathbb{T}_{1}$}\tabularnewline
 & {\large{}$-\tilde{T}_{12}$} & {\large{}0} & {\large{}$\tilde{T}_{23}$} & {\large{}$\cdots$} & {\large{}$\tilde{T}_{2N}$} & {\large{}$\mathbb{T}_{2}$}\tabularnewline
 & {\large{}$-\tilde{T}_{13}$} & {\large{}$\tilde{T}_{12}$} & {\large{}0} & {\large{}$\cdots$} & {\large{}$\tilde{T}_{2N}$} & {\large{}$\mathbb{T}_{3}$}\tabularnewline
 & {\large{}$\vdots$} & {\large{}$\vdots$} & {\large{}$\vdots$} & {\large{}$\ddots$} & {\large{}$\vdots$} & {\large{}$\vdots$}\tabularnewline
 & {\large{}$-\tilde{T}_{1N}$} & {\large{}$-\tilde{T}_{2N}$} & {\large{}$-\tilde{T}_{3N}$} & {\large{}$\cdots$} & {\large{}0} & {\large{}$\mathbb{T}_{N}$}\tabularnewline
\cline{2-7} 
\multicolumn{1}{c|}{$\sum_{i=1}^{N}\tilde{T}_{ij}$} & {\large{}$-\mathbb{T}_{1}$} & {\large{}$-\mathbb{T}_{2}$} & {\large{}$-\mathbb{T}_{3}$} & {\large{}$\cdots$} & {\large{}$-\mathbb{T}_{N}$} & {\large{}0}\tabularnewline
\end{tabular} & \setlength{\unitlength}{0.9cm}
\begin{picture}(5,4)(.5,1.8)
\put(0.9,3.3){\large{$+\tilde{T}_{ij}$}}
\put(4.3,1.2){\large{$+\tilde{T}_{jk}$}}
\put(4.3,3.3){\large{$-\tilde{T}_{ik}$}}
\put(1,3){\circle*{0.1}}
\put(4,1){\circle*{0.1}}
\put(4,3){\circle*{0.1}}
\put(1,4){\circle*{0.01}}
\put(2,4){\circle*{0.01}}
\put(3,4){\circle*{0.01}}
\put(4,4){\circle*{0.01}}
\put(5,4){\circle*{0.01}}
\put(1,3){\circle*{0.01}}
\put(2,3){\circle*{0.01}}
\put(3,3){\circle*{0.01}}
\put(4,3){\circle*{0.01}}
\put(5,3){\circle*{0.01}}
\put(1,2){\circle*{0.01}}
\put(2,2){\circle*{0.01}}
\put(3,2){\circle*{0.01}}
\put(4,2){\circle*{0.01}}
\put(5,2){\circle*{0.01}}
\put(1,1){\circle*{0.01}}
\put(2,1){\circle*{0.01}}
\put(3,1){\circle*{0.01}}
\put(4,1){\circle*{0.01}}
\put(5,1){\circle*{0.01}}
\dashline{0.2}(1,3)(4,1)
\dashline{0.2}(4,3)(4,1)
\dashline{0.2}(4,3)(1,3)
\end{picture}\tabularnewline
 & \tabularnewline
$a)$ & $b)$\tabularnewline
\end{tabular}
\par\end{centering}

\caption{$a)$ A tabular view of relationship between $\tilde{T}_{ij}$ and
$\mathbb{T}_{i}$ in equation (\ref{eq:T_from_T_ij}) we well as the
balance condition $\sum_{i}\mathbb{T}_{i}=0$. $b)$ How the signs
of the terms in $\tilde{T}_{ij}$ are added to form (\ref{eq:MAIN_Z_laughlin}).
In the figure it is assumed that $i<j<k$.\label{fig:Triangles}}
\end{figure}

In (\ref{eq:MAIN_Z_laughlin}) we have for increased readability suppressed
the common factors $N\left(\tau\right)$ and $\left(\frac{2N_{\phi}\pi^{2}}{\tau_{2}}\right)^{\frac{N_{e}}{4}}$
that come from the CFT normalization and normalization of the $\eta_{k}$
respectively.

\subsection{Numerically evaluating the Fock coefficients\label{sub:Numerically-evaluating}}

In the previous part we showed that the Laughlin state (\ref{eq:wfn_Laughlin})
could be expanded in a Fock basis with coefficients $\mathcal{Z}\left(\mathbb{T}\right)$
given by (\ref{eq:MAIN_Z_laughlin}). Unfortunately the given expression
is not particularly helpful when it comes to numerical evaluation.
The culprit is the simultaneous infinite sums over all of the $\tilde{T}_{ij}$:s.
To alleviate this problem a bit, $\mathcal{Z}\left(\mathbb{T}\right)$
can be rewritten such that the sums can be performed in a recursive
manner. This will substantially reduce the scaling of the computation.
The details of this manipulation can be found in the supplemental
material.

A word of caution should be given. Although the optimized expression
for $\mathcal{Z}\left(\mathbb{T}\right)$ can be put on a computer,
the efficiency in evaluating it is still inferior to that of simply
diagonalizing the Haldane pseudo-potential Hamiltonian. Using exact
diagonalization the coefficients for around 12 particles can be extracted
numerically, but in our current implementation of $\mathcal{Z}\left(\mathbb{T}\right)$,
only 6 or so particles can be achieved. 

Our method can as mentioned above not compete with exact diagonalization
for the Laughlin state, but that is on the other hand not the main
purpose of the expansion. An obvious advantage of our approach is
the explicit access to the $\tau\mbox{-}$dependence of the Fock coefficients,
something the exact diagonalization can not provide.

Also, the reader should also be aware that the pseudo-potential trick
only exists for the Laughlin state. For states higher in the hierarchy
there is no local Hamiltonian for which these state are the exact
zero energy eigenstates. In these cases the results in this article
is the only way we are aware of to analytically extract Fock coefficients.

There is also another setting in which analytical knowledge of these
coefficients may be of use. This is when considering the non-chiral
extensions of the chiral HH-hierarchy. For this extension the $K\mbox{-}$matrix\cite{Wen_1991}
is spit into two separate pieces as $K=\kappa-\bar{\kappa}$. This
can be done even for the Laughlin case by splitting $K=q$ as $\kappa=q+p$
and $\bar{\kappa}=p$. On the torus, the CFT methods introduced in
Ref. \cite{Hermanns_2008} and Ref. \cite{Fremling_2014} can be generalized
to handle general $K\mbox{-}$matrices in an analogous way to the
procedure on the plane\cite{Hansson_2007a,Hansson_2007b}. The wave
function that is obtained\cite{Fremling_2013} contains factors $\elliptic 1{z_{ij}}{\tau}^{q+p}\elliptic 1{\bar{z}_{ij}}{\bar{\tau}}^{p}$
such that it is not only in the LLL any more. This wave function can
be projected on the LLL analytically using a basis of coherent states\cite{Fremling_2013}.
In this projection, the coefficients $\mathcal{Z}\left(\tau\right)$
and $\bar{\mathcal{Z}}\left(\bar{\tau}\right)$ of the underlying
Laughlin states appear naturally as ingredients. The details of the
analytic projection mentioned in the above paragraph will be the subject
of a future paper.

\section{Asymptotic Behavior in the TT-limit\label{sec:TT-limit-analysis}}

In this section we will return to the main equations (\ref{eq:MAIN_psi_laughlin_Z})
and (\ref{eq:MAIN_Z_laughlin}). We will analyze these expressions
in the limit of $\tau\to\i\infty$ (or $L\to0$ for constant area),
which is what we call the thin torus (TT) limit. We are interested
in studying the TT-limit to answer fundamental questions regarding
the plasma analogy. Using techniques from conformal field theory\cite{Read_2009}
a candidate for the normalization $\mathcal{N}\left(\tau\right)$
of the Laughlin state can be obtained -- see equation (\ref{eq:Laughlin_Norm_CFT}).
The assumption that $\mathcal{N}\left(\tau\right)$ captures the full
$\tau\mbox{-}$dependence is related to the plasma analogy introduced
by Laughlin\cite{Laughlin_1983}. It states that the normalization
of the Laughlin state is the Boltzmann weight for a single component
plasma of charged particles with a logarithmic electrostatic interaction.
As this plasma is know to be screening, the braiding properties of
the quasi-particles can be deduced, assuming they are separated far
enough. On the torus, the assumption of screening properties with
respect to $\tau$ allows us to compute \eg modular group representations.
In this section we will check if the screening holds also in the TT-limit.
We will expand on this discussion in part \ref{sub:Normalization}.

This section has three parts. In the first part, \ref{sub:Asymptotic-Z},
we analytically take the TT-limit and extract the dominant Fock contributions.
In this limit only a few of the Fock coefficients will remain nonzero,
and their analytical expressions are drastically simplified. We also
extract their relative scaling as a function of $\tau$. This knowledge
will then be used in parts \ref{sub:Normalization} and \ref{sub:Viscocity}
to make statements about the corrections to the CFT normalization
and the Hall viscosity in the TT-limit.

Section \ref{sub:Normalization} concerns the true normalization of
(\ref{eq:wfn_Laughlin}) in the TT-limit. We will use the fact that
there is only one particular configuration with non-zero Fock coefficient
in the TT-limit -- the root partition -- to compute the full normalization
when $\tau\to\i\infty$. From this knowledge we compute the asymptotic
correction to the normalization $\mathcal{N}\left(\tau\right)$ proposed
by the CFT construction in Ref. \cite{Read_2009}. The result shows
that the plasma analogy does not hold in the TT-limit. 

Under the assumption that $\mathcal{N}\left(\tau\right)$ is the full
normalization, \ie\emph{ }that the unknown piece $\mathcal{N}_{0}$
contains no $\tau\mbox{-}$dependent contributions, the Hall viscosity
$\eta^{H}$ for $\psi_{L}$ can be analytically extracted. In part
\ref{sub:Viscocity} we will analyze how the viscosity changes as
we approach the TT-limit. We will argue that since it is only the
root configuration that is present in this limit, the viscosity is
trivially $\eta^{H}=\frac{1}{4}\hbar\bar{n}$ instead of $\eta^{H}=\frac{3}{4}\hbar\bar{n}$
which is the value for the square torus. This is the same viscosity
as that of a single slater determinant state. To make the claim more
robust, and also give some indications of why the viscosity is independent
of system size in the regime of small $L$, we compute the viscosity
analytically for the root configuration and the sub-dominant Fock
contribution in this limit.

\subsection{Asymptotic scaling of the Laughlin state in TT-limit\label{sub:Asymptotic-Z}}

Let us now take a look at the TT-limit for the Laughlin state. In
this section we will show that the root configuration is the dominant
configuration\cite{Su_1981,Bergholtz_2008}, and we will also give
the relative scaling of all the two particle squeezed states.

We start from (\ref{eq:MAIN_Z_laughlin}) and use the short hand notation
$A\left(N_{e}\right)$ for the Gaussian weight in $\mathcal{Z}\left(\mathbb{T}\right)$:
\begin{equation}
A\left(N_{e}\right)=\sum_{i<j<k}\left(\tilde{T}_{ij}+\tilde{T}_{jk}+\tilde{T}_{ki}\right)^{2}.\label{eq:Laughlin_weight}
\end{equation}
 We are interested in the behavior deep in the TT-regime, when $\tau\to\i\infty$.
In this regime the Fock expansion will be exponentially dominated
by the configuration of $\tilde{T}_{ij}$ that can maximize the weight
\begin{equation}
W\left(\tilde{T}_{ij}\right)=e^{\i\pi\tau\frac{1}{qN_{e}}A\left(N_{e}\right)}\prod_{i<j}\tilde{Z}_{\tilde{T}_{ij}}^{\left(q\right)}.\label{eq:W_the_Z_weight}
\end{equation}

For large $\tau$ it is straight forward to show that $\elliptic 3z{\tau}^{M}\approx1$
for $-\frac{\tau_{2}}{2}<\Im\left(z\right)<\frac{\tau_{2}}{2}$ and
as a consequence $\tilde{Z}_{\tilde{T}}^{\left(q\right)}\to\delta_{\tilde{T},\frac{q}{2}}$
in this limit. That is, $\tilde{Z}_{\tilde{T}_{jj}}^{\left(q\right)}$
has its largest value $\tilde{Z}_{\tilde{T}_{jj}}^{\left(q\right)}\to1$
(and zero otherwise) when $\tilde{T}_{ij}=\pm\frac{q}{2}$. We take
as an ansatz that this configuration will also give the maximum for
$W\left(\tilde{T}_{ij}\right)$. Explicitly, we choose the ansatz
\begin{equation}
\tilde{T}_{ij}=\frac{q}{2}\left(\theta_{j,i}-\theta_{i,j}\right),\label{eq:TT-configuration}
\end{equation}
where $\theta$ is the Heaviside function 
\[
\theta_{i,j}=\left\{ \begin{array}{cc}
1 & i<j\\
0 & i\ge j
\end{array}\right..
\]
In words: $\tilde{T}_{ij}=\frac{q}{2}$ when $i>j$ and $\tilde{T}_{ij}=-\frac{q}{2}$
when when $i<j$. This particular ansatz is favorable since it produces
the correct root partition. The electron with label $i$ will (when
$s=0$) have momentum $k_{i}=\mathbb{T}_{i}=\sum_{j}\tilde{T}_{ij}$
such that $k_{i}=qi-\frac{q}{2}\left(N_{e}+1\right)$. The value of
$A\left(N_{e}\right)$ for this configuration is 
\[
A\left(N_{e}\right)=\sum_{i<j<k}\left(\frac{q}{2}+\frac{q}{2}-\frac{q}{2}\right)^{2}=\sum_{i<j<k}\frac{q^{2}}{4}=\frac{q^{2}}{24}\left(N_{e}^{3}-3N_{e}^{2}+2N_{e}\right).
\]
 The sum $P_{N_{e}}=\sum_{i<j<k}^{N_{e}}1=\frac{1}{6}\left(N_{e}^{3}-3N_{e}^{2}+2N_{e}\right)$
is computed either as the third order polynomial that has roots $P_{0}=P_{1}=P_{2}=0$
and $P_{3}=1$ or by using Faulhaber's formula $\sum_{k=1}^{n}k^{2}=\frac{1}{6}\left(2n^{3}+3n^{2}+n\right)$.

The reference weight is thus 
\begin{equation}
W_{TT}=W\left(\frac{q}{2}\left(\theta_{j,i}-\theta_{i,j}\right)\right)=e^{\i\pi\tau\frac{q}{24}\left(N_{e}^{2}-3N_{e}+2\right)}.\label{eq:W_the_Z_weight_TT}
\end{equation}

We must still ensure that the ansatz (\ref{eq:TT-configuration})
yields a global minimum (up to permutations of the electrons) of (\ref{eq:W_the_Z_weight}),
but here we only check that it is local\footnote[1]{A global proof has not been attempted.}.
To do so we investigate changes to $A\left(N_{e}\right)$ as one of
the $\tilde{T}_{ij}$ is varied. Due to the permutation symmetry in
$A\left(N_{e}\right)$ we can without loss of generality choose $i<j$.
There is a subtlety here; as we have chosen the reference TT-configuration
to be that of (\ref{eq:TT-configuration}), the ordering of the electrons
is implicit from $1$ to $N_{e}$. Another choice of the signs in
(\ref{eq:TT-configuration}) would lead to a corresponding permutation
of the particles. Mathematically this shows up in (\ref{eq:Laughlin_weight})
in such a way that $\tilde{T}_{ij}$ is negative in some terms and
positive in some others. We can interpret a change $\tilde{T}_{ij}\to\tilde{T}_{ij}+m$
as electron $i$ moving $m$ steps to the left and electron $j$ moves
$m$ steps to the right. Thus -- because of the implicit ordering
-- (for $i<j$) for $m>0$ is a squeeze and $m<0$ is an anti-squeeze.

We start by considering the simplest case of changing only $\tilde{T}_{12}\to\tilde{T}_{12}+m$.
The effect on $W\left(\tilde{T}_{ij}\right)$ comes from two factors.
The first is the change in $A$ that is $\Delta A=\left(N_{e}-2\right)m\left(m-q\right)$
and the second is from $\tilde{Z}_{\tilde{T}_{12}}^{\left(q\right)}$.
This factor becomes 

\begin{equation}
\tilde{Z}_{\frac{q}{2}+m}^{\left(q\right)}\approx\left(\begin{array}{c}
q\\
\left|m\right|
\end{array}\right)e^{\i\pi\tau\left(\left|m\right|-\frac{m^{2}}{q}\right)},\label{eq:Z_scaling}
\end{equation}

for $\frac{q}{2}\leq m\leq\frac{q}{2}$.

By looking just at $\Delta A$ we see that $\Delta A<0$ when $0<m<q$,
which translates into the particles being squeezed. This must be a
pathological result as it would render the TT-state unstable to clustering
of electrons. However, when we also take $\tilde{Z}_{\tilde{T}_{ij}}^{\left(q\right)}$
into account we get 
\begin{equation}
W\left(\tilde{T}_{ij}\right)\approx\left(\begin{array}{c}
q\\
\left|m\right|
\end{array}\right)e^{\i\pi\tau\frac{1}{qN_{e}}\Delta W}\times W_{TT}.\label{eq:W_relative_W_TT}
\end{equation}
The combined difference $\Delta W$ is

\begin{equation}
\Delta W=-2m^{2}+2\left|m\right|q\times\left\{ \begin{array}{lr}
1 & m>0\\
N_{e}-1 & m<0
\end{array}\right.,\label{eq:1p_sqeeze_distance}
\end{equation}
which for $\left|m\right|\leq\frac{q}{2}$ is always a positive. Note
the somewhat counterintuitive behavior that $\ldots010\overleftarrow{10}00\overrightarrow{01}010\ldots$
is more penalized than $\ldots0100\overrightarrow{01}\overleftarrow{10}0010\ldots$
in the TT-limit. The former would have an excitation energy of $\Delta E=2\Delta_{3}+\Delta_{4}+\cdots$
and the latter $\Delta E=\Delta_{2}+2\Delta_{3}+\cdots$ which is
larger. The terms $\Delta_{k}$ are greater than zero and can be though
of as pseudo-potentials, although their origin is slightly different.
For simple potentials like the Coulomb potential $\Delta_{1}>\Delta_{2}>\Delta_{3}>\ldots>0$.
See \eg\emph{ }equation (19) in Ref. \cite{Bergholtz_2008} for details.
Of course, using perturbation theory, the first correction would not
be given by the energy, but by the amplitudes of the hopping terms.
In this case the hopping term for squeezing $V_{3,-1}$ is larger
that anti-squeezing $V_{3,1}$. Thus, the relative scaling of the
squeezed $m=+1$ state is $qe^{\i\pi\tau\frac{2}{N_{e}q}\left(q-1\right)}$
whereas it is $qe^{\i\pi\tau\frac{2}{N_{e}q}\left(qN_{e}-q-1\right)}$
for the anti-squeezed state. 

\begin{figure}
\begin{centering}
\includegraphics[width=0.85\columnwidth]{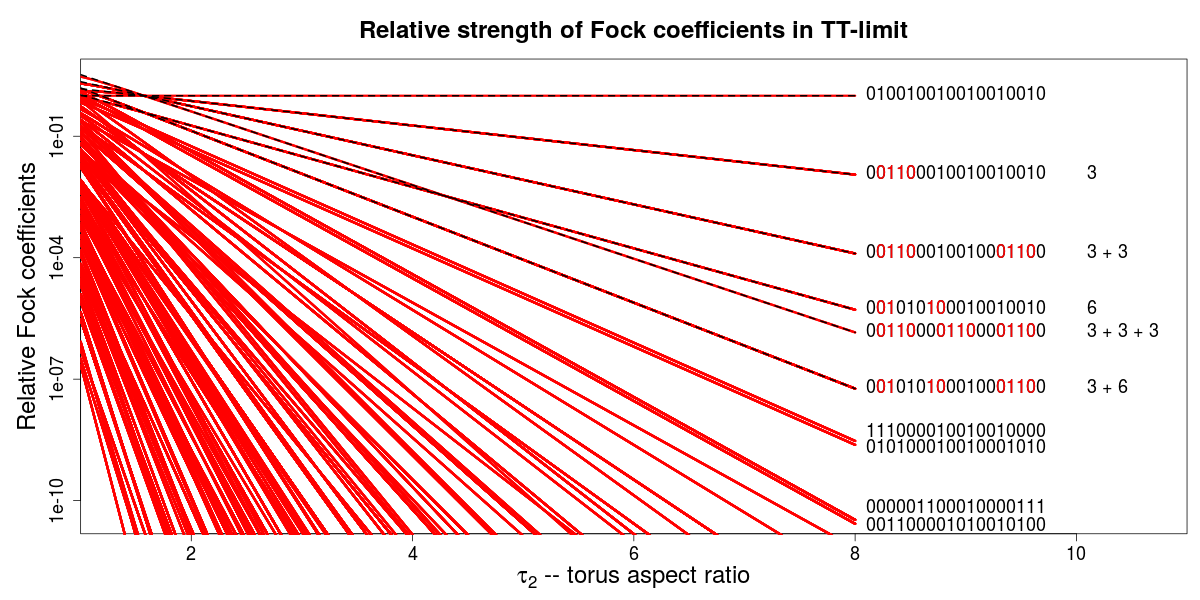}
\par\end{centering}

\caption{The relative scaling of the difference Fock coefficients for $N_{e}=6$
particles compared to the root TT- configuration $\ldots010010010010\ldots$.
The largest coefficients are labeled with the pattern of occupied
orbitals for that state. The black dashed lines correspond to analytic
scaling given by (\ref{eq:1p_sqeeze_distance}) and (\ref{eq:1P_sqeeze_Delta_distance}).
The analytic expansion fits well with the numeric data. The red text
highlights the two orbitals an electron has moved between. These states
are obtained by squeezing electrons that are $\delta=$$3$, $3+3$,
$6$, $3+3+3$, $3+6$ orbitals away. Here $\delta$ is the number
of orbitals between the squeezed electrons and all squeezes are one
step inward. The notation $a+b$ ($a+b+c$) means that two (three)
different pairs distances $a$ and $b$ ($c$) orbitals away are squeezed.
The analytic scaling is obtained by treating the two (three) squeezes
as independent. \label{fig:TT-scaling-of-coeffs}}
\end{figure}

In Figure \ref{fig:TT-scaling-of-coeffs} we see the relative scaling
of the root partition in comparison to the sub-dominant Fock-states
for $N_{e}=6$ particles. We see that the two leading sub-dominant
states are a squeezed ($0\overrightarrow{01}\overleftarrow{10}0010010010010$)
and a doubly squeezed ($0\overrightarrow{01}\overleftarrow{10}00100\overrightarrow{01}\overleftarrow{10}010$)
state. The order and scaling of these is well captured by the leading
order expansion (\ref{eq:1p_sqeeze_distance}). In the Figure it is
possible also possible to identify a triply squeezed state ($0\overrightarrow{01}\overleftarrow{10}00\overrightarrow{01}\overleftarrow{10}00\overrightarrow{01}\overleftarrow{10}0$)
and this is also well descried by (\ref{eq:1p_sqeeze_distance}) taken
three times.

Observe that as the area is fixed to $L^{2}\tau_{2}=2\pi N_{e}q$
we can rewrite the scaling difference as $e^{\i\pi\tau\frac{1}{qN_{e}}\Delta W}=e^{\frac{\i\pi\tau_{1}}{qN_{e}}\Delta W}e^{-\frac{2\pi^{2}}{L^{2}}\Delta W}$
where the scale depends only on $L$, and $\tau_{1}$ only changes
the relative phase. This shows that the relative scaling in the TT-limit
is independent of system size and depends only on $L$ instead of
$\tau$. This is actually to be expected since in the thin torus limit
the only relevant length scale is the short circumference $L$.

We can now also consider the more generic case where particle $i$
and particle $i+\Delta$ are squeezed $m$ steps. The change in $A$
this time is 

\begin{eqnarray}
\Delta A & = & \left(N_{e}-2\right)m^{2}-mq\left(N_{e}-2\Delta\right),\label{eq:Diff_B}
\end{eqnarray}

where $\Delta=j-i$. Taking into account the change from $\tilde{Z}_{\tilde{T}_{ij}}^{\left(q\right)}$
gives again (\ref{eq:W_relative_W_TT}) but this time with 
\begin{eqnarray}
\Delta W & = & -2m^{2}+2\left|m\right|q\times\left\{ \begin{array}{lr}
\Delta & m>0\\
N_{e}-\Delta & m<0
\end{array}\right..\label{eq:1P_sqeeze_Delta_distance}
\end{eqnarray}

Putting $\Delta=1$ yields the result in (\ref{eq:1p_sqeeze_distance}).
Just as with the special case $\tilde{T}_{12}$, $\Delta W>0$ for
small values of $m$. Note also that the formula for $m>0$ can be
mapped on the case $m<0$ by letting $m\to-m$ and $\Delta\to N_{e}-\Delta$.
This symmetry is natural since an inwards squeeze of $m$ steps a
distance $\Delta$ away can alternatively be seen as outward squeeze
of two particles $N_{e}-\Delta$ steps away.

Looking again at Figure \ref{fig:TT-scaling-of-coeffs} we see that
we can identify the one particle squeezed states ($0\overrightarrow{01}010\overleftarrow{10}0010010010$)
corresponding to ($\Delta=2$, $m=1$). We can also see ($0\overrightarrow{01}010\overleftarrow{10}00100\overrightarrow{01}\overleftarrow{10}0$)
which is a combination of ($\Delta=2,m=1$) and ($\Delta=1,m=1$)
acting on different pairs. The scaling of both of these are well described
by (\ref{eq:1P_sqeeze_Delta_distance}).

\subsection{$\mathcal{N}_{0}$ in the TT-limit\label{sub:Normalization}}

In this part we will investigate the possibility of TT-limit corrections
to the normalization of the Laughlin state proposed by the CFT construction
in Ref. \cite{Read_2009}. If TT-corrections turn out to be present
in the limit $N_{e}\to\infty$, it will have immediate consequences
for the validity of the plasma analogy. The plasma analogy states
-- in the form relevant here -- that the free energy of a single component
plasma in two dimensions on a torus does not depend on the geometry
of the torus, provided the area is held constant. The original plasma
analogy formulated by Laughlin states that the free energy does not
depend on the positions of test charges in the plasma, provided they
are sufficiently far apart\cite{Laughlin_1983}. In both cases the
validity of the analogy relies on the property that the plasma should
be screening.

At the end of Section (\ref{sub:The-Laughlin-wave}) we discussed
the behavior of the true normalization $\mathcal{N}_{0}$ for a small
system of $N_{e}=6$ electrons. We found by consulting Figure \ref{fig:N_0_in_tau_plane}
that as long as the torus is reasonably thick in both directions then
$\mathcal{N}_{0}$ is also constant. For $\tau\approx\i$, variations
of $\mathcal{N}_{0}$ appeared only in the fourth decimal (see Figure
\ref{fig:N_0_in_tau_plane}b). However, looking at more asymmetric
tori ($\tau_{2}>10$) we noted that $\mathcal{N}_{0}$ deviates from
being constant and develops clear $\tau\mbox{-}$dependence. This
we interpret as one of the torus handles being so small that the plasma
no longer can screen in that direction. As a consequence the plasma
stops to screen and the normalization obtained from CFT is incorrect.

We will now develop an approximate expression for the $\tau\mbox{-}$dependence
of $\mathcal{N}_{0}$ in the TT-limit. This is enabled since the expression
for $\psi_{L}$ simplifies in this limit. Due to the differences in
scaling, only one of the Fock states will survive and $\psi_{L}$
will be just a product state
\begin{equation}
\psi_{L}\to\psi_{TT}\left(z\right)=\mathfrak{F}_{\mathrm{TT}}\left(z\right).\label{eq:psi_L_TT-limit_FOck}
\end{equation}

This product state is precisely that of the root partition. Also,
since $\psi_{L}=\mathcal{N}_{0}^{-1}\psi_{s}$ the limit can be formulated
as 

\begin{eqnarray}
\psi_{L} & \to & \mathcal{N}_{0}^{-1}\mathcal{N}\left(\tau\right)\mathcal{Z}\left(\mathbb{T}\right)\sqrt{N_{e}!}\mathfrak{F}_{\mathrm{TT}}\left(z\right),\label{eq:psi_L_TT_limit_CFT}
\end{eqnarray}
 where both $\mathcal{N}\left(\tau\right)$ and $\mathcal{Z}\left(\mathbb{T}\right)$
are known analytically. Comparing (\ref{eq:psi_L_TT-limit_FOck})
and (\ref{eq:psi_L_TT_limit_CFT}) leads to the the form of $\mathcal{N}_{0}$
in the TT-limit $\left(\tau_{2}\to\infty\right)$ as 
\begin{equation}
\mathcal{N}_{0}\to\mathcal{N}_{\TT}=\mathcal{N}\left(\tau\right)\mathcal{Z}\left(\mathbb{T}\right)\sqrt{N_{e}!}.\label{eq:N0_sim_N_TT}
\end{equation}
Note that for $q=1$, $\mathcal{N}_{0}=\mathcal{N}_{\TT}$ for all
$\tau$ as in this case there exist only one Fock state, the filled
Landau level.

The expressions both for $\mathcal{N}\left(\tau\right)$ and $\mathcal{Z}\left(\mathbb{T}\right)$
do also simplify when $\tau\to\i\infty$ as all sub-dominant terms
will vanish. For instance, the Dedekind eta function $\eta\left(\tau\right)$
that appears in $\mathcal{N}\left(\tau\right)$ can be expressed as
$\eta\left(\tau\right)=e^{\i\pi\tau\frac{1}{12}}\prod_{n=1}^{\infty}\left(1-e^{\i2\pi\tau n}\right)$,
so the a asymptotic expansion is simply $\eta\left(\tau\right)\to e^{\i\pi\tau\frac{1}{12}}$.
Inserting this into (\ref{eq:Laughlin_Norm_CFT}) gives asymptotic
behavior of $\mathcal{N}\left(\tau\right)$ as

\[
\mathcal{N}\left(\tau\right)\to\frac{\tau_{2}^{\frac{qN_{e}}{4}}}{e^{\i\pi\tau\frac{1}{24}\left[qN_{e}\left(N_{e}-3\right)+2\right]}}.
\]
Similarly for $\mathcal{Z}\left(\mathbb{T}\right)$ the dominant $\tau\mbox{-}$dependent
factor of (\ref{eq:MAIN_Z_laughlin}) is given by (\ref{eq:W_the_Z_weight_TT})
and so
\begin{equation}
\mathcal{Z}\left(\mathbb{T}\right)\to\left(\frac{2N_{\phi}\pi^{2}}{\tau_{2}}\right)^{\frac{N_{e}}{4}}e^{\i\pi\tau\frac{q}{24}\left(N_{e}^{2}-3N_{e}+2\right)},\label{eq:Z_TT_scaling}
\end{equation}
 in the TT-limit. Inserting the above simplifications into (\ref{eq:N0_sim_N_TT})
gives 
\begin{equation}
\mathcal{N}_{\TT}=\sqrt{N_{e}!}\left(2qN_{e}\pi^{2}\right)^{\frac{N_{e}}{4}}\tau_{2}^{\frac{N_{e}}{4}\left(q-1\right)}e^{\i\pi\tau\frac{q-1}{12}}.\label{eq:Laughlin_norm_correction}
\end{equation}

\begin{figure}
\begin{centering}
\begin{tabular}{cc}
$a)$ & \tabularnewline
 & \includegraphics[width=0.6\columnwidth]{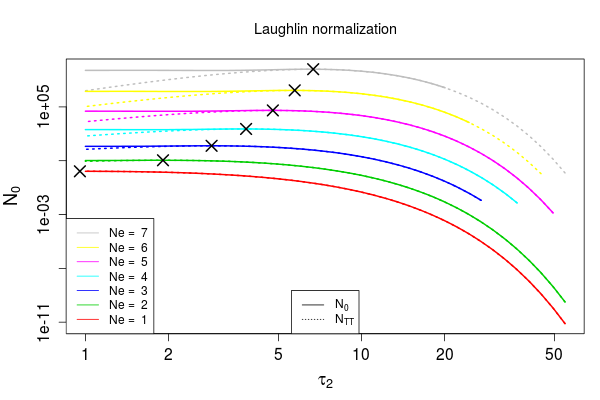}\tabularnewline
$b)$ & \tabularnewline
 & \includegraphics[width=0.6\columnwidth]{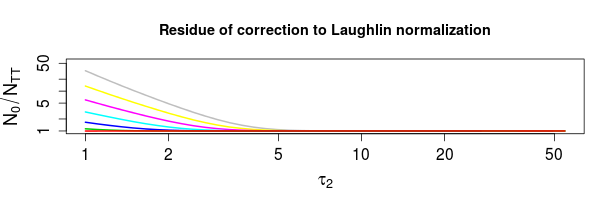}\tabularnewline
\end{tabular}
\par\end{centering}

\caption{Normalization of the Laughlin $\nu=\frac{1}{3}$ wave function defined
in (\ref{eq:wfn_Laughlin}) for rectangular tori. $a)$ Comparison
of the true normalization $\mathcal{N}_{0}$ (solid lines) to the
analytic approximation $\mathcal{N}_{\protect\TT}$ (dotted lines)
in (\ref{eq:Laughlin_norm_correction}) $b)$ The quotient $\mathcal{N}_{0}/\mathcal{N}_{\protect\TT}$
as a function of $\tau_{2}$. When $\tau_{2}>\frac{3N_{e}}{\pi}$
(marked by 'X') then $\mathcal{N}_{\protect\TT}\approx\mathcal{N}_{0}$
to very good accuracy.\label{fig:Laughlin_norm}}
\end{figure}

We should at this points ask ourselves how well $\mathcal{N}_{\TT}$
actually approximates $\mathcal{N}_{0}$, as it is obtained from a
rather crude analysis. It turns out that $\mathcal{N}_{\TT}$ does
a good job, provided $\tau$ is large enough. In Figure \ref{fig:Laughlin_norm}a
is displayed a comparison of $\mathcal{N}_{0}$ and $\mathcal{N}_{\TT}$
from $\tau_{2}=1$ and deep in the TT-regime. The solid lines display
$\mathcal{N}_{0}$ and the dotted lines $\mathcal{N}_{\TT}$. Two
things should be noted in this figure:

Firstly, from Figure \ref{fig:N_0_in_tau_plane} it was already clear
that while $\mathcal{N}_{0}$ is constant over a large region of $\tau\mbox{-}$space,
it does depend on $\tau$ for thin tori. In Figure \ref{fig:Laughlin_norm},
we now see that when this happens, \eg at $\tau_{2}\approx6$ for
$N_{e}=7$, $\mathcal{N}_{\TT}$ matches $\mathcal{N}_{0}$ very well
for all values of $N_{e}$ plotted. We thus conclude that $\mathcal{N}_{\TT}$
manages to capture the full $\tau\mbox{-}$dependence of $\mathcal{N}_{0}$
where this exists. We should however note that in the region where
$\mathcal{N}_{0}$ actually is constant, then the approximation $\mathcal{N}_{\TT}$
breaks down as it is not constant there.

\begin{figure}
\begin{centering}
\begin{tabular}{cccc}
$a)$ &  & $b)$ & \tabularnewline
 & \includegraphics[width=0.4\columnwidth]{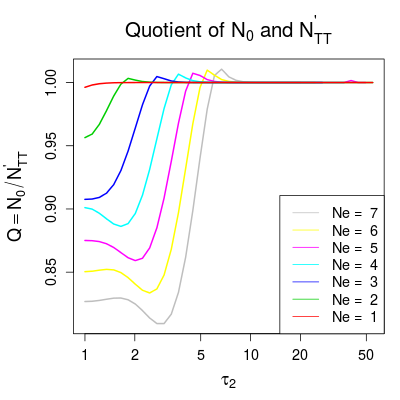} &  & \includegraphics[width=0.4\columnwidth]{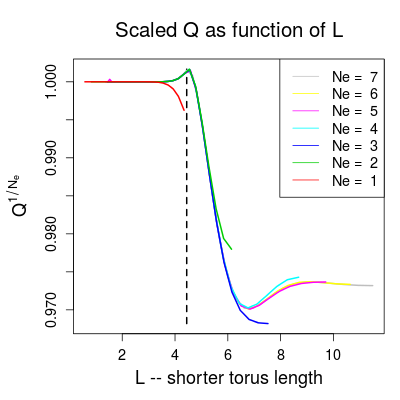}\tabularnewline
\end{tabular}
\par\end{centering}

\caption{Comparison of the modified TT-normalization $\mathcal{N}_{\protect\TT}^{\prime}$
(\ref{eq:N_TT_prime}) and the exact normalization $\mathcal{N}_{0}$
of the Laughlin $\nu=\frac{1}{3}$ wave function at rectangular tori.
$a)$ Plot of $Q=\mathcal{N}_{0}/\mathcal{N}_{\protect\TT}^{\prime}$
as a function of $\tau_{2}$. The revised normalization is accurate
to within factors of size unity. $b)$ The scaled quotient $Q^{\frac{1}{N_{e}}}$
as function of the shorter length scale $L$. This shows that $\mathcal{N}_{0}$
and $\mathcal{N}_{\protect\TT}^{\prime}$ only differ (up to small
finite size deviations) by a size independent factor that only depends
on the shorter length $L$ as $f^{N_{e}}\left(L\right)$. The dashed
line ($L=\pi\sqrt{\frac{2}{3}q}$) shows the break point between the
two regions of $\mathcal{N}_{\protect\TT}^{\prime}$.\label{fig:Laughlin_norm_residue}}
\end{figure}

Secondly, the maximum of $\mathcal{N}_{\TT}$ (at $\tau_{2}=\frac{3N_{e}}{\pi}$)
lies in a region where where $\mathcal{N}_{0}$ is still approximately
constant. Under the assumption that this will be the case also for
larger system sizes, we may refine the approximation of $\mathcal{N}_{\TT}$
in the region $1<\tau_{2}<\frac{3N_{e}}{\pi}$. We simply assume that
if $\tau_{2}\leq\frac{3N_{e}}{\pi}$ then no $\tau\mbox{-}$dependent
correction is needed and if $\tau_{2}\ge\frac{3N_{e}}{\pi}$ then
the correction should be according to (\ref{eq:Laughlin_norm_correction}).
The constant correction for $\tau_{2}\leq\frac{3N_{e}}{\pi}$ is naturally
set to the value of $\mathcal{N}_{\TT}$ at $\tau_{2}=\frac{3N_{e}}{\pi}$.
The revised approximation is thus

\begin{eqnarray}
\mathcal{N}_{\TT}^{\prime} & = & \left\{ \begin{array}{lr}
\mathcal{N}_{\TT}\left(\tau=\i\frac{3N_{e}}{\pi}\right) & 1<\tau_{2}\leq\frac{3N_{e}}{\pi}\\
\\
\mathcal{N}_{\TT}\left(\tau\right) & \tau_{2}\ge\frac{3N_{e}}{\pi}
\end{array}\right.\label{eq:N_TT_prime}\\
 & = & \sqrt{N_{e}!}\left(2qN_{e}\pi^{2}\right)^{\frac{N_{e}}{4}}\left\{ \begin{array}{lr}
\left(\frac{3N_{e}}{e\pi}\right)^{N_{e}\frac{q-1}{4}} & 1<\tau_{2}\leq\frac{3N_{e}}{\pi}\\
\\
\tau_{2}^{\frac{N_{e}}{4}\left(q-1\right)}e^{\i\pi\tau\frac{q-1}{12}} & \tau_{2}\ge\frac{3N_{e}}{\pi}.
\end{array}\right.\nonumber 
\end{eqnarray}
 We define $Q=\frac{\mathcal{N}_{0}}{\mathcal{N}_{\TT}^{\prime}}$
as the quotient between the exact normalization and the revised approximation
$\mathcal{N}_{\TT}^{\prime}$. In Figure \ref{fig:Laughlin_norm_residue}a
we show that the revised normalization $\mathcal{N}_{\TT}^{\prime}$
is correct up to factors of order unity for all $\tau$. To the best
of our knowledge this is the first estimate of the normalization of
the Laughlin state spanning both the thick torus and thin torus regimes,
although we acknowledge that a calculation for the thick cylinder
has been performed in Ref. \cite{Tokatly_2009}. We also note, in
the same figure, that $Q$ is close to 1, but often slightly bellow,
for all system sizes considered, \eg $1\gtrsim Q\geq0.8$ for $N_{e}=7$.

Further, as we argue that $L$ and not $\tau$ is the natural variable
to describe the physics, we may re-plot $Q$ as a function of $L$.
This is a reasonable action since $\tau_{2}=\frac{3N_{e}}{\pi}$ is
equivalent to $L=\pi\sqrt{\frac{2}{3}q}\approx4.44\sqrt{\frac{q}{3}}$
which is independent of $N_{e}$. This means that when $L<\pi\sqrt{\frac{2}{3}q}$
then $\mathcal{N}_{0}$ has $\tau\mbox{-}$dependence irrespective
of the system size. This is another example showing that the physics
in the TT-limit is dominated by the short torus length, and insensitive
to the overall size of the system. 

Figure \ref{fig:Laughlin_norm_residue}b plots $Q^{\frac{1}{N_{e}}}$
as a function of $L$, where we take the $N_{e}^{\mathrm{th}}$ root
to cancel the difference in scaling at different $N_{e}$. As should
be apparent, all the curves $\left[Q\left(N_{e},L\right)\right]^{\frac{1}{N_{e}}}$
have collapsed approximately on a single curve $f\left(L\right)$.
Here, $f\left(L\right)$ is an unknown function that is independent
of $N_{e}$. By graphical inspection the form of $f\left(L\right)$
seems to converge for system sizes as small as $N_{e}=5$. Also the
value of $f\left(L\right)$ seem to stabilize around $L\approx8\ell_{B}$
to $f\left(L\right)\approx0.973\left(3\right)$.

To obtain the numerical values for $\mathcal{N}_{0}$ we used exact
diagonalization of the Haldane pseudo-potential to access the properly
normalized $\psi_{L}$. This method will only work for system sizes
where we actually can perform the exact diagonalization and generate
real space samples. To see that $Q\left(N_{e},L\right)\approx f^{N_{e}}\left(L\right)$
beyond this regime we resort to estimating the norm by Monte-Carlo
(MC) and importance sampling. We then generate samples according to
the probability distribution $p\left(z\right)$ that we choose to
be the absolute square of the $\nu=1$ wave function. The MC estimate
of the norm is then 
\[
\mathcal{N}_{\mathrm{MC}}^{2}\approx\left(2\pi qN_{e}\right)^{N_{e}}\frac{1}{Z}\sum_{j=1}^{N_{\mathrm{MC}}}\frac{\left|\psi_{s}\left(z_{j}\right)\right|^{2}}{p\left(z_{j}\right)},
\]
where $Z=\sum_{j=1}^{N_{\mathrm{MC}}}\frac{1}{p\left(z_{j}\right)}$
and $N_{\mathrm{MC}}$ is the number of MC samples. The factor $\left(2\pi qN_{e}\right)^{N_{e}}$
is gives the volume integrated over, as importance sampling will only
compute the average value of $\left|\psi_{s}\right|^{2}$.

\begin{figure}
\begin{centering}
\includegraphics[width=0.6\columnwidth]{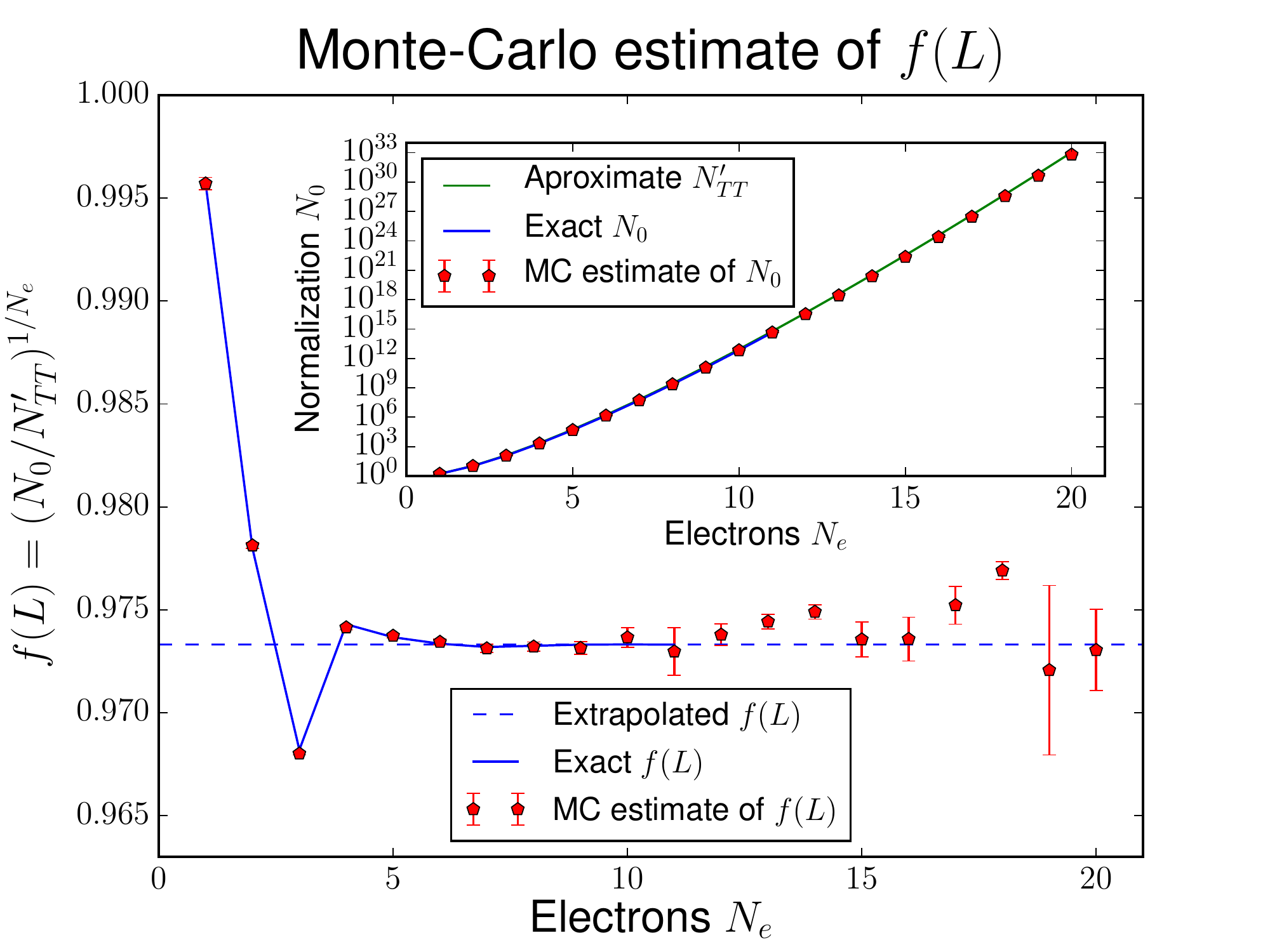}
\par\end{centering}

\caption{Monte-Carlo estimate of normalization of the Laughlin $\nu=\frac{1}{3}$
wave function defined in (\ref{eq:wfn_Laughlin}) at $\tau=\protect\i$,
for $N_{e}\cdot10^{7}$ configurations. (Inset) Comparison of the
MC normalization $\mathcal{N}_{\mathrm{MC}}$ (red) compared to the
exact $\mathcal{N}_{0}$ (blue) and the approximate $\mathcal{N}_{\protect\TT}^{\prime}$
(green) in (\ref{eq:N_TT_prime}). (Main plot) Plot of $f\left(N_{e}\right)=Q^{\frac{1}{N_{e}}}$
estimated by MC (red) compared to the exact (blue). This data suggests
that $f\left(L\right)$ approaches a constant when $L\to\infty$.
Note that as the MC calculation has a skew distribution, the more
accurate (lower) values have a larger error. \label{fig:Laughlin_norm_MC} }
\end{figure}

In Figure \ref{fig:Laughlin_norm_MC} we compare $\mathcal{N}_{\mathrm{MC}}$
to $\mathcal{N}_{0}$ and $\mathcal{N}_{\TT}^{\prime}$, for $\tau=\i$
(square torus) and up to $N_{e}=20$ electrons. Exact numeric values
for $\mathcal{N}_{0}$ exist for up to $N_{e}=11$ but beyond that
it is hard to evaluate in real space the ground state wave function
$\psi_{L}$. In the Figure \ref{fig:Laughlin_norm_MC} we see that
$\mathcal{N}_{\mathrm{MC}}$ and $\mathcal{N}_{0}$ agree (as they
should) to within MC errors for all system sizes where we can compute
$\mathcal{N}_{0}$ exactly. In the main figure we see that $Q\left(N_{e},L\right)\approx f^{N_{e}}\left(L\right)$
seems to hold beyond where we have exact data for $\mathcal{N}_{0}$.
We note that $\mathcal{N}_{\mathrm{MC}}$ has a tendency to overestimate
$\mathcal{N}_{0}$, which is why the lower values (\eg $N_{e}=19$)
has a much larger error than the higher ones (\eg $N_{e}=18$). We
conclude that using Monte Carlo strengthens the picture that $Q\left(N_{e},L\right)\approx f^{N_{e}}\left(L\right)$,
and that $f\left(L>8\ell_{B}\right)\approx0.973$ is constant. Another
way of stating the same is that $\mathcal{N}_{\TT}^{\prime}$ at $\tau\approx\i$
is correct up to a factor $0.973^{N_{e}}$, and sub-leading multiplicative
corrections must be polynomial in $N_{e}$.

\subsection{Modular Transformation Phases in the TT-limit\label{sub:Modular-Transformation-Phases}}

\begin{figure}
\begin{centering}
\includegraphics[width=0.6\columnwidth]{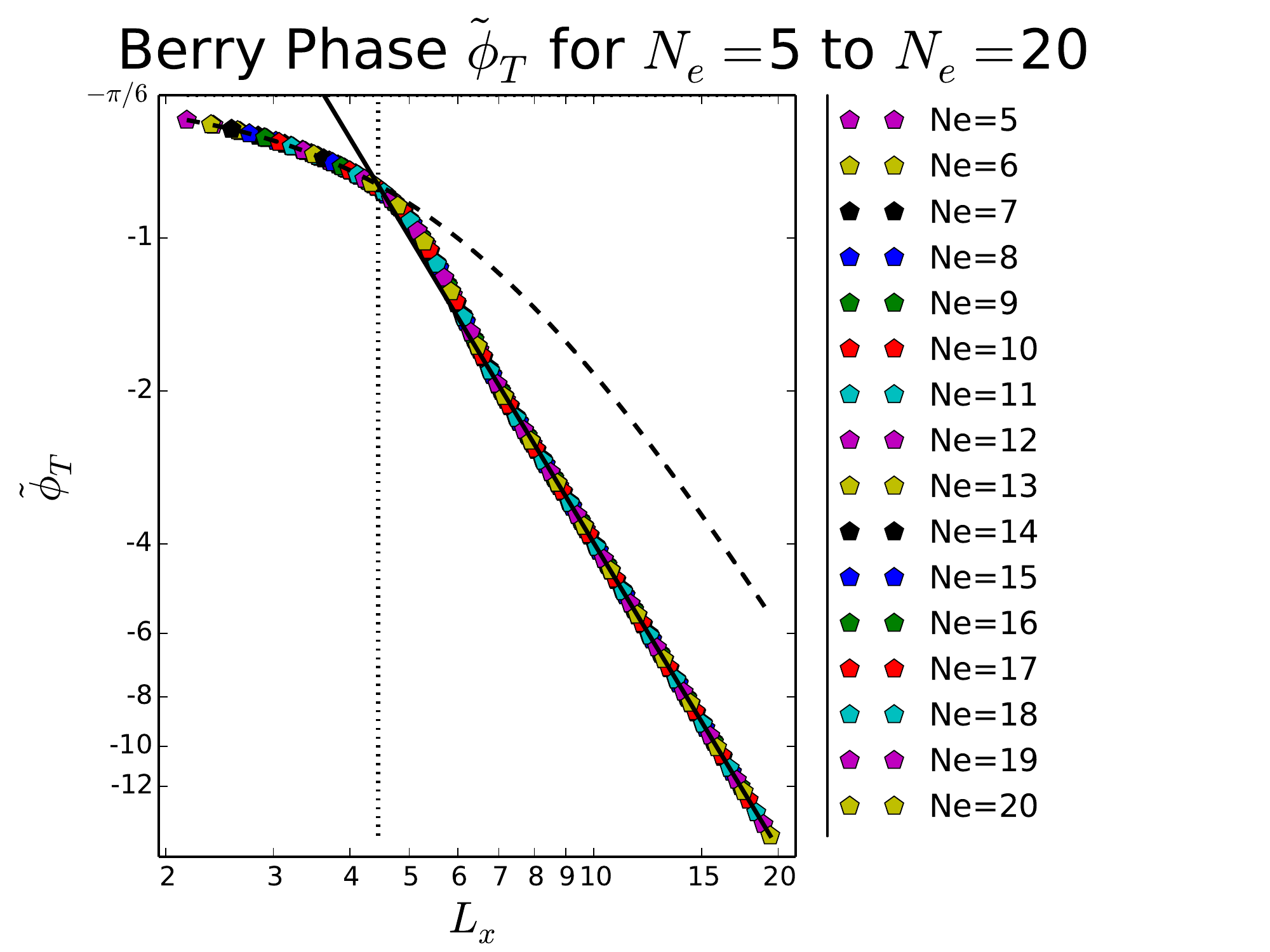}
\par\end{centering}

\caption{$\tilde{\phi}_{T}$ plotted as function of $L$, such that all data
collapses on the curves $-\frac{L^{2}}{8\pi}$ (dotted) and $-\frac{L^{2}}{24\pi}-\frac{\pi}{6}$
(dashed). The vertical (dotted) line marks $L=\pi\sqrt{\frac{2}{3}q}$
.\label{fig:Berry_T_by_Lx}}
\end{figure}

We are now in a position to answer what happened with the Berry phase
$\phi_{T}$ in the thermodynamic limit $\tau_{2}\to\infty$ in Figure
\ref{fig:Modular_T_transforms}. In the TT-limit we know that $\psi_{L}$
is just a slater determinant, so we know that it is on the form used
to obtain equation (\ref{eq:Berry_T_CFT}) and (\ref{eq:Berry_S_CFT}),
but his time with $P=\frac{N_{e}}{4}$. However we also know that
$\psi_{s}$ needs to be normalized by the factor $\mathcal{N}_{\TT}$
in (\ref{eq:Laughlin_norm_correction}) which contains a factor of
$e^{\i\pi\tau\frac{q-1}{12}}$. As a result, the the normalized $\psi_{L}^{\prime}=\frac{\psi_{s}}{\left|\mathcal{N}_{\TT}\right|}$
actually has the form $\tau_{2}^{P}e^{-\i\pi\tau_{1}\frac{q-1}{12}}\hat{\psi}\left(\left\{ z\right\} ,\tau\right)$.
The Berry phase picked up by this state is therefore 
\begin{equation}
\phi_{T}^{\left(TT\right)}=-\frac{N_{e}}{4\tau_{2}}-\pi\frac{q-1}{12}.\label{eq:Berry_T_TT}
\end{equation}
 Now, using the relation $L^{2}\tau_{2}=2\pi qN_{e}$ we can rewrite
the two curves (\ref{eq:Berry_T_CFT}) and (\ref{eq:Berry_T_TT})
as 
\[
\phi_{T}^{\left(CFT\right)}=-\frac{qL^{2}}{24\pi}\qquad\mbox{and}\qquad\phi_{T}^{\left(TT\right)}=-\frac{L^{2}}{24\pi}-\pi\frac{q-1}{12}.
\]
 Thus plotting $\phi_{T}$ in units of $L$ instead of $\tau_{2}$
all the data for different $N_{e}$ should now collapse onto the same
curve. Indeed, looking at Figure \ref{fig:Berry_T_by_Lx} this is
precisely what we see. Also, the crossover from $\phi_{T}^{\left(CFT\right)}$
to $\phi_{T}^{\left(TT\right)}$ happens again around $L=\pi\sqrt{\frac{2}{3}q}$. 

\subsection{Hall viscosity in the TT-limit\label{sub:Viscocity}}

In this section we will investigate the consequence for the Hall viscosity\cite{Avron_1995}
when there is $\tau\mbox{-}$dependence in $\mathcal{N}_{0}$. From
studying $\tilde{\phi}_{T}$ in Figure (\ref{fig:Berry_T_by_Lx})
we know that the proper parameter that determines $\tilde{\phi}_{T}$
is $L$ instead of $\tau$. It would thus seem likely that $L$ would
also determine the behavior of the viscosity $\bar{s}$.

\begin{figure}
\begin{centering}
\includegraphics[width=0.7\columnwidth]{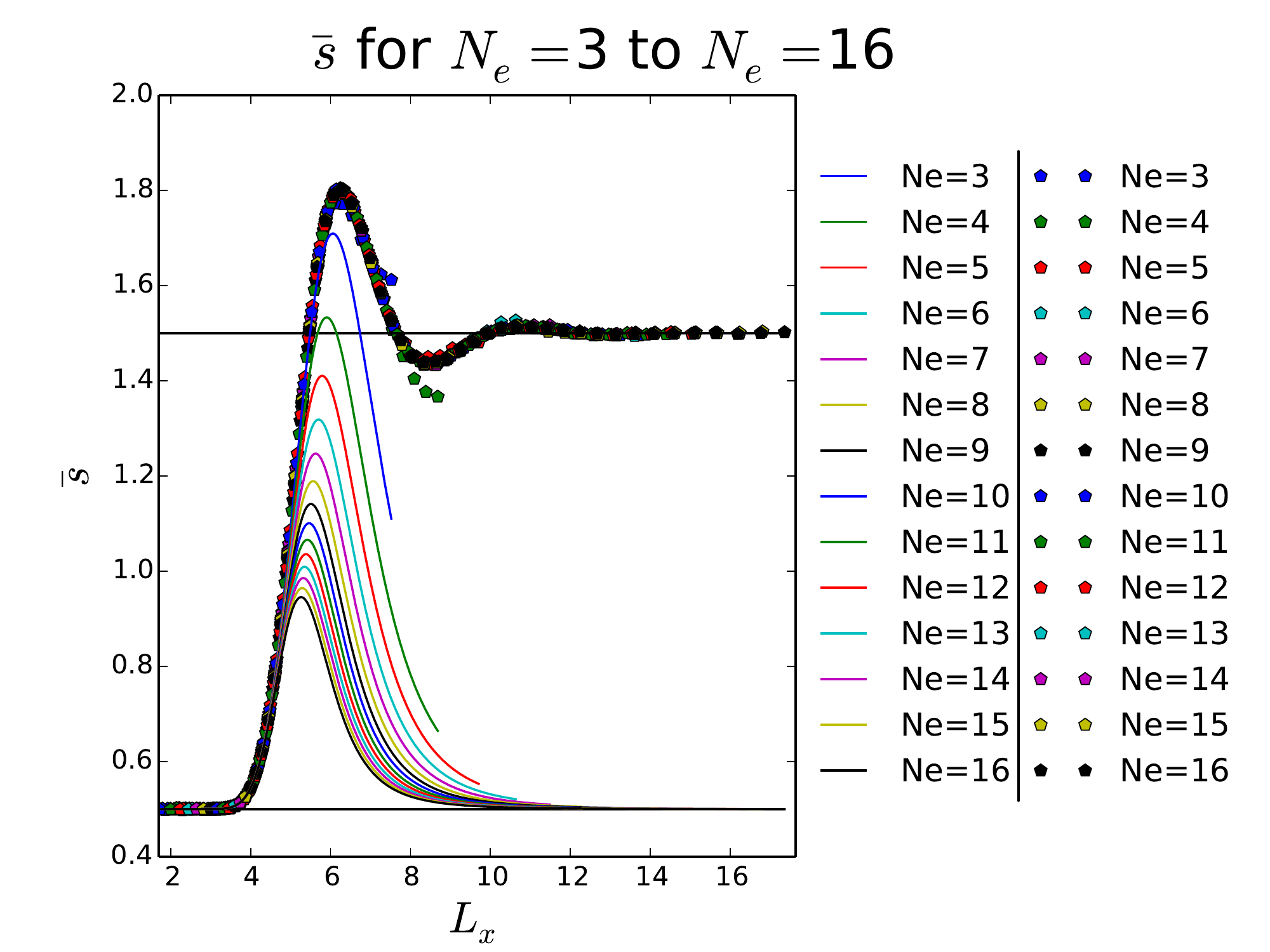}
\par\end{centering}

\caption{(Data Points) The viscosity (measured as effective average spin $\bar{s}$)
of the Laughlin state as function of $L$ for different $N_{e}$.
Note that the curves are almost identical for all $N_{e}$ even for
as small systems as $N_{e}=3$, hinting that the physics is dictated
by the shorter length scale $L$. For $L\gtrsim10\ell_{B}$ the viscosity
has stabilized at $\bar{s}=\frac{3}{2}$, which is the expected thermodynamic
value.\protect \\
(Solid Lines) Approximation of $\bar{s}$ using (\ref{eq:s_in_TT_limit})
for different values of $N_{e}$ as a function of $L$. Note that
this function has the right features of rising from $\bar{s}=\frac{1}{2}$
along the same curve almost independent of $N_{e}$. For $L\lesssim5$
the analytic and numerical results in the plot are in good agreement.
For $L\gtrsim5$ the curves start to deviate as the approximation
breaks down. \protect \\
\label{fig:vicocity_in_TT_lmit}}
\end{figure}

An example of this is shown in Figure \ref{fig:vicocity_in_TT_lmit},
where the data from Figure (\ref{fig:Berry_Curvature}) is replotted
as a function of $0<L<\sqrt{2\pi qN_{e}}$. The different curves extend
to different $L$, since the torus area increases with system size.
Two things should be noted here. Firstly, all the curves of $\bar{s}$
for different $N_{e}$ are approximately identical when plotted against
$L$. This again lends support to the interpretation that it is $L$
that is important for the physics. Secondly, as the thin torus is
approached -- $L\to0$ -- the value of $\bar{s}$ drops to $\bar{s}=\frac{1}{2}$
from the square torus value of $\bar{s}=\frac{3}{2}$. This indicates
that there are extra Berry phase contributions from the normalization
$\mathcal{N}_{0}$ and is direct evidence that the plasma is not screening
in the TT-limit. This last observation leads naturally to the need
for systematic corrections to $\mathcal{N}_{0}$, as a function of
$\tau$.

Using the results from section \ref{sub:Asymptotic-Z} and equation
(\ref{eq:N_TT_prime}) we can approximately compute the viscosity
of the Laughlin state in the thin torus region and in the square torus
region separately. We first consider $\psi=\frac{1}{\mathcal{N}_{\TT}^{\prime}}\psi_{s}$,
where $\mathcal{N}_{\TT}^{\prime}$ is constant for $L>\pi\sqrt{\frac{2}{3}q}$,
giving $\bar{s}=\frac{q}{2}$ as above. However for $L<\pi\sqrt{\frac{2}{3}q}$
there is an now an extra contribution $\tau_{2}^{\frac{N_{e}}{4}\left(q-1\right)}$.
Thus, according to (\ref{eq:Effective_sbar}) the viscosity will not
be $\bar{s}=\frac{q}{2}$ as expected from the plasma analogy. Instead
we see that the viscosity is $\bar{s}=\frac{q}{2}-\frac{q-1}{2}=\frac{1}{2}$.
This is the same viscosity as that of the fully filled LLL, or more
generally any product state of LL orbitals. In hindsight this result
is simple to understand physically by remembering that the Laughlin
state is the exact zero energy eigenstate of the Haldane pseudo-potential
Hamiltonian\cite{Haldane_1983}. In the extreme TT-limit all hopping
elements vanish and the Hamiltonian is purely electrostatic. The Hamiltonian
is therefore diagonal and the ground state has only one non-zero Fock
coefficient; that of the root partition. This wave function $\mathfrak{F}_{\TT}\left(z\right)$
has precisely the non-holomorphic factor $\tau_{2}^{\frac{N_{e}}{4}}$
which gives $\bar{s}=\frac{1}{2}$.

For completeness we mention the case of $q=1$, which is an unnormalized
$\nu=1$ wave function. In this case $\mathcal{N}_{0}=\mathcal{N}_{\TT}$
exactly as there is only on Fock state. Thus 
\begin{eqnarray*}
\mathcal{N}_{0} & = & \mathcal{N}\left(\tau\right)\mathcal{Z}\left(\mathbb{T}\right)\sqrt{N_{e}!}\\
 & = & \frac{\sqrt{N_{e}!}\left(2N_{\phi}\pi^{2}\right)^{\frac{N_{e}}{4}}}{\eta\left(\tau\right)^{\frac{N_{e}\left(N_{e}-3\right)}{2}+1}}\times\sum_{\left\{ \tilde{T}_{ij}\in\mathbb{Z}+\frac{1}{2}\right\} }e^{\i\pi\tau\sum_{i<j}\tilde{T}_{ij}^{2}}e^{-\i\pi\tau\frac{1}{N_{e}}\sum_{i}\mathbb{T}_{i}^{2}}e^{\i\pi\sum_{i<j}\tilde{T}_{ij}},
\end{eqnarray*}
 which is purely holomorphic as all $\tau_{2}$ factors in $\mathcal{Z}\left(\mathbb{T}\right)$
and $\mathcal{N}\left(\tau\right)$ cancel. Thus we see that $q=1$
has $\bar{s}=\frac{1}{2}$ for all values of $\tau$ -- as it should,
since it is only a single slater determinant. This is also an example
that shows that to compute the Hall viscosity analytically it is essential
to have control over the \emph{non-holomorphic $\tau\mbox{-}$}dependence
of the wave function. None of the holomorphic factors will affect
the value of the viscosity, which allows for the possibility that
$\mathcal{N}_{0}$ can have holomorphic dependence on $\tau$ without
affecting the viscosity of the state.

\subsubsection{Approximating the TT-limit viscosity}

In section \ref{sub:Asymptotic-Z} we computed the relative scaling
of the root partition and the one-pair squeezed states. In this section
we will now compute the viscosity for the Laughlin state in the TT-limit
under the approximation that only the root configuration and the leading
one-pair squeezed states contribute. We will obtain a viscosity that
is not dependent on $\tau_{1}$, and where $L$ instead of $\tau_{2}$
is the relevant length scale. As the approximation is quite crude,
we will not be able to reproduce the full features of $\bar{s}_{\mathrm{eff}}$
with this analysis, but we will see good agreement when $L\lesssim5$.
Due to the difference in scaling between the root partition and the
sub-dominant contributions the viscosity will reduce to $\bar{s}_{\mathrm{eff}}=\frac{1}{2}$
in the TT-limit.

We thus consider the root partition of the $q=3$ Laughlin state $\ket{TT}=\ldots001001001\ldots$
as well as the sub-dominant configuration $\ket{i,i+1}=\ldots00100\overrightarrow{01}\overleftarrow{10}001001\ldots$.
The number of nearest neighbor one-pair squeezed states are $N_{e}$.
From (\ref{eq:1p_sqeeze_distance}) we get the relative scaling between
these terms as $a\left(\tau\right)=3e^{\i\pi\tau\frac{4}{3N_{e}}}$
. The normalized asymptotic state can thus be written as
\begin{equation}
\ket{\tau}=N\left(\tau\right)\left(\ket{TT}+a\left(\tau\right)\sum_{i=1}^{N_{e}}\ket{i,i+1}\right),\label{eq:TT-single-hopping}
\end{equation}
 where $N^{-2}\left(\tau\right)=1+N_{e}9e^{-\pi\tau_{2}\frac{8}{3N_{e}}}$
is the normalization. To compute the viscosity we use 
\[
\bar{s}=\frac{1}{2}-\i\frac{2\tau_{2}^{2}}{N_{e}}\sum_{\mathbf{k}}\left(\partial_{\tau_{1}}\overline{a_{\mathbf{k}}}\cdot\partial_{\tau_{2}}a_{\mathbf{k}}-\partial_{\tau_{2}}\overline{a_{\mathbf{k}}}\cdot\partial_{\tau_{1}}a_{\mathbf{k}}\right),
\]

which is an (for $\bar{s}$ instead of $\eta_{H}$) adapted version
of equation (2.99) in Ref. \cite{Read_2011}. When applied to $\ket{\tau}$
and identifying $a_{0}=N\left(\tau\right)$ and $a_{i}=N\left(\tau\right)a\left(\tau\right)$
we obtain
\begin{eqnarray}
\bar{s}\left(\tau\right) & = & \frac{1}{2}-\i\frac{2\tau_{2}^{2}}{N_{e}}\left[\left(\partial_{\tau_{1}}\bar{N}\cdot\partial_{\tau_{2}}N-\partial_{\tau_{2}}\bar{N}\cdot\partial_{\tau_{1}}N\right)\vphantom{\sum_{i=1}^{N_{e}}\left(\partial_{\tau_{1}}\left(Na^{\star}\right)\right)}\right.\nonumber \\
 &  & \left.+\sum_{i=1}^{N_{e}}\left(\partial_{\tau_{1}}\left(Na^{\star}\right)\cdot\partial_{\tau_{2}}\left(Na\right)-\partial_{\tau_{2}}\left(Na^{\star}\right)\cdot\partial_{\tau_{1}}\left(Na\right)\right)\right]\nonumber \\
 & = & \frac{1}{2}-\i2\tau_{2}^{2}\left(\partial_{\tau_{1}}\left(Na^{\star}\right)\partial_{\tau_{2}}\left(Na\right)-\partial_{\tau_{2}}\left(Na^{\star}\right)\cdot\partial_{\tau_{1}}\left(Na\right)\right)\\
 & = & \frac{1}{2}+\frac{64e^{\frac{8\pi\tau_{2}}{3N_{e}}}\pi^{2}\tau_{2}^{2}}{N_{e}^{2}\left(e^{\frac{8\pi\tau_{2}}{3N_{e}}}+N_{e}\right)^{2}}=\frac{1}{2}+\frac{2304\pi^{4}e^{\frac{16\pi^{2}}{L^{2}}}}{L^{4}\left(e^{\frac{16\pi^{2}}{L^{2}}}+9N_{e}\right)^{2}},\label{eq:s_in_TT_limit}
\end{eqnarray}
 which apart from a factor of $N_{e}$ in the denominator only depends
on $L$. The contribution from the first term ($a_{0}$) in (\ref{eq:TT-single-hopping})
disappears completely as $\partial_{\tau_{1}}N=0$. Note that $\bar{s}\left(\tau\right)$
does not depend on $\tau_{1}$ and only on $L$. This is an encouraging
result as it tells us that the viscosity is insensitive to skewing
of the torus geometry. Again a natural result if we expect the physics
to be dictated by the shorter length scale $L$. 

In Figure \ref{fig:vicocity_in_TT_lmit} we plot (\ref{eq:s_in_TT_limit})
as a function of $L$ and $N_{e}$. We observe that $\bar{s}\left(\tau\right)$
rises from $\bar{s}=\frac{1}{2}$ almost independently of $N_{e}$.
If we set (artificially) $N_{e}=0$ we get the idealized upper curve
$\bar{s}\left(\tau\right)=\frac{1}{2}+\frac{2304\pi^{4}}{L^{4}}e^{-\frac{16\pi^{2}}{L^{2}}}$
which starts to deviate from $\bar{s}=\frac{1}{2}$ at about $L\gtrsim4$.
We can with this construction show explicitly that in the deep TT-limit
we will approach $\bar{s}=\frac{1}{2}$ as $\frac{e^{-\frac{16\pi^{2}}{L^{2}}}}{L^{4}}\to0$
when $L\to0$.

Now, if we compare to the exact viscosity in the same figure we see
that there is good quantitative agreement between (\ref{eq:s_in_TT_limit})
and the viscosity of the full Laughlin state when $L\lesssim5$. For
$L$ larger than this value the analytic approximation for the viscosity
breaks down. We stress that we should not expect perfect agreement
for all $L$ with this simple model. The reasons are several. Firstly,
when $\tau_{2}\to1$ there are going to be increasingly more states
that are relevant, which we ignore here. Secondly, the relative scaling
in (\ref{eq:1P_sqeeze_Delta_distance}) is only the leading expansion
of relative scaling, and will also break down when $\tau_{2}\to1$.

We summarize this section by stating that the viscosity in the TT-limit
is not $\bar{s}=\frac{q}{2}$ as expected from the Read conjecture.
However, the $L\mbox{-}$dependence of $\bar{s}$ is -- up to small
corrections -- independent of system size. Numerical studies of \eg\emph{
}$\nu=\frac{2}{5}$ suggests that this holds true for, at least some,
more complicated filling fractions\cite{Moran_2015}.

\section{Fock Expansion of the Hierarchy States\label{sec:Hierachy-expand}}

In this section we will generalize the results obtained in Section
\ref{sec:Laughlin-Expansion} for the chiral Haldane-Halperin hierarchy
states recently constructed on the torus using CFT techniques\cite{Hermanns_2008,Fremling_2014}.
The Laughlin wave function fits in to a much larger framework of trial
wave functions, namely the Haldane-Halperin hierarchy of incompressible
states\cite{Haldane_1983,Halperin_1983}. These states have been realized
using correlation functions in CFT on both plane\cite{Hansson_2007a,Hansson_2007b},
sphere\cite{Kvorning_2013} and torus\cite{Hermanns_2008,Fremling_2014}.
In order to make this section reasonably self contained we review
the basic structure of the torus Haldane-Halperin hierarchy\cite{Haldane_1983,Halperin_1983}
wave functions. We here summarize the construction of the Haldane-Halperin
hierarchy on the torus and refer to Ref. \cite{Fremling_2014} and
references therein for more details.

The topological information regarding the quantum Hall fluids are
encoded in the Wen-Zee $K\mbox{-}$matrix\cite{Wen_1991}. The matrix
is symmetric with integer entries and has the same dimensionality
$n$ as the level of the hierarchy state. The $K\mbox{-}$matrix formalism
describes $n$ different layers with $N_{\alpha}$ particles in layer
$\alpha$. The distribution of particles in the different layers are
determined by the requirement that the quantum Hall droplet should
be homogeneous. Under this requirement, the group sizes are given
by $N_{\phi}\sum_{\beta}K_{\alpha\beta}^{-1}=N_{\alpha}$. The filling
fraction is readily extracted as $\nu=\sum_{\alpha\beta}K_{\alpha\beta}^{-1}$.\footnote[1]{In
this article we are exclusively working in the basis where $t=\left(1,\ldots,1\right)$.}

For the Jain series\cite{Jain_2007_Book} at $\nu=\frac{n}{mn+1}$
, the $K\mbox{-}$matrix is $n\mbox{-}$dimensional and has a particularly
simple form $K_{\alpha\beta}=m+\delta_{\alpha\beta}$. The Jain series
is a subset of the chiral hierarchy, which will be considering in
this work. As examples, the $K\mbox{-}$matrices for the Laughlin
$\nu=\frac{1}{q}$ states and the $\nu=\frac{2}{5}$ Jain state are
given by $K=q$ and $K=\left(\begin{array}{cc}
3 & 2\\
2 & 3
\end{array}\right)$.

Electronic wave function for the HH-hierarchy are constructed using
correlation functions in CFT. The correlation functions consist of
insertions of $n$ different types of electronic operators, one for
each layer, and a neutralizing background charge operator. Many-particle
wave functions with well defined single particle boundary conditions
are identified with the conformal blocks of primary operators in the
CFT. At filling fraction $\nu=\frac{p}{q}$, the many-particle wave
functions can be written as

\begin{equation}
\psi_{s}\left(z\right)=\mathcal{N}\left(\tau\right)e^{\i\pi\tau N_{\phi}\sum_{i}y_{i}}\prod_{i<j}^{N_{e}}\elliptic 1{\frac{z_{ij}}{L}}{\tau}^{K_{ij}}\mathcal{F}_{s}\left(z,\tau\right).\label{eq:Wfn_CFT}
\end{equation}
 Here $K_{ij}$ should be interpreted as $K_{\alpha\beta}$ for the
groups $\alpha$, $\beta$ that $i,j$ belong to. Just as with the
Laughlin state, the label $s$ enumerates the $q$ different degenerate
states that has to exist when the denominator of $\nu=\frac{p}{q}$
is $q$. Also, in analogy to the Laughlin case, these states can be
transformed into each other by rigid translations of all the particles.
The CFT construction also gives a proposal for the normalization factor
$\mathcal{N}\left(\tau\right)$.

Note that for two particles in the same group, the anti-symmetry properties
of (\ref{eq:Wfn_CFT}) are dictated by the Jastrow factor $\prod_{i<j}\elliptic 1{z_{ij}}{\tau}^{K_{ij}}$
alone and is not influenced by the CoM piece $\mathcal{F}_{s}\left(z,\tau\right)$.
This means that each group is fully anti-symmetric within the group.
The same is not true for particles in different groups. 

The chiral hierarchy has the property that all the edge modes travel
in the same direction. This manifests itself in that all the eigenvalues
of $K$ are positive. This enables us to parametrize $K$ using charge
vectors $\mathbf{q}_{\alpha}$. These are real $n\mbox{-}$dimensional
vectors such that $\mathbf{q}_{\alpha}\cdot\mathbf{q}_{\beta}=K_{\alpha\beta}$.
The vectors $\mathbf{q}_{\alpha}$ span an infinite $n\mbox{-}$dimensional
charge lattice $\Gamma$. The charge lattice $\Gamma$ is important
for the construction of the CoM function. The CoM function is given
by 
\begin{equation}
\mathcal{F}_{s}\left(z,\tau\right)=\sum_{\mathbf{q}\in\Gamma}e^{\i\pi\tau\left(\mathbf{q}+\mathbf{h}_{s}\right)^{2}}e^{\i2\pi\left(\mathbf{q}+\mathbf{h}_{s}\right)\left(\mathbf{Z}+\mathbf{t}_{s}\right)},\label{eq:CoM_CFT}
\end{equation}

where $\mathbf{Z}=\frac{1}{L}\sum_{i}\mathbf{q}_{i}z_{i}$. The parameters\textbf{
$\mathbf{h}_{s}$} and\textbf{ }$\mathbf{t}_{s}$ carry the many-body
momentum index. They\textbf{ }are chosen such that for periodic boundary
conditions $e^{\i2\pi\mathbf{t}_{s}\cdot\mathbf{q}_{i}}=e^{\i2\pi\mathbf{h}_{s}\cdot\mathbf{q}_{i}}=\left(-1\right)^{N_{\phi}+K_{ii}}$
and $e^{\i2\pi\mathbf{h}_{s}\cdot\frac{\mathbf{Q}}{N_{\phi}}}=e^{\i2\pi s\frac{p}{q}}\left(-1\right)^{p\left(N_{\phi}+K_{11}\right)}$.
In the last equation appears the total charge $\mathbf{Q}=\sum_{i}^{N_{e}}\mathbf{q}_{i}$
of the combined charge vectors. To ensure that all the particles are
at the same flux the homogeneity condition 
\begin{equation}
\mathbf{Q}\cdot\mathbf{q}_{j}=\sum_{j=1}^{N_{e}}K_{ij}=N_{\phi},\label{eq:Homogeneity_condition}
\end{equation}
 has to apply. This can also be formulated in terms of $K_{\alpha\beta}$
as the homogeneity condition that dictates the group sizes; $\sum_{\beta}K_{\alpha\beta}N_{\beta}=N_{\phi}$.
To obtain the physical wave function from the CFT conformal blocks,
equation (\ref{eq:Wfn_CFT}) has to be anti-symmetrized and external
translation operators have to be added as

\begin{equation}
\tilde{\psi}_{s}=\mathcal{A}\left\{ \prod_{\alpha=1}^{n}\mathbb{D}_{\left(\alpha\right)}^{\alpha-1}\psi_{s}\right\} ,\label{eq:Wfn_CFT_Full}
\end{equation}
 where the $\mathcal{A}$ stands for anti-symmetrization. The operator
$\mathbb{D}_{\left(\alpha\right)}$ is the torus counterpart of the
external derivatives on the plane $\mathbb{D}_{\left(\alpha\right)}^{\mathrm{Plane}}=\prod_{i_{\alpha}\in I_{\alpha}}\partial_{z_{i_{\alpha}}}$.
The anti-symmetrization in the above equation is there to ensure that
all particles are indistinguishable. For the Laughlin state in (\ref{eq:wfn_Laughlin})
the $\mathcal{A}$ is redundant as that state is by construction an
anti-symmetric single-component wave function. Note also that $\mathcal{A}$
acts primarily between groups as all the particles within the same
group are by construction anti-symmetrized. For a detailed discussion
about the construction of $\mathbb{D}_{\left(\alpha\right)}$, $\psi_{s}$
and $\tilde{\psi}_{s}$ we refer to \cite{Fremling_2014} and references
therein.

\subsection{Main Results and Formulas\label{sub:Main-Results-Hierarchy}}

The Fock expansion for the chiral hierarchy in (\ref{eq:Wfn_CFT})
can be obtained by straight forward generalization of the method used
to analyze the Laughlin state. As such, also the form and interpretation
of the different terms are very similar. Thus, the more general (\ref{eq:Wfn_CFT})
can be written as 

\begin{equation}
\psi_{s}=\mathcal{N}\left(\tau\right)\sum_{\left\{ \tilde{T}_{ij}\in\mathbb{Z}+\frac{K_{ij}}{2}\right\} }\mathcal{Z}\left(\tilde{T}_{ij}\right)\sum_{\left\{ m_{\alpha}\right\} }\prod_{i=1}^{N_{e}}\zeta_{k_{i}}\left(z_{i}\right).\label{eq:MAIN_HH_expansion}
\end{equation}
The structure is again analogous to that of (\ref{eq:MAIN_psi_laughlin_Z}).
The main difference is that there is an extra summation over the $n$
indexes $m_{\alpha}$, $\alpha=1,\ldots,n$. These indexes $m_{\alpha}$
come from the CoM function (\ref{eq:CoM_CFT}) and are here brought
to the form where the they only appear in the definition of the momentum
$k_{i}$ 
\begin{equation}
k_{i}=\mathbb{T}_{i}+N_{\phi}\sum_{\alpha}m_{\alpha}\delta_{i,N_{\alpha}}+s.\label{eq:momentum_HH}
\end{equation}

As (\ref{eq:MAIN_HH_expansion}) is written, the sums over $m_{\alpha}$
can immediately be performed to produce $\sum_{\left\{ m_{\alpha}\right\} }\prod_{\alpha}\zeta_{k_{N_{\alpha}}}\left(z_{N_{\alpha}}\right)=\left(\frac{2N_{\phi}\pi^{2}}{\tau_{2}}\right)^{\frac{n}{4}}\prod_{\alpha}\eta_{k_{N_{\alpha}}}\left(z_{N_{\alpha}}\right)$.
However we write (\ref{eq:MAIN_HH_expansion}) with this summation
unperformed as it only affects $n$ of the $N_{e}$ factors in $\prod_{i=1}^{N_{e}}\zeta_{k_{i}}\left(z_{i}\right)$. 

This time, $\mathcal{Z}\left(\tilde{T}_{ij}\right)$ is not invariant
under $\mathbb{T}_{i}\to\mathbb{T}_{i}+N_{\phi}$ which means that
we can not write (\ref{eq:MAIN_HH_expansion}) on the form $\sum_{\left\{ \mathbb{T}_{i}\right\} }\tilde{\mathcal{Z}}\left(\mathbb{T}_{i}\right)\prod_{i=1}^{N_{e}}\eta_{k_{i}}\left(z_{i}\right)$
with a simple expression for $\tilde{\mathcal{Z}}\left(\mathbb{T}_{i}\right)$.
This is not to say that it cannot be done, only that $\tilde{\mathcal{Z}}\left(\mathbb{T}_{i}\right)$
will not have the simple structure that $\mathcal{Z}\left(\mathbb{T}_{i}\right)$
in (\ref{eq:MAIN_Z_laughlin}) has. The $\mathbb{T}_{i}$ still obey
the same balance condition $\sum_{i}\mathbb{T}_{i}=0$ as earlier.
The total momentum is just as above $K_{\mathrm{total}}=\sum_{i}k_{i}=N_{e}s\,\mathrm{mod}\,N_{\phi}$.

The generalized weight is now

\begin{eqnarray}
\mathcal{Z}\left(\tilde{T}_{ij}\right) & = & \exp\left\{ \i\pi\tau\left(\sum_{i<j}\frac{\tilde{T}_{ij}^{2}}{K_{ij}}-\frac{1}{N_{\phi}}\sum_{i}\mathbb{T}_{i}^{2}\right)\right\} \prod_{i<j}\left\{ e^{\i\pi\tilde{T}_{ij}}\tilde{Z}_{\tilde{T}_{ij}}^{\left(K_{ij}\right)}\right\} ,\label{eq:MAIN_Z_HH}
\end{eqnarray}

which has the same type of structure as (\ref{eq:MAIN_Z_laughlin}).
Again the $\tilde{T}_{ij}$ are anti-symmetric just as for the Laughlin
state. Similarly the factor $e^{\i\pi\tilde{T}_{ij}}$ is responsible
of for the anti-symmetry under the exchange of two $\mathbb{T}_{i}$
from the same group. 

Note that, as the different groups of electrons now are distinguishable,
$\tilde{\mathcal{Z}}\left(\mathbb{T}_{i}\right)$ is not proportional
to the Fock coefficients, which was the case for the Laughlin state.
In fact the states in the hierarchy needs to go through (\ref{eq:Wfn_CFT_Full})
to become physical electronic wave functions for a single component
state. As a result, the coefficients given by $\tilde{\mathcal{Z}}\left(\mathbb{T}_{i}\right)$
would \emph{not} even be the Fock coefficients, but rather the components
that enter into them.

\subsection{TT-limit scaling for HH-states}

In Section \ref{sec:TT-limit-analysis} the asymptotic scaling of
the Fock coefficients where extracted for the Laughlin state. A similar
analysis should be possible also for the general hierarchy states
using (\ref{eq:MAIN_Z_HH}). Just as with the Laughlin state the dominant
state should be the TT-state\cite{Tao_1983,Bergholtz_2008}. This
is the state where the electrons are maximally separated along the
long dimension of the torus. The analysis is made more complicated
by the fact that $\mathcal{Z}\left(\tilde{T}_{ij}\right)$ is not
invariant under $\mathbb{T}_{i}\to\mathbb{T}_{i}+N_{\phi}$ which
gives more than one weight that corresponds to the root partition.

We can however, without performing that full analysis, shed some light
on an observation made in an earlier paper. In Ref. \cite{Fremling_2014}
it was noted that in comparison to the Coulomb ground state at $\nu=\frac{2}{5}$,
some of the translation operators that appear in (\ref{eq:Wfn_CFT_Full})
gives better overlaps than others. For instance, in the TT-limit,
the translation operator $D_{1,0}$ gives the best overlap with coulomb,
whereas $D_{0,1}$ yields an overlap that is almost zero. See Fig.
3 of that paper. The understanding we had was that $D_{1,0}$ constituted
a smaller translation and therefore resembled a derivative more than
$D_{0,1}$ did. From a technical point of view we can now also say
that since the TT-limit is dominated by the single slater determinant
of the root partition the effect of $D_{1,0}$ and $D_{0,1}$ are
drastically different. All Fock-states are eigenstates of the operator
$D_{1,0}$, and so it will only give rise to a phase factor. The $D_{0,1}$
operator on the other hand will shift all the electrons in one group
relative to all the others, which changes the configuration. As an
example, for $\nu=\frac{2}{5}$, the TT-configuration is$\ldots010100101001010\ldots$.
When acted on by $D_{1,0}$ it will just pick up a phase and becomes
$e^{\i2\pi\sum_{i}\frac{k_{i}}{N_{\phi}}}\ldots010100101001010\ldots$.
Under the action of $D_{0,1}$ however, it becomes $\ldots001100011000110\ldots$
which is a configuration with almost no overlap with the Coulomb ground
state.

\section{Summary}

In this paper we have rewritten the Laughlin state in the momentum
Fock-basis. The techniques developed here can also be applied to the
full chiral Haldane-Halperin hierarchy wave functions on the torus,
although the results are not as easily interpreted as for the Laughlin
wave function. We have further re-expressed the Fock-coefficients
of the Laughlin state in a recursive form for faster numerical evaluation. 

We used the analytic expressions for the Fock coefficients to study
the Laughlin state deep in the TT-limit. We did this by analyzing
the asymptotic $\tau\mbox{-}$dependence of the Fock coefficients.
Using this asymptotic expansion we have analytically computed viscosity
in the TT-limit as well as the leading $\tau\mbox{-}$dependence on
the correction to the normalization. We found a correction to the
proposed CFT-normalization that is accurate up to a factor $f^{N_{e}}\left(L\right)$. 

Our analytical approximation for the viscosity show qualitative but
not quantitative agreement with the numerically obtained viscosity.
We conclude that the parameter that dictates the viscosity in the
TT-limit is the short length scale $L$ and not $\tau$. Thus, any
state sufficiently deep into the TT-limit will always have the ``trivial''
viscosity of $\bar{s}=\frac{1}{2}$. Or findings suggests that the
plasma is not screening in the TT-limt. This is supported by the need
to correct the normalization by a $\tau_{2}$ dependent factor when
$L\lesssim4.4\ell_{B}$. This results thus warns that Hall viscosity
can not be used as a probe to detect topological information on this
asymmetric geometry. We still believe, however, that the plasma analogy
holds in the thermodynamic limit, both for the Laughlin state and
for the rest of the hierarchy states, and we have given some numerical
evidence that this is the case in this paper. We further believe that
you can study the viscosity as a function of $L$ to determine whether
a topological phase has stabilized or not. By thermodynamic limit,
we here mean the limit where both the torus axes are much larger than
the magnetic length.

One of the advantages of studying the TT-limit is that it can be exactly
solved while still being adiabatically connected to the bulk state
at $\tau=\i$\cite{Bergholtz_2006,Bergholtz_2008}. The gap to excitations
remains open when tuning $\tau$ between $\tau=\i$ and $\tau=\i\infty$
and the charges of the quasi-particles are the same. What we have
seen however in this paper is that some other properties does change,
such as viscosity, but there are other quantities that also get altered.
One such is the exclusion statistic\cite{Haldane_1991} for the quasi-particles
which changes\cite{Kardell_2011} at about the same $L$ as the viscosity.
This serves as a reminder that the TT-physics is in many cases the
same as in the bulk, but that there are notable exceptions that one
needs to be weary of.

\subsection*{Acknowledgments}

I would like to thank Thors Hans Hansson, Joost Slingerland and Eddy
Ardonne for many helpful comments on the manuscript. The exact diagonalization
was carried out using the Hammer\footnote[1]{http://www.thphys.nuim.ie/hammer}
package. This work was supported by the Swedish Science Research Council
and Science Foundation Ireland Principal Investigator Award 12/IA/1697.
I also wish to acknowledge the SFI/HEA Irish Centre for High-End Computing
(ICHEC) for the provision of computational facilities and support.\\

\bibliographystyle{alpha}
\bibliography{References}

\appendix

\section{A recursive formula for $Z_{n}^{\left(N\right)}$\label{App:Z_n_recursion}}

In this appendix we list some of the properties of the $\vartheta\mbox{-}$function
structure factor $Z_{k}^{\left(N\right)}$. This object appears in
the exponentiation of the generalized $\vartheta\mbox{-}$function
as

\begin{eqnarray}
\ellipticgeneralized abz{\tau}^{N} & = & \sum_{T\in\mathbb{Z}}e^{\i\pi\tau\frac{1}{N}\left(k+aN\right)^{2}}e^{\i2\pi\left(k+aN\right)\left(z+b\right)}\tilde{Z}_{k}^{\left(N\right)}.\label{eqapp:Theta_Z_expansion}
\end{eqnarray}

We define $Z_{k}^{\left(N\right)}$ as 
\[
Z_{k}^{\left(N\right)}=\sum_{\left\{ t_{i}\right\} =-\infty}^{\infty}e^{\i\pi\tau\sum_{i=1}^{^{N}}\tilde{t}_{i}^{2}},
\]
 where $\sum_{i}\tilde{t}_{i}=0$ and $\sum_{i}t_{i}=k$ and $t_{i}=\tilde{t}_{i}+\frac{k}{N}$.
This enables us to rewrite it as 

\begin{equation}
Z_{k}^{\left(N\right)}=\sum_{\left\{ t_{i}\right\} ^{N-1}=-\infty}^{\infty}e^{\i\pi\tau\sum_{i=1}^{^{N}}\left(t_{i}-\frac{k}{N}\right)^{2}}\label{eq:Compact_Z_n}
\end{equation}
 where $\sum_{i}t_{i}=k$. By letting $t_{i}\to-t_{i}$ we easily
see that $Z_{-k}^{\left(N\right)}=Z_{k}^{\left(N\right)}$. As one
of the $k_{i}$ is linearly dependent on the others, we may without
loss of generality we choose it to be $t_{N}=k-\sum_{i=1}^{N-1}t_{i}$,
such that $\frac{k}{N}-t_{N}=\frac{k}{N}-k+\sum_{i=1}^{N-1}t_{i}=\sum_{i=1}^{N-1}t_{i}-\frac{k}{N}\left(N-1\right)$.
We may now rewrite $Z_{k}^{\left(N\right)}$ as depending on $N-1$
linearly independent variables $\left\{ t_{i}\right\} ^{N-1}$ as

\[
Z_{k}^{\left(N\right)}=\sum_{\left\{ t_{i}\right\} ^{N-1}=-\infty}^{\infty}e^{\i\pi\tau\sum_{i=1}^{N-1}\left(t_{i}-\frac{k}{N}\right)^{2}}e^{\i\pi\tau\left(T_{N-1}-\frac{k}{N}\left(N-1\right)\right)^{2}},
\]
 where we introduced the dummy variable $T_{N-1}=\sum_{i=1}^{N-1}t_{i}$.
We note that $Z_{k}^{\left(N\right)}=Z_{k+N}^{\left(N\right)}$ as
the shift $k\to k+N$ is reverted by the re-summation $t_{i\neq N}\to t_{i\neq N}+1$,
\ie\emph{ }$T_{N-1}\to T_{N-1}+N-1$. We now note that we may extract
a reduced form of (\ref{eq:Compact_Z_n}) by identifying $k$ with
$T_{N-1}$ and $N$ with $N-1$. We then get 
\[
Z_{k}^{\left(N\right)}=\sum_{t_{N-1}=-\infty}^{\infty}e^{\i\pi\tau N\left(N-1\right)\left(\frac{T_{N-1}}{N-1}-\frac{k}{N}\right)^{2}}Z_{T_{N-1}}^{\left(N-1\right)}.
\]
 Since $T_{N-1}\to T_{N-1}+N-1$ and $k\to k+N$ cancel each other
we can rewrite the recursion using $\vartheta\mbox{-}$functions as

\begin{eqnarray}
Z_{k}^{\left(N\right)} & = & \sum_{t\in\mathbb{Z}_{N-1}}Z_{t}^{\left(N-1\right)}\ellipticgeneralized{\frac{t}{N-1}-\frac{k}{N}}00{N\left(N-1\right)\tau}\label{eq:Z_N_recursive}
\end{eqnarray}

with the root function $Z_{t}^{\left(1\right)}=1$. 

\global\long\def\i{\imath}

\global\long\def\plll{\mathcal{P}_{\mathrm{LLL}}}

\global\long\def\TT{\mathrm{TT}}

\global\long\def\T{\mathcal{T}}

\global\long\def\S{\mathcal{S}}

\global\long\def\Td{\mathcal{T}\mbox{-}}

\global\long\def\Sd{\mathcal{S}\mbox{-}}

\global\long\def\elliptic#1#2#3{\vartheta_{#1}\left(\left.#2\vphantom{#3}\right|#3\right)}

\global\long\def\ellipticweight#1#2{\kappa\left[\left.#1\vphantom{#2}\right|#2\right]}

\global\long\def\shortelliptic#1#2{\vartheta_{#1}\left(#2\right)}

\global\long\def\ellipticgeneralized#1#2#3#4{\vartheta\left[\begin{array}{c}
 #1\\
 #2 
\end{array}\right]\left(\left.#3\vphantom{#4}\right|#4\right)}

\global\long\def\ket#1{\left|#1\right\rangle }

\global\long\def\bra#1{\left\langle #1\right|}

\global\long\def\braket#1#2{\left\langle #1\left|\vphantom{#1}#2\right.\right\rangle }

\global\long\def\ketbra#1#2{\left|#1\vphantom{#2}\right\rangle \left\langle \vphantom{#1}#2\right|}

\global\long\def\braOket#1#2#3{\left\langle #1\left|\vphantom{#1#3}#2\right|#3\right\rangle }

\global\long\def\mbf#1{\mathbf{#1}}

\newpage{}

\title{Supplementary material:\\
Success and failure of the plasma analogy for Laughlin states on a
torus }

\newcommand{\zlaughlin}[0]{(main-text:23)} 

\newcommand{\psiz}[0]{(main-text:22)} 

\newcommand{\wfnlaughlin}[0]{(main-text:7)} 

\newcommand{\fockbasis}[0]{(main-text:5)} 

\newcommand{\basiswfn}[0]{(main-text:1)} 

\newcommand{\fouerierterms}[0]{(main-text:3)} 

\newcommand{\foueriersum}[0]{(main-text:4)} 

\newcommand{\thetaexp}[0]{(main-text:25)} 

\newcommand{\ZHH}[0]{(main-text:72)} 

\newcommand{\HHexp}[0]{(main-text:70)} 

\newcommand{\CFTCoM}[0]{(main-text:67)}

\section{Expanding Laughlin in a Fock Basis\label{sec:Laughlin-Expansion-1}}

In this section we give the details on how to rewrite the real space
Laughlin wave function \wfnlaughlin\, in the single particle orbital
basis of \fockbasis, \ie the Fock basis and arrive at equation \zlaughlin.

The key to constructing the Fock expansion is to extract the Fourier
components $e^{\i2\pi kz}$ from \wfnlaughlin\, as they will directly
map onto the Fourier components of \basiswfn. In order to do so,
these components need to be extracted from the Jastrow factors as
well as the center of mass factor. At this stage of the construction
one may identify all the factors containing $z$ with a corresponding
$\zeta_{k}\left(z\right)$. Reconstructing the full single particle
orbitals $\eta_{k}\left(z\right)$ is then done by massaging the sums
over $k$.

\subsection{Fourier expanding the Laughlin state\label{sub:Expand-Jastrow}}

In this part we give the steps to reach \zlaughlin. When Fourier
expanding the Jastrow factors we will use an expansion where the Fourier
factor is explicit. The generalized $\vartheta$-function to the power
$M$ can be expanded as in \thetaexp\, where $Z_{T}^{\left(M\right)}$
is a factor that encodes all the information about the $M^{\mathrm{th}}$
power. We have the relations $Z_{T+M}^{\left(M\right)}=Z_{T}^{\left(M\right)}$
and by mirror symmetry, $\elliptic 3z{\tau}=\elliptic 3{-z}{\tau}$
\footnote[1]{Here $\elliptic 3z{\tau}=\ellipticgeneralized 00z{\tau}$
is the third Jacobi theta function.}, function also $Z_{T}^{\left(M\right)}=Z_{-T}^{\left(M\right)}$.

Applying the expansion \thetaexp\, to the full Jastrow factor in
\wfnlaughlin\, gives the form 

\begin{equation}
\prod_{i<j}\elliptic 1{z_{ij}}{\tau}^{q}=\sum_{\left\{ T_{ij}\in\mathbb{Z}\right\} }e^{\i\pi\tau\sum_{i<j}\frac{1}{q}\left(T_{ij}+\frac{q}{2}\right)^{2}}e^{\i2\pi\sum_{i<j}\left(T_{ij}+\frac{q}{2}\right)\left(\frac{z_{ij}}{L}+\frac{1}{2}\right)}\prod_{i<j}Z_{T_{ij}}^{\left(q\right)},\label{eq:jastrow}
\end{equation}
 where with each pair $z_{ij}$ has been associated with a summation
index $T_{ij}$. As $z_{ij}$ is anti-symmetric in its indices $z_{ij}=-z_{ji}$
we will require that $T_{ij}$ be antisymmetric too. We choose a skewed
anti-symmetry condition $T_{ij}+T_{ji}=-q$ to make the anti-symmetry
for $\tilde{T}_{ij}$ trivial. To facilitate further calculations
we introduce the non-integer variable $\tilde{T}_{ij}=T_{ij}+\frac{q}{2}$,
which has symmetry properties $\tilde{T}_{ij}=-\tilde{T}_{ji}$. From
this follows that $\tilde{T}_{ii}=0$ for all $i$\footnote[2]{Formally
$\tilde{T}_{ii}$ is never defined, as $z_{ii}$ does not appear in
the Jastrow factor (\ref{eq:jastrow}).}. We also introduce the shifted
$Z$-weight $\tilde{Z}_{\tilde{T}_{ij}}^{\left(q\right)}=\tilde{Z}_{\tilde{T}_{ij}-q}^{\left(q\right)}=Z_{T_{ij}}^{\left(q\right)}$
which has the same symmetries as $Z_{T_{ij}}^{\left(q\right)}$ and
especially obeys $\tilde{Z}_{\tilde{T}_{ij}}^{\left(q\right)}=\tilde{Z}_{\tilde{T}_{ji}}^{\left(q\right)}$.
We also introduce the variable 
\begin{equation}
\mathbb{T}_{i}=\sum_{j=1}^{N_{e}}\tilde{T}_{ij},\label{eq:bb_T}
\end{equation}
 to simplify the $z$-dependent factor $\sum_{i<j}\tilde{T}_{ij}z_{ij}=\sum_{i,j}^{N_{e}}z_{i}\tilde{T}_{ij}=\sum_{i}^{N_{e}}\mathbb{T}_{i}z_{i}$.
Here $\mathbb{T}_{i}$ is thus related to the total momentum of particle
$i$. Note that because of the anti-symmetry of $\tilde{T}_{ij}$
there is a balance condition on $\mathbb{T}_{i}$ giving 
\begin{equation}
\sum_{i=1}^{N_{e}}\mathbb{T}_{i}=\sum_{i,j=1}^{N_{e}}\tilde{T}_{ij}=0.\label{eq:T_Balance_contition}
\end{equation}
Putting all of these things together, the Jastrow factor can be written
as

\begin{eqnarray}
\prod_{i<j}\elliptic 1{z_{ij}}{\tau}^{q} & = & \sum_{\left\{ \tilde{T}_{ij}\in\mathbb{Z}+\frac{q}{2}\right\} }e^{\i\pi\tau\sum_{i<j}\frac{\tilde{T}_{ij}^{2}}{q}}e^{\i\pi\sum_{i<j}\tilde{T}_{ij}}e^{\i2\pi\sum_{i}\mathbb{T}_{i}\frac{z_{i}}{L}}\prod_{i<j}\tilde{Z}_{\tilde{T}_{ij}}^{\left(q\right)}.\label{eq:Jastrow_expansion-copy}
\end{eqnarray}
 Note that we have managed to extract the phase factor $e^{\i2\pi\mathbb{T}_{i}\frac{z_{i}}{L}}$
for each coordinate separately. The price we have paid is the introduction
of the $\tilde{T}_{ij}$ which label the interdependence of the different
momentum components. The balance condition (\ref{eq:T_Balance_contition})
ensures that the wave function will be in a well defined momentum
sector.

We may in a similar fashion as for the Jastrow factor also express
the CoM function in the relevant Fourier components. To be able to
encompass any type of boundary conditions we work with the more general
form $\mathcal{F}_{h,t}\left(\left\{ z\right\} ,\tau\right)=\ellipticgeneralized{\frac{h}{q}}t{qZ}{q\tau}$.
At the end of the calculation we can set $h=t=s$ and recover the
result in \psiz. 

Using this parametrization the CoM piece is rewritten as

\begin{equation}
\mathcal{F}_{h,t}\left(\left\{ z\right\} ,\tau\right)=\sum_{m}e^{\i\pi\tau q\left(m^{2}+2m\frac{h}{q}+\frac{h^{2}}{q^{2}}\right)}e^{\i2\pi\left(\sum_{i}mq\frac{z_{i}}{L}+mt+\sum_{i}h\frac{z_{i}}{L}+\frac{ht}{q}\right)}.\label{eq:CoM_expansion}
\end{equation}

Putting together the expansions (\ref{eq:Jastrow_expansion-copy}),
(\ref{eq:CoM_expansion}) and the Gaussian factor gives us the wave
function
\begin{eqnarray}
\psi_{h,t} & = & e^{\i\pi\tau N_{\phi}\sum_{i}y_{i}^{2}}\sum_{\left\{ \tilde{T}_{ij}\in\mathbb{Z}+\frac{q}{2}\right\} }e^{\i\pi\tau\sum_{i<j}\frac{\tilde{T}_{ij}^{2}}{q}}e^{\i\pi\sum_{i<j}\tilde{T}_{ij}}\prod_{i<j}\tilde{Z}_{\tilde{T}_{ij}}^{\left(q\right)}\nonumber \\
 &  & \times\sum_{m}e^{\i\pi\tau q\left(m+\frac{h}{q}\right)^{2}}e^{\i2\pi\left(\mathbb{T}_{i}+mq+h\right)\sum_{i}\frac{z_{i}}{L}}e^{\i2\pi\left(mt+\frac{ht}{q}\right)}.\label{eq:Psi_expanded}
\end{eqnarray}
 The next step is to extract the single particle wave functions \basiswfn\,
from (\ref{eq:Psi_expanded}) by forming basis state factors of the
form \fouerierterms. Me may now collect the factors containing $z_{i}$
and $y_{i}^{2}$ into $\zeta_{k_{i}}\left(z_{i}\right)$ using \fouerierterms.
This gives rise to the substitution 

\[
e^{\i\pi\tau N_{\phi}\sum_{i}y_{i}^{2}}e^{\i2\pi k_{i}\sum_{i}\frac{z_{i}}{L}}=e^{-\i\pi\tau\frac{1}{N_{\phi}}\sum_{i}k_{i}^{2}}\prod_{i=1}^{N_{e}}\zeta_{k_{i}}\left(z_{i}\right)
\]
 where 
\begin{equation}
k_{i}=\mathbb{T}_{i}+mq+h.\label{eq:momentum_TT_relations-copy}
\end{equation}

We can now see that the role that $h$ plays is to ensure that $k_{i}$
is an integer. The parameter $t$ plays a similar role but for the
momentum in the $y\mbox{-}$direction. For periodic boundary conditions
then $k_{i}\in\mathbb{Z}$, which means that $\mathbb{T}_{i}+h$ must
be an integer. Since the offset on $\mathbb{T}_{i}$ is $\sum_{j\neq1}\frac{q}{2}=\left(N_{e}-1\right)\frac{q}{2}$
(the term $\tilde{T}_{ii}=0$ does not contribute) the same offset
has to apply for $h$. Expanding the counter weight term $e^{-\i\pi\tau N_{\phi}\sum_{i}k_{i}^{2}}$
gives 
\begin{eqnarray*}
e^{-\i\pi\tau\frac{1}{N_{\phi}}\sum_{i}k_{i}^{2}} & = & e^{-\i\pi\tau\frac{1}{N_{\phi}}\sum_{i}\mathbb{T}_{i}^{2}}e^{-\i\pi\tau\frac{1}{q}\left(mq+h\right)^{2}},
\end{eqnarray*}
 where we used (\ref{eq:T_Balance_contition}). Putting these terms
back gives

\begin{eqnarray}
\psi_{h,t} & = & \sum_{\left\{ \tilde{T}_{ij}\in\mathbb{Z}+\frac{q}{2}\right\} }\sum_{m}e^{\i\pi\tau\sum_{i<j}\frac{\tilde{T}_{ij}^{2}}{q}}e^{-\i\pi\tau\frac{1}{N_{\phi}}\sum_{i}\mathbb{T}_{i}^{2}}e^{\i\pi\sum_{i<j}\tilde{T}_{ij}}\nonumber \\
 &  & \times e^{\i2\pi\left(mt+\frac{ht}{q}\right)}\prod_{i<j}\tilde{Z}_{\tilde{T}_{ij}}^{\left(q\right)}\prod_{i=1}^{N_{e}}\zeta_{k_{i}}\left(z_{i}\right).\label{eq:psi_fock_laughlin}
\end{eqnarray}

Before we proceed let us stop and inspect the current state of $\psi_{h,t}$.
The expression (\ref{eq:psi_fock_laughlin}) already has some of the
desired structure. For instance we have managed to isolate the factors
$\zeta_{k_{i}}\left(z_{i}\right)$ which are the building blocks of
the basis states in \foueriersum. In order to get to \psiz\, though,
we need to perform parts of the sums over $\tilde{T}_{ij}$. This
is what we will do in the next section.

\subsection{Fock coefficient for the Laughlin state\label{sub:Expand-Laughlin}}

Here we give the extra steps needed to arrive at the Laughlin state
\zlaughlin. The remaining task is to make $k_{i}$ independent of
$m$ such that $k_{i}=\mathbb{T}_{i}+h$. 

Note that for periodic boundary conditions we mush choose $t$ and
$h$ as the half-integer $t=q\left(N_{e}-1\right)\frac{1}{2}$ and
$h=q\left(N_{e}-1\right)\frac{1}{2}+\mathbb{Z}$. This will work for
both fermions and bosons and in the latter case $h$ and $t$ would
always be integers.

We start by grouping (\ref{eq:psi_fock_laughlin}) as

\begin{eqnarray}
\psi_{h,t} & = & \sum_{\left\{ \mathbb{T}_{i}\in\mathbb{Z}+q\frac{N_{e}-1}{2}\right\} }\sum_{m}\mathcal{Z}\left(\mathbb{T}\right)e^{\i2\pi mt}\prod_{i=1}^{N_{e}}\zeta_{k_{i}}\left(z_{i}\right).\label{eq:psi_fock_Z}
\end{eqnarray}

where 
\begin{equation}
\mathcal{Z}\left(\mathbb{T}\right)=\left(\frac{2N_{\phi}\pi^{2}}{\tau_{2}}\right)^{\frac{N_{e}}{4}}\sum_{\left\{ \tilde{T}_{ij}\in\mathbb{Z}+\frac{q}{2}\right\} }e^{\i\pi\tau\frac{1}{q}\sum_{i<j}\tilde{T}_{ij}^{2}}e^{-\i\pi\tau\frac{1}{N_{\phi}}\sum_{i}\mathbb{T}_{i}^{2}}e^{\i\pi\sum_{i<j}\tilde{T}_{ij}}\prod_{i<j}\tilde{Z}_{\tilde{T}_{ij}}^{\left(q\right)}.\label{eq:Z_Laughlin_coeff}
\end{equation}

Here we have dropped an overall factor of $e^{\i2\pi\frac{ht}{q}}$.
This is the same $\mathcal{Z}\left(\mathbb{T}\right)$ that appears
in \zlaughlin, and the $\tilde{T}_{ij}$ are still subject to the
constraint (\ref{eq:bb_T}). By extracting $\mathcal{Z}\left(\mathbb{T}\right)$
we have formulated (\ref{eq:psi_fock_Z}) in a very suggestive form
reminiscent of \psiz, only that the summation is a little off. For
instance have the extra factor of $e^{\i2\pi mt}$, a sum over $m$
and the sum over $\mathbb{T}_{i}$ is infinite. We remove $e^{\i2\pi mt}$
by shifting all the $\mathbb{T}_{i}$ by $\mathbb{T}_{i}\to\mathbb{T}_{i}-qm$
except for $\mathbb{T}_{N_{e}}$ which by the constraint (\ref{eq:T_Balance_contition})
goes to $\mathbb{T}_{N_{e}}\to\mathbb{T}_{N_{e}}+\left(N_{e}-1\right)qm$.
This is performed in practice by shifting the sums over the $\tilde{T}_{ij}$
as $\tilde{T}_{ij}\to\tilde{T}_{ij}+qm\left(\delta_{i,N_{e}}-\delta_{j,N_{e}}\right)$
. Under this transformation the momenta change as
\[
k_{i}\to\mathbb{T}_{i}+h+\delta_{i,N_{e}}N_{\phi}m
\]
 and $\mathcal{Z}\left(\mathbb{T}\right)\to\mathcal{Z}\left(\mathbb{T}\right)e^{i\pi\left(N_{e}-1\right)qm}$.
This extra phase extracted will exactly cancel the factor $e^{\i2\pi mt}$
present in (\ref{eq:psi_fock_Z}). We also split $\mathbb{T}_{i}$
as $\mathbb{T}_{i}\to\mathbb{T}_{i}+r_{i}N_{\phi}$ where now $\mathbb{T}_{i}\in\mathbb{Z}_{N_{\phi}}$
and $r_{i}\in\mathbb{Z}$, such that $r_{i}$ also obeys the balance
condition $\sum_{i}r_{i}=0$. It can be shown that $\mathcal{Z}\left(\mathbb{T}_{i}+r_{i}N_{\phi}\right)=\mathcal{Z}\left(\mathbb{T}_{i}\right)$
is independent of $r_{i}$. Finally shifting $m\to m-r_{N_{e}}$ enables
us to complete the sum \foueriersum\, to obtain the expression \psiz,
where $k_{i}=\mathbb{T}_{i}+h\,\mathrm{mod}\,N_{\phi}$. It can be
shown, as a sanity check, that $\mathcal{Z}\left(\mathbb{T}\right)$
is fully anti-symmetric in th interchange of its variables $\mathbb{T}_{i}$.

\section{A Recursion Formula for \textmd{\normalsize{}$\mathcal{Z}$\label{sec:Reccusion}}}

In section \ref{sec:Laughlin-Expansion-1} we showed that the Laughlin
state \wfnlaughlin could be expanded in a Fock basis with coefficients
$\mathcal{Z}\left(\mathbb{T}\right)$ given by \zlaughlin. Unfortunately
the given expression is not particularly helpful when it comes to
numerical evaluation. The culprit is the simultaneous infinite sums
over all of the $\tilde{T}_{ij}$:s. To alleviate this problem a bit
we will in this section rewrite $\mathcal{Z}\left(\mathbb{T}\right)$
such that the sums can be performed in a recursive manner. This will
substantially reduce the scaling of the computation.

This section is divided into three parts. The first part, \ref{sub:Recursion_Z},
deals with rewriting $\mathcal{Z}\left(\mathbb{T}\right)$ for $N_{e}$
particles such that it depends recursively on the components for $N_{e}-1$
particles. The remaining two parts, further manipulates the expression
for $\mathcal{Z}\left(\mathbb{T}\right)$ for enhanced numerical efficiency.

A word of caution should be given. Although the final expression (\ref{eq:Z_recursive_formulation_final})
can be put on a computer, the efficiency in evaluating it is still
inferior to that of simply diagonalizing the Haldane pseudo-potential
Hamiltonian. Using exact diagonalization the coefficients for around
12 particles can be extracted numerically, but in our current implementation
of (\ref{eq:Z_recursive_formulation_final}), only 6 or so particles
can be achieved. The reader looking at equations (\ref{eq:Z_recursive_formulation_final})
and (\ref{eq:Lambda_form}) may be worried that they are wrong or
will render errors when implemented numerically. We have however validated
our code against the exactly diagonalized Haldane pseudo-potential
Hamiltonian and have perfect agreement for all tested cases. 

With the given word of caution in mind, the reader should also be
aware that the pseudo-potential trick only exists for the Laughlin
state. For states higher in the hierarchy there is no local Hamiltonian
for which these state are the exact zero energy eigenstates. In these
cases the results in this article is the only way we are aware of
to analytically extract Fock coefficients.

\subsection{Recursion in $N_{e}$\label{sub:Recursion_Z}}

In this part, we write $\mathcal{Z}\left(\mathbb{T}\right)$ in a
recursive formulation such that the coefficients for $N$ particles
depends on the coefficients for $N-1$ particles. For analytical purposes
the Gaussian exponential piece of $\mathcal{Z}\left(\mathbb{T}\right)$
can be written in different ways as

\[
A\left(N\right)=\sum_{i<j}\tilde{T}_{ij}^{2}-\frac{1}{N}\sum_{i}\mathbb{T}_{i}^{2}=\frac{1}{N}\sum_{i<j<k}\left(\tilde{T}_{ij}+\tilde{T}_{jk}+\tilde{T}_{ki}\right)^{2}=\sum_{i<j}^{N}\left(\tilde{T}_{ij}-\frac{\mathbb{T}_{i}-\mathbb{T}_{j}}{N}\right)^{2}.
\]

Especially the two last expression shows explicitly that $\mathcal{Z}\left(\mathbb{T}\right)$
is a converging sum, but they do not particularly help when numerically
evaluating $\mathcal{Z}\left(\mathbb{T}\right)$. In order to find
a numerically tractable formulation of $\mathcal{Z}\left(\mathbb{T}\right)$
we will develop a recursive construction in $N$. For this purpose
we introduce some more notation. We write $\mathcal{Z}_{\mathbb{T}}^{\left(N\right)}$
as the Fock coefficients for $N$ electrons. We also write $\mathbb{T}_{i}^{\left(N\right)}=\sum_{j=1}^{N}\tilde{T}_{ij}$
to keep track of which electrons are being included in the momentum
$\mathbb{T}_{i}^{\left(N\right)}$ at level $N$. This induces the
relation between different system sizes as

\begin{equation}
\tilde{T}_{iN}=\mathbb{T}_{i}^{\left(N\right)}-\mathbb{T}_{i}^{\left(N-1\right)}.\label{eq:Layer_separation}
\end{equation}
By extracting all the pieces from $\mathcal{Z}_{\mathbb{T}}^{\left(N\right)}$
that belong in $\mathcal{Z}_{\mathbb{T}}^{\left(N-1\right)}$ it is
possible to recursively formulate the former in terms of the latter.
This step is straight forward as any factors containing $\tilde{T}_{ij}$
where both $i,j<N$ can be moved to $\mathcal{Z}_{\mathbb{T}}^{\left(N-1\right)}$
immediately. The remaining pieces are $\tilde{T}_{iN}$ which can
be rewritten according to (\ref{eq:Layer_separation}) such that the
recursive form of $\mathcal{Z}_{\mathbb{T}}^{\left(N\right)}$ is

\begin{eqnarray}
\mathcal{Z}_{\mathbb{T}}^{\left(N\right)} & = & \mathcal{S}\left(N\right)\sum_{\left\{ \mathbb{T}_{i}^{\left(N-1\right)}\in\mathbb{Z}+q\frac{N-2}{2}\right\} }e^{\i\pi\tau q\Delta\left(N\right)}e^{\i\pi\sum_{i}^{N-1}\mathbb{T}_{i}^{\left(N\right)}}\mathcal{Z}_{\mathbb{T}}^{\left(N-1\right)}\cdot\prod_{i}^{N-1}Z_{\mathbb{T}_{i}^{\left(N\right)}-\mathbb{T}_{i}^{\left(N-1\right)}}^{\left(q\right)}.\label{eq:Z_recursive_formulation}
\end{eqnarray}
 In the above equation we have introduced the difference between the
Gaussian exponentials as
\begin{equation}
\Delta\left(N\right)=\frac{1}{q^{2}}A\left(N\right)-\frac{1}{q^{2}}A\left(N-1\right)=\left(N-1\right)N\sum_{i=1}^{N-1}\left(\frac{\mathbb{T}_{i}^{\left(N-1\right)}}{q\left(N-1\right)}-\frac{\mathbb{T}_{i}^{\left(N\right)}}{q\left(N\right)}\right)^{2}-\frac{\mathbb{T}_{N}^{\left(N\right)2}}{q^{2}N}.\label{eq:Delta_Ne}
\end{equation}
 The prefactor $\S\left(N\right)=\left(\frac{2q\pi^{2}}{\tau_{2}}\right)^{\frac{1}{4}}\frac{N^{\frac{N}{4}}}{\left(N-1\right)^{\frac{N-1}{4}}}$
is the quotient between the two factors $\left(\frac{2qN\pi^{2}}{\tau_{2}}\right)^{\frac{N}{4}}$
and $\left(\frac{2q\left(N-1\right)\pi^{2}}{\tau_{2}}\right)^{\frac{N-1}{4}}$.
Already this re-formulation has reduced the computational complexity
of $\mathcal{Z}^{\left(N\right)}$ from order $\mathcal{O}\left(e^{\frac{1}{2}N^{2}}\right)$,
since each pair of indexes had a sum $\tilde{T}_{ij}$, to a still
hard but more humble $\mathcal{O}\left(e^{N}\right)$. This is of
course provided that the results further down the recursion can be
stored and reused. To further facilitate the evaluation we seek a
formulation that as much as possible disentangles the different $\mathbb{T}_{i}^{\left(N-1\right)}$
sums from each other. This will the purpose of the next part.

\subsection{Orthogonalizing the sum in $\Delta\left(N\right)$\label{sub:Ortogonalizing_Z}}

With the form of $\Delta\left(N\right)$ in (\ref{eq:Delta_Ne}) we
can compute $\mathcal{Z}_{\mathbb{T}}^{\left(N\right)}$ recursively.
There are two computational problems that need to be remedied though.
The first is the form of $\Delta\left(N\right)$, as the different
sums over $\mathbb{T}_{j}^{\left(N-1\right)}$ are not separable,
\ie they contain cross terms $\mathbb{T}_{i}\mathbb{T}_{j}$. These
cross terms reduce the numerical efficiently as all the sums need
to be evaluated simultaneously. The second problem is that the sums
over $\mathbb{T}^{\left(N-1\right)}$ are infinite whereas $\mathcal{Z}_{\mathbb{T}}^{\left(N-1\right)}$
are defined modulo $q\left(N-1\right)$. We thus seek to rewrite $\Delta\left(N\right)$
such that this modularity is explicit. In what follows we will remedy
both of these problems in turn.

To reduce the notational complexity we define $\frac{\mathbb{T}_{i}^{\left(N-1\right)}}{q\left(N-1\right)}=T_{i}$
and $\frac{\mathbb{T}_{i}^{\left(N\right)}}{qN}=R_{i}$ such that
(\ref{eq:Delta_Ne}) can be written as

\begin{equation}
\pi\left(N\right)=\frac{\Delta\left(N\right)}{N\left(N-1\right)}=\sum_{i=1}^{N-1}\left(T_{i}-R_{i}\right)^{2}-\frac{R_{N}^{2}}{N-1}.\label{eq:Delta_Ne_reduced}
\end{equation}
 Note that $R_{i}$ and $T_{i}$ -- just as $\mathbb{T}_{i}^{\left(N\right)}$
-- are subject to neutrality conditions $\sum_{i}^{N}R_{i}=0$ and
$\sum_{i}^{N-1}T_{i}=0$ and herein lies the difficulty. The neutrality
condition means that (\ref{eq:Delta_Ne_reduced}) is not diagonal
in neither $R_{i}$ or $T_{i}$. The first order of business is to
rewrite (\ref{eq:Delta_Ne_reduced}) on an explicitly positive definite
form. The details of the rewriting are in \ref{App:Delta_N_e} and
the result is 

\begin{equation}
\pi\left(M+2\right)=\sum_{n=1}^{M}w_{n}\left(\sum_{j=1}^{M}v_{j}^{\left(n\right)}T_{j}-\frac{1}{\lambda_{n}}\sum_{j=1}^{M+1}v_{j}^{\left(n\right)}R_{j}\right)^{2}.\label{eq:eta_positive_definite}
\end{equation}

We chose to use $M$ here instead of $N-2$ to empathize that there
are $M$ independent variables $T_{j}$. The vectors $\nu^{\left(n\right)}$
are 

\begin{equation}
v_{j}^{\left(n\right)}=\left\{ \begin{array}{ll}
1 & j\leq n\\
-n & j=n+1\\
0 & j>n+1
\end{array}\right.,\label{eq:Eigenvectors-copy}
\end{equation}
 and are eigenvectors to the $M\times M$ matrix $\mathcal{M}_{ij}=1+\delta_{ij}$.
The matrix $\mathcal{M}_{ij}$ has eigenvalues $\lambda_{n}=1+M\delta_{n,M}$.
The squared norm of the vectors $v^{\left(n\right)}$ is $s_{n}=n+n^{2}\left(1-\delta_{M,n}\right)$
which enters (\ref{eq:eta_positive_definite}) as $w_{n}=\frac{\lambda_{n}}{s_{n}}=\frac{\left(1+M\delta_{n,M}\right)^{2}}{n\left(n+1\right)}$.

As (\ref{eq:eta_positive_definite}) stands, the sums over $T_{j}$
are still entangled. Note that there exists no integer shifts of the
$T_{i}$ that will diagonalize the sum. The simple argument for this
is because $\det\left(v_{j}^{\left(n\right)}\right)=M!$, which means
that the inverse of $v_{j}^{\left(n\right)}$ is not integer valued.
We note however, that the sum $\sum_{j=1}^{M}v_{j}^{\left(n\right)}T_{j}$
can be manipulated to the form $\left(1+n\right)T_{n}-nT_{n+1}$ ($n<M$)
or $T_{M}$ ($n=M$) by the simultaneous shifts $T_{j}\to T_{j}-T_{j-1}$
for $1<j\leq M$. We will use this fact in the next part to reduce
the complexity in evaluating $\mathcal{Z}_{\mathbb{T}}^{\left(N\right)}$.

\subsection{From $\pi\left(M+2\right)$ to $\vartheta\mbox{-}$functions}

In order to arrive at an efficient computation we must make contact
with the periodicity of $\left(N-1\right)q$ in $\mathcal{Z}_{\mathbb{T}}^{\left(N-1\right)}$.
This periodicity coincides with $T_{i}\to T_{i}+\mathbb{Z}$. For
this purpose we therefore extract exactly this piece from $T_{i}$
by splitting of the fractional pieces as $T_{i}\to T_{i}+\frac{\mathbb{T}_{i}^{\left(M+1\right)}}{q\left(M+1\right)}$.
Here $T_{i}\in\mathbb{Z}$ and $\mathbb{T}_{i}^{\left(M+1\right)}\in\mathbb{Z}_{q\left(M+1\right)}+\frac{qM}{2}$.
We proceed by simultaneously shifting $T_{j}\to T_{j}-T_{j-1}$ for
$1<j\leq M$. The positive definite squares are now

\begin{eqnarray}
\pi\left(M+2\right) & = & \sum_{n=1}^{M}w_{n}\left(\left(n+1\right)T_{n}-nT_{n+1}-\frac{D_{n}}{\lambda_{n}}\right)^{2}+w_{M}\left(T_{M}-\frac{D_{M}}{\lambda_{M}}\right)^{2},\label{eq:pi_reduced_T_T}
\end{eqnarray}

where $D_{n}$ is the combination of $R_{i}$ and $\mathbb{T}_{i}^{\left(M+1\right)}$
giving

\begin{equation}
D_{n}=-\sum_{j=1}^{M}v_{j}^{\left(n\right)}\frac{\lambda_{n}\mathbb{T}_{i}^{\left(M+1\right)}}{q\left(M+1\right)}+\sum_{i=1}^{M+1}v_{j}^{\left(n\right)}\frac{\mathbb{T}_{i}^{\left(M+2\right)}}{q\left(M+2\right)}.\label{eq:Def_of_D_n}
\end{equation}
 We further manipulate (\ref{eq:pi_reduced_T_T}) by separating $T_{n}$
as $T_{n}\to nl_{n}+q_{n}$, where $l_{n}\in\mathbb{Z}$ and $q_{n}\in\mathbb{Z}_{n}$.
For the term with $n=1$ this trivially becomes $T_{1}\to l_{1}$.
On top of this we remove the cross terms between $l_{n}$ and $l_{n+1}$
by shifting $l_{n}\to l_{n}+l_{n+1}$ starting from $n=1$ and going
upward to $n=M-1$. More details are given in \ref{app:Sum_to_Theta}.
The expression for $\pi\left(M+2\right)$ now becomes

\begin{eqnarray}
\pi\left(M+2\right) & = & \sum_{n=1}^{M}\left(n+1\right)n\left(l_{n}+\frac{q_{n}}{n}-\frac{q_{n+1}}{\left(n+1\right)}-\frac{D_{n}}{n\left(n+1\right)}\right)^{2},\label{eq:Recusrsive q_n}
\end{eqnarray}

where we have introduced the dummy index $q_{M+1}=0$ to write the
whole expression as one sum. The sums over $l_{n}$ can now be performed
to produce $\vartheta$-functions, such that the full recursive product
is 

\begin{eqnarray}
\mathcal{Z}_{\mathbb{T}}^{\left(N\right)} & = & \S\left(N\right)\sum_{\left\{ \mathbb{T}_{i}^{\left(N-1\right)}\in\mathbb{Z}_{q\left(N-1\right)}+\frac{q}{2}\left(N-2\right)\right\} }\Lambda\left(\mathbb{T}^{\left(N\right)},\mathbb{T}^{\left(N-1\right)}\right)\label{eq:Z_recursive_formulation_final}\\
 &  & \times e^{\i\pi\sum_{i}^{N-1}\mathbb{T}_{i}^{\left(N\right)}}\mathcal{Z}_{\mathbb{T}}^{\left(N-1\right)}\cdot\prod_{i}^{N-1}Z_{\mathbb{T}_{i}^{\left(N\right)}-\mathbb{T}_{i}^{\left(N-1\right)}}^{\left(q\right)}.\nonumber 
\end{eqnarray}
 The introduced weight is 
\begin{equation}
\Lambda\left(\mathbb{T}^{\left(N\right)},\mathbb{T}^{\left(N-1\right)}\right)=\sum_{\left\{ q_{n}\in\mathbb{Z}_{n}\right\} }^{N-2}\prod_{n=1}^{N-2}\ellipticgeneralized{\frac{q_{n}}{n}-\frac{q_{n+1}}{n+1}-\frac{D_{n}}{n\left(n+1\right)}}00{\left(N-1\right)N\left(n+1\right)nq\tau}.\label{eq:Lambda_form}
\end{equation}

The $\vartheta\mbox{-}$factors in (\ref{eq:Lambda_form}) converge
fast since $\exp\left(\i\pi\left(N-1\right)N\left(n+1\right)nq\tau\right)$
will usually be a small number. Also the sum over $\sum_{\left\{ q_{n}\in\mathbb{Z}_{n}\right\} }^{N-2}$
looks naively as is contains a number of terms which will scale as
$\left(N-2\right)!$. However, because of the structure $\frac{q_{n}}{n}-\frac{q_{n+1}}{n+1}$
it can be decomposed into $\sum_{n=1}^{N-1}n\left(n+1\right)=\mathcal{O}\left(N^{3}\right)$
evaluations of $\vartheta\mbox{-}$functions.

\section{Fock Expansion of the Hierarchy States\label{sec:Hierachy-expand-1}}

In this part we give the steps to reach \ZHH. The construction will
parallel the construction for the Laughlin states albeit with more
structure. We will here only point out the main differences and display
important steps in the expansion.

\subsection{Fourier expanding the hierarchy\label{sub:Expand-HH}}

Just as for the Laughlin state we begin with the Jastrow factors.
Again we use \thetaexp to expand the $\vartheta_{1}\mbox{-}$functions.
The expansion reads 

\begin{eqnarray}
\prod_{i<j}\elliptic 1{z_{ij}}{\tau}^{K_{ij}} & = & \sum_{\left\{ \tilde{T}_{ij}\in\mathbb{Z}+\frac{K_{ij}}{2}\right\} }e^{\i\pi\tau\sum_{i<j}\frac{\tilde{T}_{ij}^{2}}{K_{ij}}}e^{\i\pi\sum_{i<j}\tilde{T}_{ij}}e^{\i2\pi\sum_{i}\mathbb{T}_{i}\frac{z_{i}}{L}}\prod_{i<j}\tilde{Z}_{\tilde{T}_{ij}}^{\left(K_{ij}\right)},\label{eq:Jastrow_expansion}
\end{eqnarray}

where just as before $\tilde{T}_{ij}=-\tilde{T}_{ji}$. Note that
since the pair $z_{ij}$ now comes with exponent $K_{ij}$, there
are different types of $\tilde{Z}_{\tilde{T}_{ij}}^{\left(K_{ij}\right)}$
present. However, they still obey the same equations $\tilde{Z}_{\tilde{T}_{ij}}^{\left(K_{ij}\right)}=\tilde{Z}_{\tilde{T}_{ij}-K_{ij}}^{\left(K_{ij}\right)}=Z_{T_{ij}}^{\left(K_{ij}\right)}$
and $\tilde{Z}_{\tilde{T}_{ij}}^{\left(K_{ij}\right)}=\tilde{Z}_{\tilde{T}_{ji}}^{\left(K_{ij}\right)}$.
Note that $\tilde{Z}_{\tilde{T}_{ji}}^{\left(K_{ij}\right)}$ has
its maximum TT-value at $\tilde{Z}_{\frac{K_{ij}}{2}}^{\left(K_{ij}\right)}$.
Similarly as before, we have also introduced the variable $\mathbb{T}_{i}=\sum_{j=1}^{N_{e}}\tilde{T}_{ij},$
in accordance with (\ref{eq:bb_T}). The balance condition on $\mathbb{T}_{i}$
is the same as earlier $\sum_{i}\mathbb{T}_{i}=0.$ The balance conditions
(\ref{eq:T_Balance_contition}) ensures that the wave function will
be in a well defined momentum sector.

We may in a similar fashion as for the Jastrow factor also express
the CoM function in the relevant Fourier components. This part is
more complicated compared to Laughlin case there is more structure
in the CoM term \CFTCoM. 

For this purpose we parametrize the charge lattice vector $\mathbf{q}$
as $\mathbf{q}=\sum_{\alpha}m_{\alpha}\mathbf{q}_{\alpha}$ where
$\left\{ \mathbf{q}_{\alpha}\right\} $ span the charge lattice $\Gamma$.
Since we are only considering the abelian chiral hierarchy it suffices
to use the one-dimensional parameterization $\mathbf{h}=h\mathbf{h}_{0}$
and $\mathbf{t}=t\mathbf{h}_{0}$. At the end of the calculation we
can set $h=t=s$ and recover the result in \psiz\, and \ZHH. Here
$\mathbf{h}_{0}=\frac{\mathbf{Q}}{N_{\phi}}$ has the property $\mathbf{h}_{0}\cdot\mathbf{q}_{\alpha}=1$
and $\mathbf{h}_{0}^{2}=\frac{N_{e}}{N_{\phi}}=\frac{p}{q}$. This
parametrization is generic enough to capture all the states in the
abelian chiral hierarchy. Note that $q\mathbf{h}_{0}\in\Gamma$ which
means that all $q$ degenerate states at $\nu=\frac{p}{q}$ are obtained
by letting $s$ take values from 1 to $q$. For states with higher
degeneracy than $q$, such as the Halperin 331-state with $K=\left(\begin{array}{cc}
3 & 1\\
1 & 3
\end{array}\right)$, a more elaborate parametrization will be necessary. For more detailed
properties of $\mathbf{h}_{0}$ and $\Gamma$ see \eg Appendix B
of Ref. \cite{Fremling_2014}. 

Using this parametrization the center off mass piece \CFTCoM\, can
after expansion be written as

\begin{eqnarray}
\mathcal{F}_{h,t}\left(\left\{ z\right\} ,\tau\right) & = & \sum_{\left\{ m_{\alpha}\right\} }e^{\i\pi\tau\left[\sum_{\alpha,\beta}m_{\beta}m_{\alpha}K_{\alpha\beta}+2\sum_{\alpha}m_{\alpha}h+h^{2}\frac{N_{e}}{N_{\phi}}\right]}\label{eq:CoM_eqpansion}\\
 &  & \times e^{\i2\pi\left[\sum_{i}\sum_{\alpha}m_{\alpha}K_{\alpha i}\frac{z_{i}}{L}+\sum_{\alpha}m_{\alpha}t+\sum_{i}h\frac{z_{i}}{L}+ht\frac{N_{e}}{N_{\phi}}\right]}.\nonumber 
\end{eqnarray}

Again, when we combine the expansions (\ref{eq:Jastrow_expansion}),
(\ref{eq:CoM_eqpansion}) and the Gaussian we may isolate the single
particle wave function building blocks \fouerierterms. We substitute
$e^{\i2\pi k_{i}\frac{z_{i}}{L}}$ for $\zeta_{k_{i}}\left(z_{i}\right)$,
but this time there is more structure in the momentum relation

\begin{equation}
k_{i}=\mathbb{T}_{i}+\sum_{\alpha}m_{\alpha}K_{\alpha i}+h.\label{eq:momentum_TT_relations}
\end{equation}

As earlier we see that both $h$ ensures that $k_{i}$ is an integer.
For periodic boundary conditions then $k_{i}\in\mathbb{Z}$, which
means that $\mathbb{T}_{i}+h$ must be an integer. Since the offset
on $\mathbb{T}_{i}$ is $\sum_{j\neq i}\frac{K_{ij}}{2}=\frac{N_{\phi}-K_{ii}}{2}$
(the term $\tilde{T}_{ii}=0$ does not contribute) the same offset
has to apply for $h$. We are here assuming that all $K_{\alpha\alpha}$
have the same parity, a reasonable assumption equivalent to demanding
that either all particles are bosons or all particles are fermions.

Substituting for $\zeta_{k_{i}}\left(z_{i}\right)$ and expanding
the counter weight term $e^{-\i\pi\tau N_{\phi}\sum_{i}k_{i}^{2}}$
using (\ref{eq:T_Balance_contition}) and $\mathbf{Q}\cdot\mathbf{q}_{j}=N_{\phi}$
the expansion reads
\begin{eqnarray}
\psi_{h,t} & = & \sum_{\left\{ \tilde{T}_{ij}\in\mathbb{Z}+\frac{K_{ij}}{2}\right\} }\sum_{\left\{ m_{\alpha}\right\} }e^{\i\pi\tau\sum_{\alpha,\beta}m_{\beta}m_{\alpha}K_{\alpha\beta}}\nonumber \\
 &  & \times e^{\i\pi\tau\sum_{i<j}\frac{\tilde{T}_{ij}^{2}}{K_{ij}}}e^{-\i\pi\tau\frac{1}{N_{\phi}}\sum_{i}\left(\mathbb{T}_{i}+\sum_{\alpha}m_{\alpha}K_{\alpha i}\right)^{2}}\nonumber \\
 &  & \times e^{\i2\pi ht\frac{N_{e}}{N_{\phi}}}e^{\i2\pi\sum_{\alpha}m_{\alpha}t}e^{\i\pi\sum_{i<j}\tilde{T}_{ij}}\prod_{i<j}\tilde{Z}_{\tilde{T}_{ij}}^{\left(K_{ij}\right)}\prod_{i=1}^{N_{e}}\zeta_{k_{i}}\left(z_{i}\right).\label{eq:psi_fock_completed}
\end{eqnarray}

In this expression we now have extracted the $\zeta_{k_{i}}$ for
the individual particles. Not also that $h$ no longer is a parameter
in the expansion. This is to be expected as the only difference between
the $q$ different states is the labeling of the Fock coefficients.

\subsection{Resummation of $\tilde{T}_{ij}$}

We now continue and perform some manipulations on the sums of $\tilde{T}_{ij}$.
The aim of these manipulations is to collect the $m_{\alpha}$ in
$k_{i}$ on specific particles. We note that $\sum_{i}k_{i}=N_{\phi}\sum_{\alpha}m_{\alpha}+N_{e}h$
such that the $m_{\alpha}$ could be used to complete the construction
of the basis functions $\eta_{k}\left(z\right)$. It can be shown
that the shifts 

\begin{eqnarray}
\tilde{T}_{ij} & \to & \tilde{T}_{ij}+K_{ij}\left(S_{i}-S_{j}\right)\label{eq:Shift_transformation}\\
m_{\alpha} & \to & m_{\alpha}+\sum_{i_{\alpha}\in I_{\alpha}}S_{i_{\alpha}},\nonumber 
\end{eqnarray}
 will send $k_{i}\to k_{i}+N_{\phi}S_{i}$ while leaving the rest
of (\ref{eq:psi_fock_completed}) invariant. This shows that it is
in principle possible to construct the $\eta_{k}\left(z\right)$ for
all the particles by shifting the sums in appropriate ways.

We will however here choose a slightly different shift on the summation.
For this purpose we shift the $\tilde{T}_{ij}$ as

\[
\tilde{T}_{ij}\to\tilde{T}_{ij}+K_{ij}\sum_{\alpha}m_{\alpha}\left(\delta_{i,N_{\alpha}}-\delta_{j,N_{\alpha}}\right).
\]
 This transformation only affects the indices for the ``last'' particle
in each group, $N_{\alpha}$\footnote[1]{Note here that we abuse
notation and let $N_{\alpha}$ be the index of the last particle in
group $\alpha$ instead of the size of group $\alpha$.}, and has
the effect that $k_{i}\to k_{i}=\mathbb{T}_{i}+N_{\phi}\sum_{\alpha}m_{\alpha}\delta_{i,N_{\alpha}}+h$.
In effect it puts all the $m_{\alpha}$ on only one particle per group.
This transformation does nothing to $\tilde{Z}_{\tilde{T}_{ij}}^{\left(K_{ij}\right)}$
as it is periodic under $\tilde{T}_{ij}\to\tilde{T}_{ij}+K_{ij}$.

An extra phase of $\exp\left(\i\pi\sum_{\alpha}m_{\alpha}\left(N_{\phi}-K_{\alpha\alpha}\right)\right)$
is picked up which precisely cancels the existing phase $\exp\left(\i\pi\sum_{\alpha}m_{\alpha}t\right)$.
There are extra factors picked up from $e^{\i\pi\tau\sum_{i<j}\frac{\tilde{T}_{ij}^{2}}{K_{ij}}}$
and $e^{-\i\pi\tau\frac{1}{N_{\phi}}\sum_{i}\left(\mathbb{T}_{i}+\sum_{\alpha}m_{\alpha}K_{\alpha i}\right)^{2}}$
which come together in such a ways are to precisely cancel all $m_{\alpha}$
dependence in the Fock weight. Taking these things together we have
the expression

\begin{eqnarray}
\psi_{h,t} & = & \sum_{\left\{ \tilde{T}_{ij}\in\mathbb{Z}+\frac{K_{ij}}{2}\right\} }e^{\i\pi\tau\sum_{i<j}\frac{\tilde{T}_{ij}^{2}}{K_{ij}}}e^{-\i\pi\tau\frac{1}{N_{\phi}}\sum_{i}\mathbb{T}_{i}^{2}}\nonumber \\
 &  & \times e^{\i\pi\sum_{i<j}\tilde{T}_{ij}}\prod_{i<j}\tilde{Z}_{\tilde{T}_{ij}}^{\left(K_{ij}\right)}\sum_{\left\{ m_{\alpha}\right\} }\prod_{i=1}^{N_{e}}\zeta_{k_{i}}\left(z_{i}\right),\label{eq:psi_fock_reduced}
\end{eqnarray}
 where we have also dropped the constant phase $e^{\i2\pi ht\frac{N_{e}}{N_{\phi}}}$.
Note that as $m_{\alpha}$ is not part of the weight, the sums over
$m_{\alpha}$ can be performed to produce $n$ of the $N_{e}$ $\eta_{k}\left(z\right)$
needed to construct the Fock coefficients. In coming this far we can
now identify the factor making up $\mathcal{Z}$ in \ZHH, and the
rest is \HHexp given in the main text.

\section{From (\ref{eq:Delta_Ne_reduced}) to (\ref{eq:eta_positive_definite})\label{App:Delta_N_e}}

In this appendix we give a detailed derivation leading to equation
(\ref{eq:eta_positive_definite}) in the main text. We start from
equation (\ref{eq:Delta_Ne_reduced}) in the main text: 
\begin{equation}
\pi\left(N\right)=\sum_{i=1}^{N-1}\left(T_{i}-R_{i}\right)^{2}-\frac{R_{N}^{2}}{N-1}.\label{eqapp:Delta_Ne_reduced}
\end{equation}
 Note that $R_{i}$ and $T_{i}$ also are subject to neutrality conditions
$\sum_{i}^{N}R_{i}=0$ and $\sum_{i}^{N-1}T_{i}$, which means that
(\ref{eqapp:Delta_Ne_reduced}) is not diagonal in neither $R_{i}$
or $T_{i}$. We begin by rewriting (\ref{eqapp:Delta_Ne_reduced})
on an explicitly positive definite form with only $N-2$ terms. Since
the number of linearly independent variables $T_{i}$ variables are
$N-1$ we also define $M=N-2$ to keep track of this number.

We proceed in two steps. First we set set $R_{i}=0$ and diagonalize
only $T_{i}$. After that we insert $R_{i}$ again and deduce the
full expression.

\subsection{Simplification $R_{i}=0$}

We start by defining the simplified 

\begin{equation}
\pi_{0}\left(M+2\right)=\left.\pi\left(M+2\right)\right|_{R_{i}=0}=\sum_{i=1}^{M+1}T_{i}^{2}=\sum_{i=1}^{M}T_{i}^{2}+\left(\sum_{i=1}^{M}T_{i}\right)^{2},\label{eq:eta_M_reduced}
\end{equation}
where only the $T_{i}$ are present. In matrix form this would correspond
to 
\begin{equation}
\pi_{0}\left(M+2\right)=\sum_{i,j=1}^{M}T_{j}\mathcal{M}_{ij}T_{j},\label{eq:Matrix_formulation}
\end{equation}
 with the matrix $\mathcal{M}_{ij}=1+\delta_{ij}$. The eigenvalues
of that matrix are $\lambda_{n}=1+M\delta_{M,n}$. The eigenvectors
can be constructed by noting that $v^{\left(M\right)}=1$ is the eigenvector
to $\lambda_{M}$ with eigenvalue $M+1$. The rest of the vectors
form an degenerate space where all vectors have eigenvalue $1$. The
full space of eigenvectors can therefore be parametrized as 
\begin{equation}
v_{j}^{\left(n\right)}=\left\{ \begin{array}{ll}
1 & j\leq n\\
-n & j=n+1\\
0 & j>n+1
\end{array}\right..\label{eq:Eigenvectors}
\end{equation}
 The squared norm of these states are 
\begin{eqnarray*}
s_{n} & = & n+n^{2}\left(1-\delta_{M,n}\right).
\end{eqnarray*}
 In terms of these $\pi_{0}\left(M+2\right)$ can be rewritten as
\begin{equation}
\pi_{0}\left(M+2\right)=\sum_{n=1}^{M}w_{n}\left(\sum_{j=1}^{M}v_{j}^{\left(n\right)}T_{j}\right)^{2}.\label{eq:Positive squares}
\end{equation}
 where $w_{n}=\frac{\lambda_{n}}{s_{n}}=\frac{\left(1+M\delta_{n,M}\right)^{2}}{n\left(n+1\right)}$.
We also remind ourselves of the special case $M=1$ in which $\pi_{0}\left(M+2\right)=2T_{1}^{2}$.

\subsection{Restoring the factors of $R_{i}$}

We now restore the parameters $R_{i}$. We can \emph{not }however
put them back by simply shifting $T_{i}\to T_{i}-R_{i}$, as the term
$\left(\sum_{i=1}^{M}T_{i}\right)^{2}$ would be handled in the wrong
way. Instead, we deduce the values of these terms by studying the
cross terms $T_{i}R_{j}$. Let's take a look again at $\pi\left(M+2\right)$
in (\ref{eqapp:Delta_Ne_reduced}) but this time writing out all the
linearly independent terms of both $T_{i}$ and $R_{i}$:

\begin{equation}
\pi\left(M+2\right)=\sum_{i=1}^{M}\left(T_{i}-R_{i}\right)^{2}+\left(\sum_{i=1}^{M}T_{i}+R_{M+1}\right)^{2}-\frac{\left(\sum_{i=1}^{M+1}R_{i}\right)^{2}}{M+1}.\label{eq:eta_M_R_resored}
\end{equation}

To match this we make the ansatz that all of the $R_{i}$ factors
can be accommodated in the positive definite squares as 
\begin{equation}
\pi\left(M+2\right)=\sum_{n=1}^{M}w_{n}\left(\sum_{j=1}^{M}v_{j}^{\left(n\right)}T_{j}-A_{n}\right)^{2}.\label{eq:Positive_square_ansatz}
\end{equation}

We here assume that 
\[
A_{n}=\sum_{j=1}^{M+1}a_{j}^{\left(n\right)}R_{j},
\]
 is some linear combination of $R_{j}$. By comparing the two expressions
for (\ref{eq:eta_M_R_resored}) and (\ref{eq:Positive_square_ansatz})
we deduce the coefficients $a_{j}^{\left(n\right)}.$ By expanding
the squares in (\ref{eq:eta_M_R_resored}) we have 

\begin{equation}
\pi\left(M+2\right)=\pi_{0}\left(M+2\right)+2\sum_{i=1}^{M}T_{i}\left(R_{M+1}-R_{i}\right)+\sum_{i=1}^{M+1}R_{i}^{2}-\frac{\left(\sum_{i=1}^{M+1}R_{i}\right)^{2}}{M+1}.\label{eq:eta_M-R_resored_expanded}
\end{equation}

Doing the same expansion for (\ref{eq:Positive_square_ansatz}) gives 

\begin{equation}
\pi\left(M+2\right)=\pi_{0}\left(M+2\right)-2\sum_{n=1}^{M}w_{n}A_{n}\sum_{j=1}^{M}v_{j}^{\left(n\right)}T_{j}+\sum_{n=1}^{M}w_{n}A_{n}^{2}.\label{eq:Positive_quare_ansatz_expanded}
\end{equation}
 As the two expressions for $\pi\left(M+2\right)$ should be equal
irrespective of the values of $T_{i}$ we get equations for $A_{n}$
that are 
\begin{eqnarray}
\sum_{n=1}^{M}A_{n}v_{i}^{\left(n\right)}w_{n} & = & R_{i}-R_{M+1}\label{eq:Equations for A_i}\\
\sum_{n=1}^{M}w_{n}A_{n}^{2} & = & \sum_{i=1}^{M+1}R_{i}^{2}-\frac{\left(\sum_{i=1}^{M+1}R_{i}\right)^{2}}{M+1}.\label{eq:Equation_for_A_i_sum}
\end{eqnarray}
 The upper line comes from the linear terms in $T_{i}$ and the lower
equation is the constant piece. Equation (\ref{eq:Equations for A_i})
is really a system system of equations $M$ unknowns and $M$ equations.
The solutions are

\begin{eqnarray*}
A_{n} & = & \frac{1}{\lambda_{n}}\sum_{i=1}^{M+1}v_{i}^{\left(n\right)}R_{i}
\end{eqnarray*}

Then, putting $A_{n}$ back into the ansatz we arrive at we expression

\[
\pi\left(M+2\right)=\sum_{n=1}^{M}w_{n}\left(\sum_{j=1}^{M}v_{j}^{\left(n\right)}T_{j}-\frac{1}{\lambda_{n}}\sum_{i=1}^{M+1}v_{i}^{\left(n\right)}R_{i}\right)^{2},
\]
 with is expression (\ref{eq:eta_positive_definite}) in the main
text.

\section{From (\ref{eq:pi_reduced_T_T}) to (\ref{eq:Recusrsive q_n})\label{app:Sum_to_Theta}}

In this section we give the detail of how to reach equation (\ref{eq:Recusrsive q_n})
from (\ref{eq:pi_reduced_T_T}). We may rewrite it slightly such that
it is

\begin{eqnarray}
\pi\left(M+2\right) & = & \sum_{n=1}^{M}\frac{1}{n\left(n+1\right)}\left(\left(n+1\right)T_{n}-nT_{n+1}\delta_{n\neq M}-D_{n}\right)^{2}.\label{eqapp:Explicit squares_massage}
\end{eqnarray}
In the expression above we have abused notation and wrote $\delta_{n\neq M}=1-\delta_{n,M}$.
Note that each squared term only contains two $T_{i}$ terms, but
that they come with different prefactors. As the context of this equation
in insensitive to integer changes in $T_{i}$, we may shift the summation
at will without needing to consider other external factors. We proceed
by splitting sums over $T_{n}$ in two pieces at$T_{n}=nl_{n}+q_{n}$
where $q_{n}\in\mathbb{Z}_{n}$ and $l_{n}\in\mathbb{Z}$. For $n=1$
this trivially means $T_{1}=l_{1}$. Doing this results in 

\[
\pi\left(M+2\right)=\sum_{n=1}^{M}\frac{1}{n\left(n+1\right)}\left(\left(n+1\right)n\left(l_{n}-l_{n+1}\delta_{n\neq M}\right)+\left(\left(n+1\right)q_{n}-nq_{n+1}\delta_{n\neq M}\right)-D_{n}\right)^{2}.
\]

The facts that are summed to infinity, \ie $l_{n}$ now come with
the same prefactor $n\left(n+1\right)$. We can thus eliminate one
of the terms by shifting $l_{n}\to l_{n}+l_{n+1}$ starting from $n=1$
and going upward to $n=M-1$. This will yield the expression

\[
\pi\left(M+2\right)=\sum_{n=1}^{M}\left(n+1\right)n\left(l_{n}+\frac{q_{n}}{n}-\frac{q_{n+1}}{n+1}\delta_{n\neq M}-\frac{D_{n}}{n\left(n+1\right)}\right)^{2}.
\]
 By treating $q_{M+1}$ as a dummy index $q_{M+1}=0$ we can remove
the $\delta_{n\neq M}$ and obtain the homogeneous looking expression

\begin{eqnarray}
\pi\left(M\right) & = & \sum_{n=1}^{M}\left(n+1\right)n\left(l_{n}+\frac{q_{n}}{n}-\frac{q_{n+1}}{\left(n+1\right)}-\frac{D_{n}}{n\left(n+1\right)}\right)^{2}\label{eqapp:Recusrsive q_n}
\end{eqnarray}
 which is (\ref{eq:Recusrsive q_n}) in the main text.
\end{document}